\documentclass{jfm}
\usepackage{newtxtext}
\usepackage{amsmath,bm}
\usepackage{newtxmath}
\usepackage{graphicx}
\graphicspath{{figures/}}
\usepackage{tikz}
\usepackage{natbib}
\usepackage{hyperref}
\hypersetup{
	colorlinks = true,
	urlcolor   = blue,
	citecolor  = blue,
}
\usepackage{lineno}

\usepackage{booktabs}
\usepackage{subfigure}
\usepackage{multirow}
\usepackage{array}
\newcolumntype{P}[1]{>{\centering\arraybackslash}p{#1}}

\title[Instabilities in flow through and around a circular array of cylinders]
{Instabilities in flow through and around a circular array of cylinders}

\author[H. Zhang, Y. Yang, G. Wang, and M. Zhang]%
{
Huaibao Zhang$^1$, Yongliang Yang$^2$, Guangxue Wang$^1$ \and Mengqi Zhang$^{3,*}$
}
\affiliation{
	$^1$School of Aeronautics and Astronautics, Sun Yat-sen University, Shenzhen, 518107, PR China \\
	$^2$School of Mechanical Engineering, Nanjing University of Science and Technology,
Nanjing 210094, PR China     \\
	$^3$Department of Mechanical Engineering, National University of Singapore, 9 Engineering Drive 1, 117575 Singapore 
	\\[\affilskip]
}

\date{\today}

\begin{document}
\maketitle

\begin{abstract}
This paper presents results of two-dimensional direct numerical simulations (DNS) and global linear stability analyses (based on mean flow and base flow) of a viscous incompressible flow past a circular array of cylinders with six-fold rotational symmetry. 
Six cylinder arrays, with varied patch density $\phi = N_c (d/D)^2$ (with $N_c$ cylinders of diameter $d$ within a patch of diameter $D$) is investigated by adjusting the $N_c$ and its arrangement. The simulations cover a wide parameter space, with $\phi$ ranging from $0.043$ to $0.315$, and the free-stream flow Reynolds numbers ($Re_D  = U_{\infty} D / \nu = < 300$ based on $D$ and the uniform incoming velocity $U_{\infty}$ and the kinematic viscosity $\nu$).
We focus on the onset of vortex shedding with variant $\phi$, since the onset of global instabilities in such arrays has not been discussed earlier in the literature.

For the patch diameters and solid volume fractions considered here,three distinct regimes are identified: (I) a low-$\phi$ regime where cylinders behave nearly independently, the flow is stable, forming steady wake without vortex street; 
(II) an intermediate-$\phi$ regime where $Re_c$ varies logarithmically with $\phi$, resembling a porous medium; and (III) a high-$\phi$ regime where $Re_c$ approaches that of a solid cylinder ($\phi = 1$).

Then we talk about the   wave maker and the structural sensitivity analysis.... which reveals that the core instability mechanism concentrates in the array wake, and the shear layer of the array.


\end{abstract}


\section{Introduction}
Flow through and around an array of rigid cylinders is a problem of broad relevance in engineering and environmental fluid mechanics, with applications ranging from offshore structures and heat exchangers to wind turbines and vegetated canopies (\cite{Zong_Nepf_2012,Zhou_Venayagamoorthy_2019}).
Such configurations exhibit complex interactions between individual cylinder wakes and the collective flow around the array, leading to regimes that transition from isolated bluff-body behavior to solid-body-like dynamics as the solid fraction $\phi$ increases.
Previous studies on a circular array of rigid cylinders have largely focused on force statistics, time-averaged flow features, and turbulence structure, under steady or unsteady incident flow on moderate to high $Re_D$ based on the diameter of the cylinder array $D$.
However, the flow dynamics and the onset of instabilities at low $Re_D$ have not been explored. 
In the following, we will investigate the fluid dynamics and the onset of instabilities
for a viscous incompressible flow past a two-dimensional circular array of cylinders with its axis perpendicular to the incoming flow, by employing direct numerical simulations (DNS) and global linear stability analyses based on mean flow and base flow.

\subsection{Flow through and around a circular array of rigid cylinders}

A few previous studies have used a group of cylinders to model a finite porous body and investigated the flow through and around the group since the it provides a fundamental perspective to understand the of flow through porous patches of more complex shape or the dispersion and mixing processes in a generic porous media(~\cite{Woods_2025}).

isolated bodies can be placed in staggered, square-grid or random arrangements. (different shapes)
In heat exchanger, the array shape can be rectangular in 2-D(~\cite{tang2021effect,tang2022opposing}), and cube in 3-D, or even arbitrary(~\cite{EAMES_HUNT_BELCHER_2004}). 
The choice of circular in 2-D for the array shape is critical at high solid fractions since rectangular arrays potentially generate behaviour that is not typical of
many industrial problems. For this reason we chose a circular array with a structured
placement of circular cylinders, as used by~\cite{NICOLLE2011,Zong_Nepf_2012, Chang_Constantinescu_2015, Taddei_Manes_Ganapathisubramani_2016, Zhou_Venayagamoorthy_2019, klettner2019effect}.
The study selects cylinders as the constituent components of the arrays, since it has been intensively investigated and the flow characteristics are well-understood (~\cite{williamson1996vortex, Zdravkovich1998}).

\cite{NICOLLE2011} performed 2-D direct numerical simulations (DNS) at moderate $Re_D = UD/\nu =2100$ for a circular array of cylinders, where $U$ is the mean velocity of the flow impinging the patch, $D$ is the diameter of the patch and $\nu $ is the kinematic viscosity.
Depending on the canopy density, three regimes can be identified: for low $\phi (<0.05)$, cylinders shed independent wakes; for intermediate $\phi (0.05–0.15)$, a steady shear layer forms around the array; and for high $\phi (>0.15)$, the array behaves like a solid body.
The drag coefficient $C_D$ found in these simulations increases with increasing density and converges to the $C_D$ of a solid cylinder with the same diameter as the patch D for higher densities. Moreover, they report that for intermediate densities, the value of $C_D$ is constant. Similar results can be found in \cite{Chang_Constantinescu_2015}.

\cite{Chang_Constantinescu_2015}, using three-dimensional large-eddy simulations (LES) at $Re_D = 10,000$, found only two regimes—vortex shedding for $\phi > 0.1$ and its suppression for $\phi < 0.05$—with no intermediate steady regime, underscoring the importance of three-dimensional effects at higher Reynolds numbers.
In contrast to \cite{NICOLLE2011}, \cite{Chang_Constantinescu_2015} did not find a constant region for the CD at intermediate densities. In fact, the drag coefficient was found to increase monotonically with no intermediate plateau. Notwithstanding this discrepancy, the magnitude of $C_D$ they report was comparable with the values obtained in \cite{NICOLLE2011}.

The intermediate regime identified by \cite{NICOLLE2011} is also characterized experimentally by \cite{Zong_Nepf_2012}. Their study of a circular cylinder array in a shallow channel quantified the distinct \textit{steady wake} region that forms directly behind the patch when $0.05 < \Phi < 0.15$. 
The Reynolds number is $Re_D = 22000$ and $41000 $.
They showed that the length of this steady wake ($L_1$) increases with patch porosity and can be predicted by modeling the growth of the plane shear layers emanating from the patch edges. Furthermore, they demonstrated that the velocity within this steady wake ($U_1$) is a function of solid volume fraction ($\Phi$) only for their tested configurations, linking it to the interior velocity of an infinitely long porous patch. The formation of a patch-scale von Kármán vortex street is delayed until the end of this steady wake, and its strength diminishes as $\Phi$ decreases.

While many studies on porous patches have assumed uniform or shallow-channel flows—often modeling the patch as taller than the flow depth—real-world obstacles are frequently immersed within thick turbulent boundary layers, such as wind turbines in the atmosphere or vegetation in rivers, \cite{Taddei_Manes_Ganapathisubramani_2016} experimentally studied circular porous patches made of cylinders immersed in a turbulent boundary layer. They found that the drag coefficient $C_D$ increases with patch density $\phi$, reaching a value higher than that of a solid cylinder of the same size. This surprising result was explained by three flow mechanisms: lateral bleeding, which widens the wake; vertical bleeding, which shifts the top shear layer upward; and trailing-edge bleeding, which affects velocity recovery. Their work highlights that porous patches can generate more drag than solid obstacles when immersed in boundary layers, providing insight for modeling drag in environmental and engineering flows.

\cite{Zhou_Venayagamoorthy_2019} used large-eddy simulation to study a suspended cylindrical canopy patch in deep water. Their work varied patch density $\phi = N_c(d/D)^2$ and aspect ratio $AR = h/D$, revealing how three-dimensional flow redistribution—both local bleeding through the patch and global diversion around it—depends on patch geometry. As $\phi$ increases, lateral and vertical bleeding strengthen while streamwise bleeding weakens. Taller patches (larger $AR$) produce stronger vertical bleeding velocities, but divert a smaller fraction of entering flow downward, due to their larger lateral surface area. Bleeding enlarges the effective wake dimensions $h_e$ and $W_e$, altering the flow partition between horizontal and vertical bypass. The drag coefficient $C_D$ increases with $\phi$ to a plateau ($\phi \approx 0.3$–$0.5$) then drops sharply for the solid case ($\phi=1$), consistent with experiments by \cite{Taddei_Manes_Ganapathisubramani_2016}. Dense porous patches thus exert more drag than solid ones of equal frontal area. In contrast, $C_D$ increases monotonically with $AR$ due to weakened bottom shear layer entrainment and enhanced bleeding. This study highlights the previously overlooked role of $AR$ in the drag and flow adjustment of suspended porous obstacles.

However, in many environmental and industrial settings, flows are inherently unsteady and inhomogeneous—for example, due to upstream turbulence, wave–current interactions, or vortex shedding from nearby bodies. The effect of such unsteady inflow on porous arrays remains underexplored. \cite{klettner2019effect} addressed the unsteady incident flow by studying the interaction between a von Kármán vortex street (generated by an upstream cylinder) and a downstream circular cylinder array across a wide range of solid fractions. Using 2-D DNS at $Re_D = 2100$ and inviscid rapid distortion theory (RDT) for high-Reynolds-number extrapolation, they showed that the combined effects of inviscid blocking and viscous drag modify the upstream velocity and strain fields, suppress vortex penetration at high $\phi$, and alter force statistics on both individual cylinders and the array as a whole. Their work highlights the significant differences between steady and unsteady incident flows, particularly in terms of tangential velocity amplification, vortex–array interaction, and the correlation of fluctuating forces.

In summary, it can be concluded that previous studies focused on moderate and high $Re$, and on flow structure and forces, etc..
In this work we will focus on the wake patterns and transition of the flow around 2-D cylinder array, for $Re \le 400$, mainly $Re \le 100$.

\subsection{Global instability analysis}

LSA provides a rigorous framework for identifying the onset of unsteady flow regimes and their underlying linear mechanisms. 
Early global LSA studies of bluff-body wakes established the linear global-mode framework and its connection to wake transition (\cite{jackson1987finiteelement, zebib1987stability,noack1994global}). These works showed how global eigenmodes and their eigenvalues characterize the onset of self-sustained oscillations in bluff-body flows and motivate using a linearized operator about a chosen reference state.

Two reference states are commonly used in global LSA: the steady base flow (the converged stationary solution of the Navier–Stokes equations) and the nonlinear time-averaged mean flow. Base-flow LSA accurately identifies the linear instability mechanism and typically yields reliable growth-rate predictions for small perturbations, while mean-flow LSA has been shown to predict the saturated oscillation frequency of nonlinear limit cycles more accurately (e.g. \cite{barkley2006linear,sipp2007global}). Comparative studies (e.g. \cite{sipp2007global,manticlugo2014selfconsistent}) highlight that mean-flow LSA can systematically underestimate linear growth rates even when it correctly captures frequencies, so the two approaches are best interpreted together to reconcile linear predictions with fully nonlinear dynamics.

Structural-sensitivity and wavemaker analyses identify the regions of the flow most receptive to perturbations and most effective for control (\cite{giannetti2007structural,chomaz2005global,Liu2016}). These sensitivity maps, combined with base- and mean-flow eigenmodes, provide a powerful route to design localized modifications of the base flow (or localized actuators) that stabilize or modify the global instability (see \cite{giannetti2007structural,sipp2013characterization}).

While these tools have been widely applied to isolated cylinders and some confined geometries, their application to cylinder arrays, especially in circular configurations, remains limited. 
In the present study we therefore compute global eigenvalues and leading modes about both the base flow and the time-averaged mean flow, and we evaluate structural-sensitivity (wavemaker) maps to identify patch regions most receptive to base-flow modification for instability control. Results are presented as spectra, mode visualizations, and sensitivity maps, and are compared to nonlinear DNS to assess which reference state better predicts growth rates and frequencies for the cylinder-array geometries considered.

Despite these advances, the fundamental stability mechanisms governing the transition from steady to unsteady flow in circular cylinder arrays remain poorly understood. Previous studies have primarily focused on documenting force statistics and flow patterns at moderate to high Reynolds numbers (e.g., \cite{NICOLLE2011,Chang_Constantinescu_2015,Zhou_Venayagamoorthy_2019}), but they cannot explain the origin of the unsteadiness itself. Specifically, it remains unclear whether the observed periodic shedding emerges from a global Hopf bifurcation of the entire array—where the coupled system undergoes a synchronous instability—or simply reflects the superposition of independent vortex shedding from individual cylinders within the array. This distinction is critical: the former implies a collective, array-scale instability mechanism, while the latter suggests that each cylinder behaves as an isolated bluff body. Conventional force-based diagnostics cannot resolve this ambiguity because they describe the consequences of unsteadiness rather than its origins.

Global linear stability analysis (LSA) offers a rigorous framework to address these questions. By examining the linearised dynamics about the base state, LSA can identify the precise Reynolds number at which the flow first becomes unstable ($Re_c$), determine the spatial structure of the unstable mode, and reveal whether the instability is localised to individual cylinders or emerges as a coherent, array-scale eigenmode (\cite{jackson1987finiteelement,zebib1987stability,noack1994global}). Moreover, structural sensitivity analysis—combining direct and adjoint eigenmodes—can pinpoint the 'wavemaker' region where the instability originates (\cite{giannetti2007structural,chomaz2005global,Luchini2014a}), providing insight into the physical mechanisms driving the transition. Previous applications of these tools to single cylinders and simple geometries have proven their power in uncovering instability mechanisms (\cite{barkley2006linear,sipp2007global,manticlugo2014selfconsistent}), yet to date, no systematic LSA has been performed for circular cylinder arrays across a range of solid fractions $\phi$, leaving a fundamental gap in our understanding of how discrete microstructure governs the onset of unsteadiness.

This gap has practical implications: predicting unsteady loads, wake interactions, and scalar transport in applications such as offshore structures, wind farms, heat exchangers, and vegetated flows requires knowledge of both the stability threshold and the underlying instability mechanism (\cite{Zdravkovich1998,williamson1996vortex}). The present study addresses this gap by combining DNS with global LSA (based on both base flow and mean flow) and structural sensitivity analysis to investigate the instability of flow through a circular array of cylinders over a wide range of $\phi$ (0.016–0.315) and Reynolds numbers ($Re < 300$, mainly $\le 100$). We focus on the following questions:

\begin{itemize}
\item How does the critical Reynolds number $Re_c$ for global instability vary with solid fraction $\phi$, and what functional form describes this dependence across different regimes?
\item Does the instability manifest as an array-scale global mode, or does it arise from the superposition of localised cylinder shedding?
\item Where are the structural sensitivity (wavemaker) regions located, and how do they evolve with $\phi$ and $Re$, revealing the core of the instability mechanism?
\end{itemize}

By answering these questions, this work aims to provide a fundamental understanding of stability in multi-body flows within a circular array, bridging the gap between discrete body interactions and solid-body behaviour, and establishing a predictive framework for transition in such systems.


\textbf{TODO} `` The paper is organised as follows. Section 2 introduces the configuration of 2-D
circular array of solid cylinders, the boundary conditions, the governing equations, i.e. nonlinear NS equations and their corresponding linearised equations about the base state (mean flow or base flow) and the numerical method. Section 3 provides a detailed verification step
of nonlinear and linear numerical methods. In § 4 we show the results and discuss the
base states, cylinder wake patterns, critical Re, nonlinear DNS bifurcation scenario, global
eigenmodes and effects of AR and Re on this flow. Finally, the results are summarised in
§ 5 and conclusions are provided ''


\section{Problem formulation}\label{problemformulation}

In the present study, we consider the stabilities of two-dimensional incompressible flow through and around the cylinder array. Since the flow around a circular cylinder has been intensively investigated and the flow characteristics are well-understood, the present study selects cylinders as the constituent components of the arrays (\cite{Zdravkovich1998}).
We align the individual cylinders of the array to have the same circular geometry such that as the solid fraction $\phi$ approaches unity, the array approaches a solid cylinder.

\subsection{Geometry}

\begin{figure}
	\begin{center}
		\subfigure[$N_c = 7$, $\phi = 0.016$ \label{fig:57}{}]{
	\resizebox*{3.cm}{!}{\includegraphics{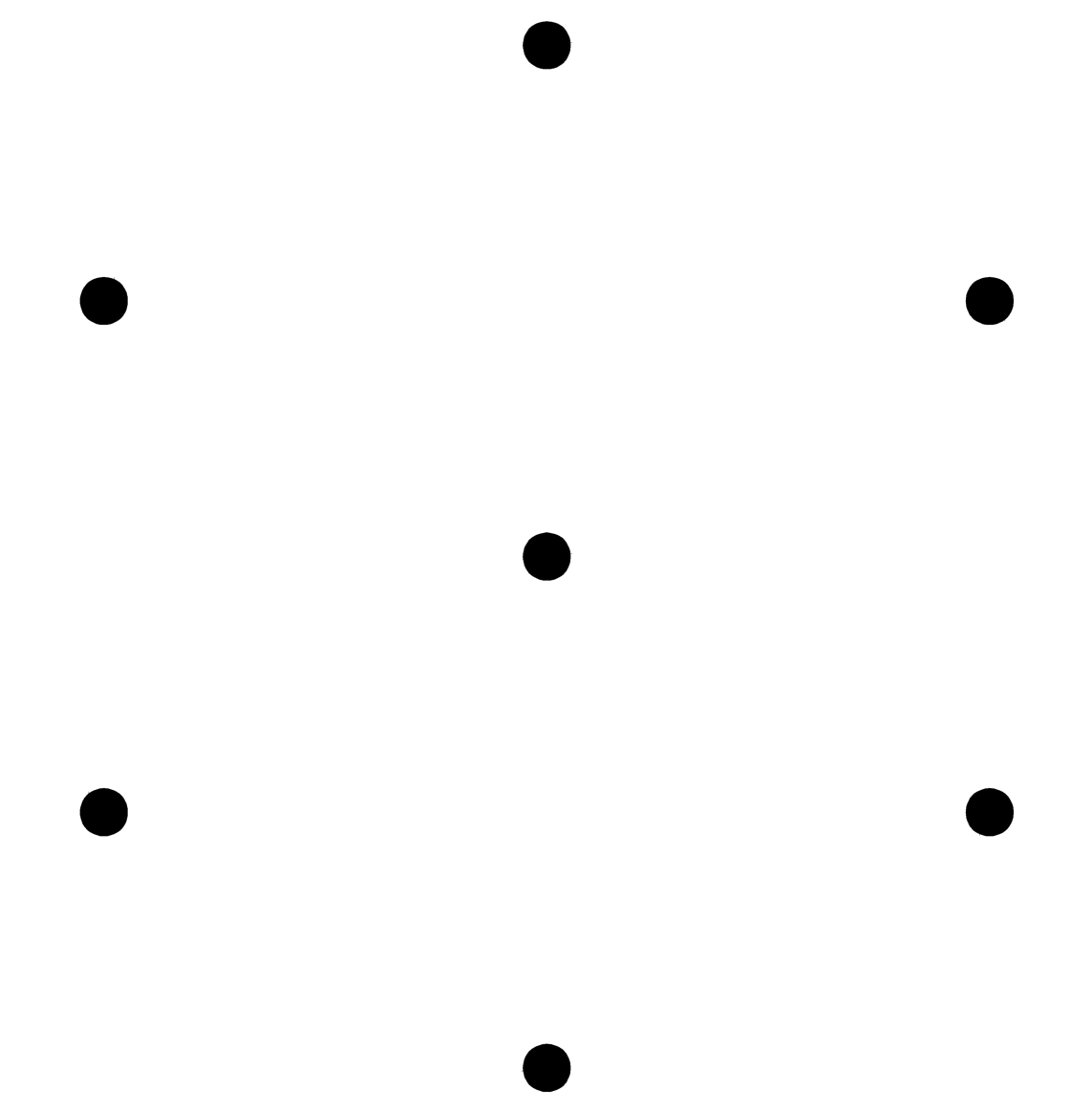}}}\hfill
		\subfigure[$N_c = 19$, $\phi = 0.043$ \label{fig:56}{}]{
	\resizebox*{3.cm}{!}{\includegraphics{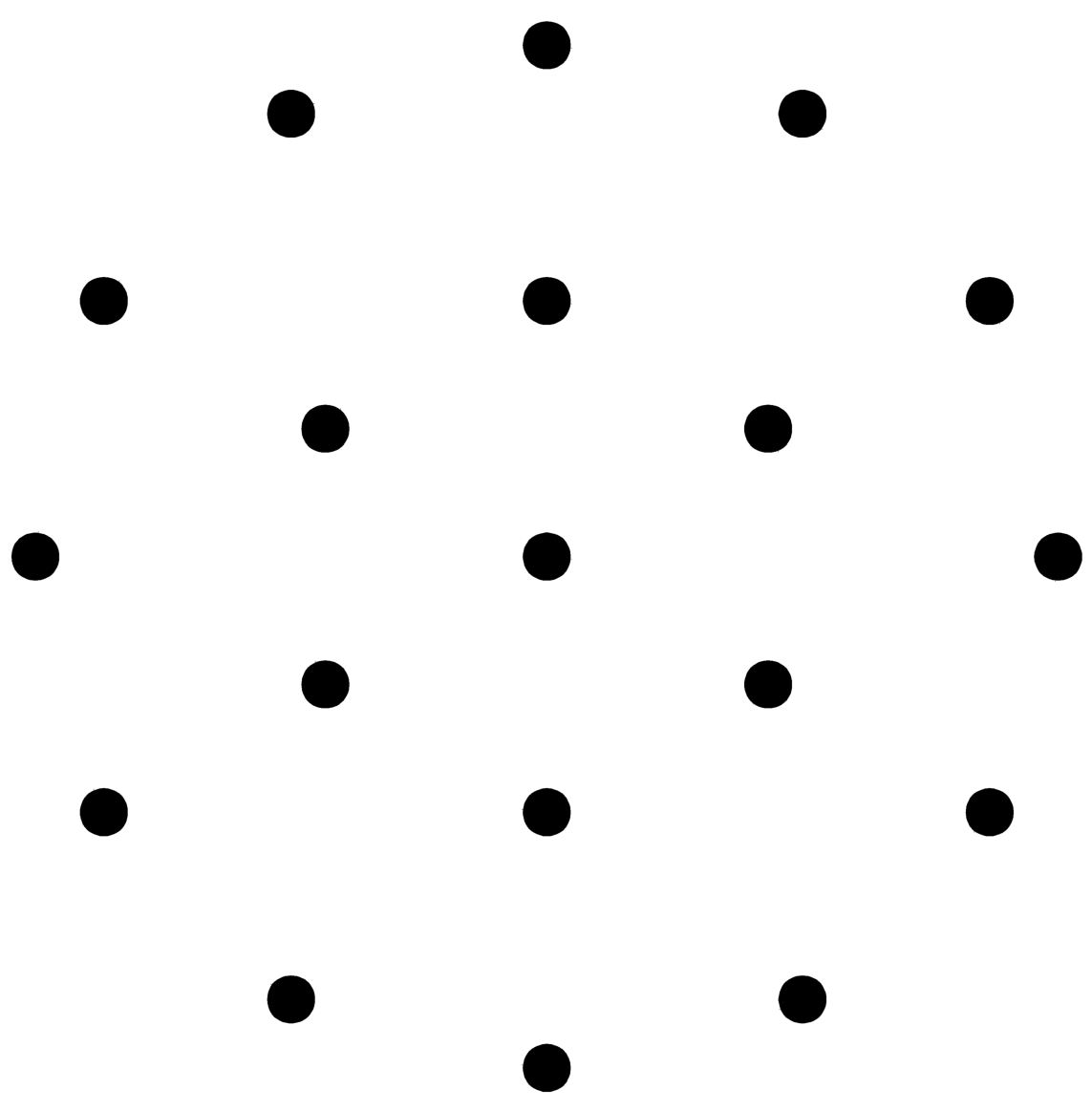}}}\hfill		
		\subfigure[$N_c = 37$, $\phi = 0.084$ \label{fig:55}{}]{
	\resizebox*{3.cm}{!}{\includegraphics{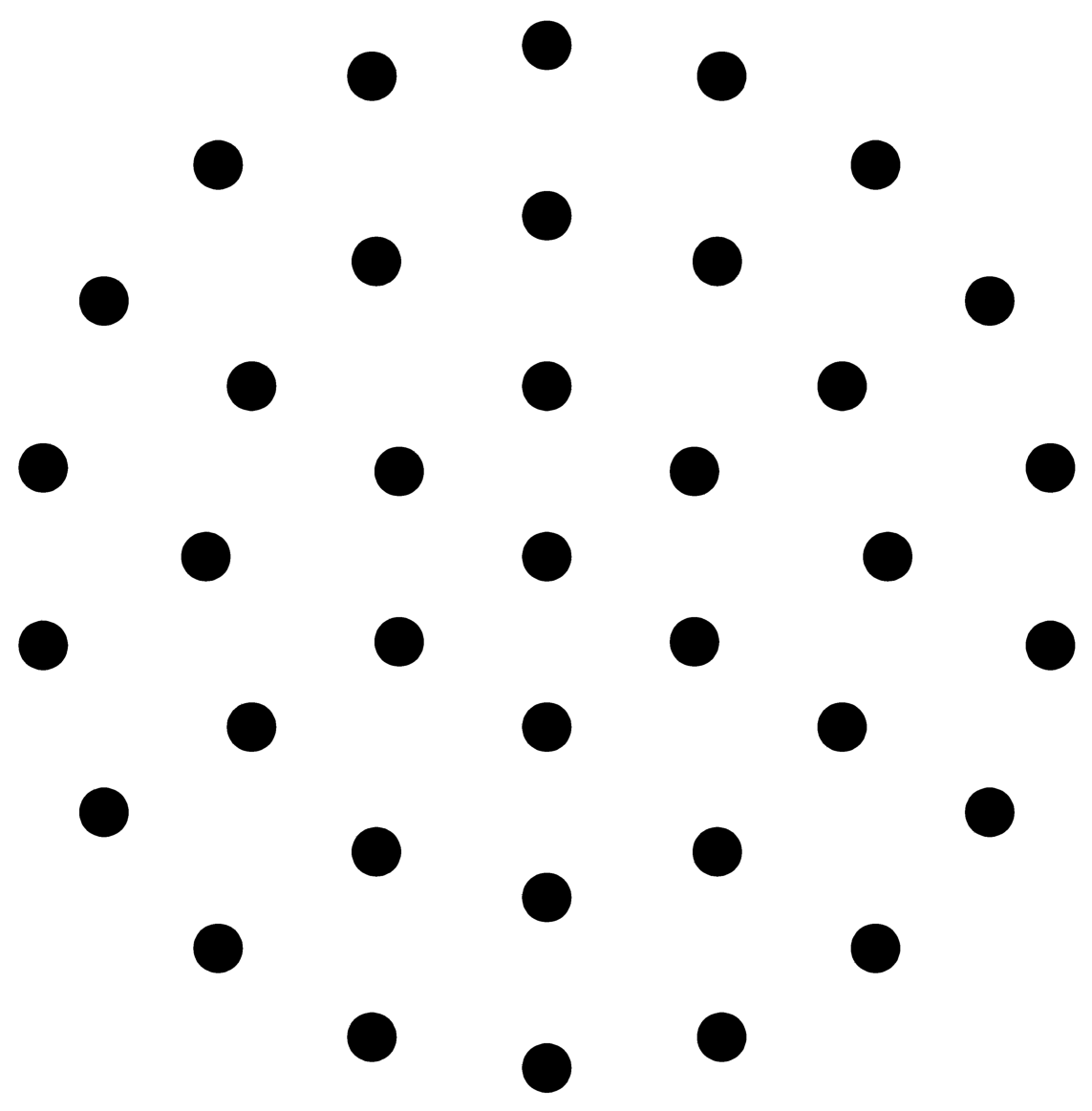}}}\hfill
		\subfigure[$N_c = 61$, $\phi = 0.138$ \label{fig:54}{}]{
	\resizebox*{3.cm}{!}{\includegraphics{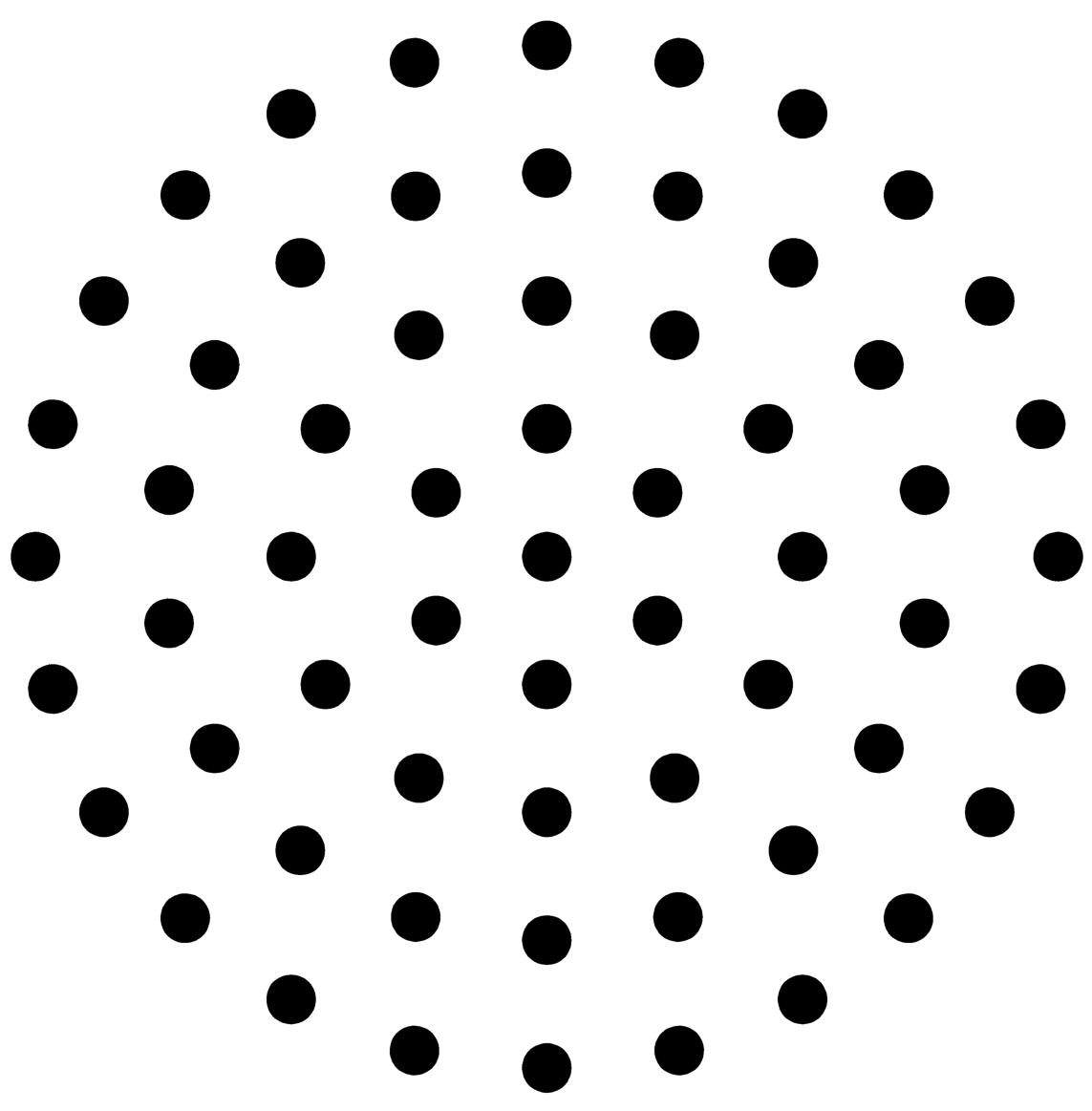}}} \\
		\subfigure[$N_c = 97$, $\phi = 0.220$ \label{fig:53}{}]{
	\resizebox*{3.cm}{!}{\includegraphics{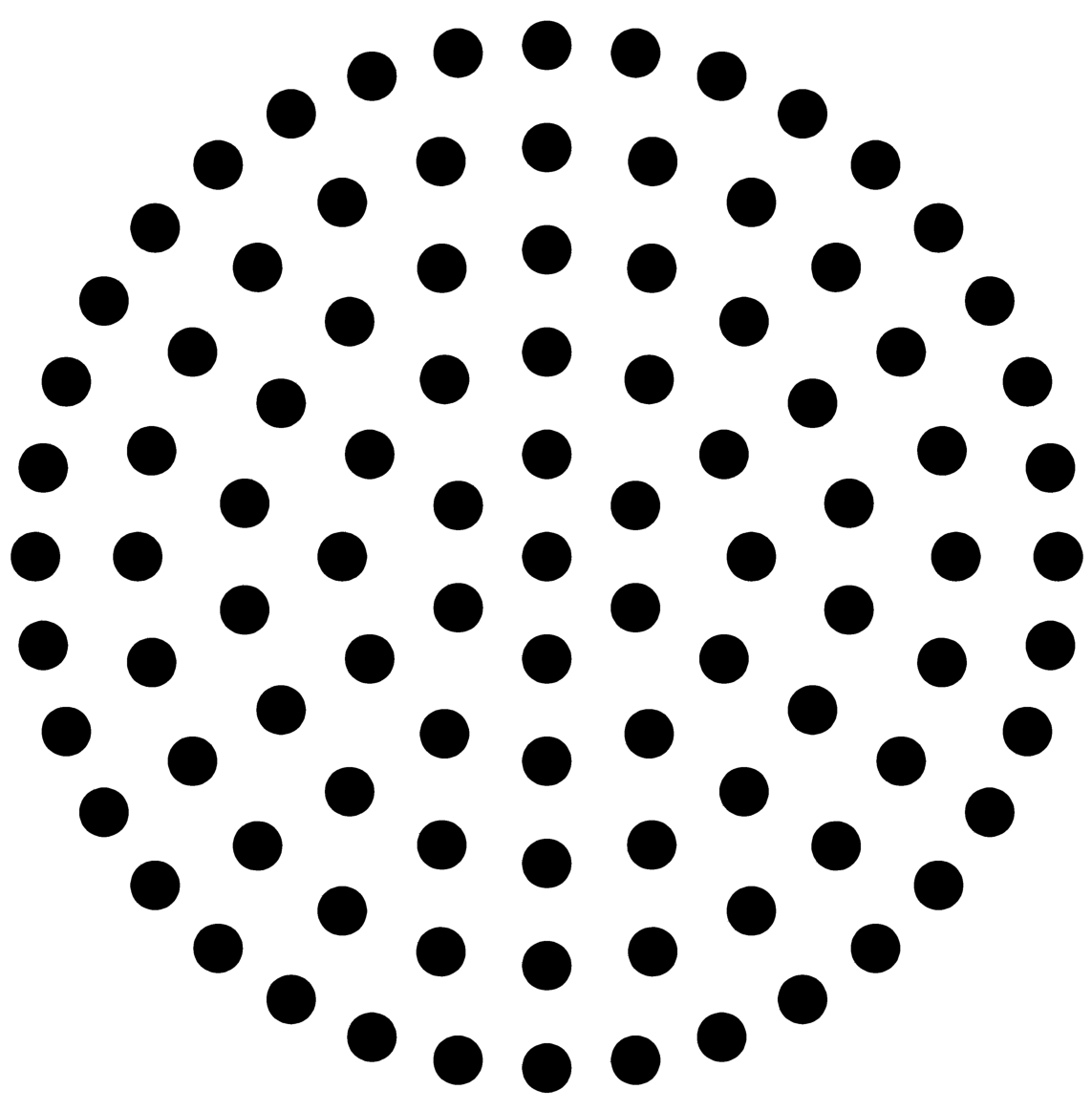}}} \hfill		
		\subfigure[$N_c = 139$, $\phi = 0.315$ \label{fig:52}{}]{
			\resizebox*{3.cm}{!}{\includegraphics{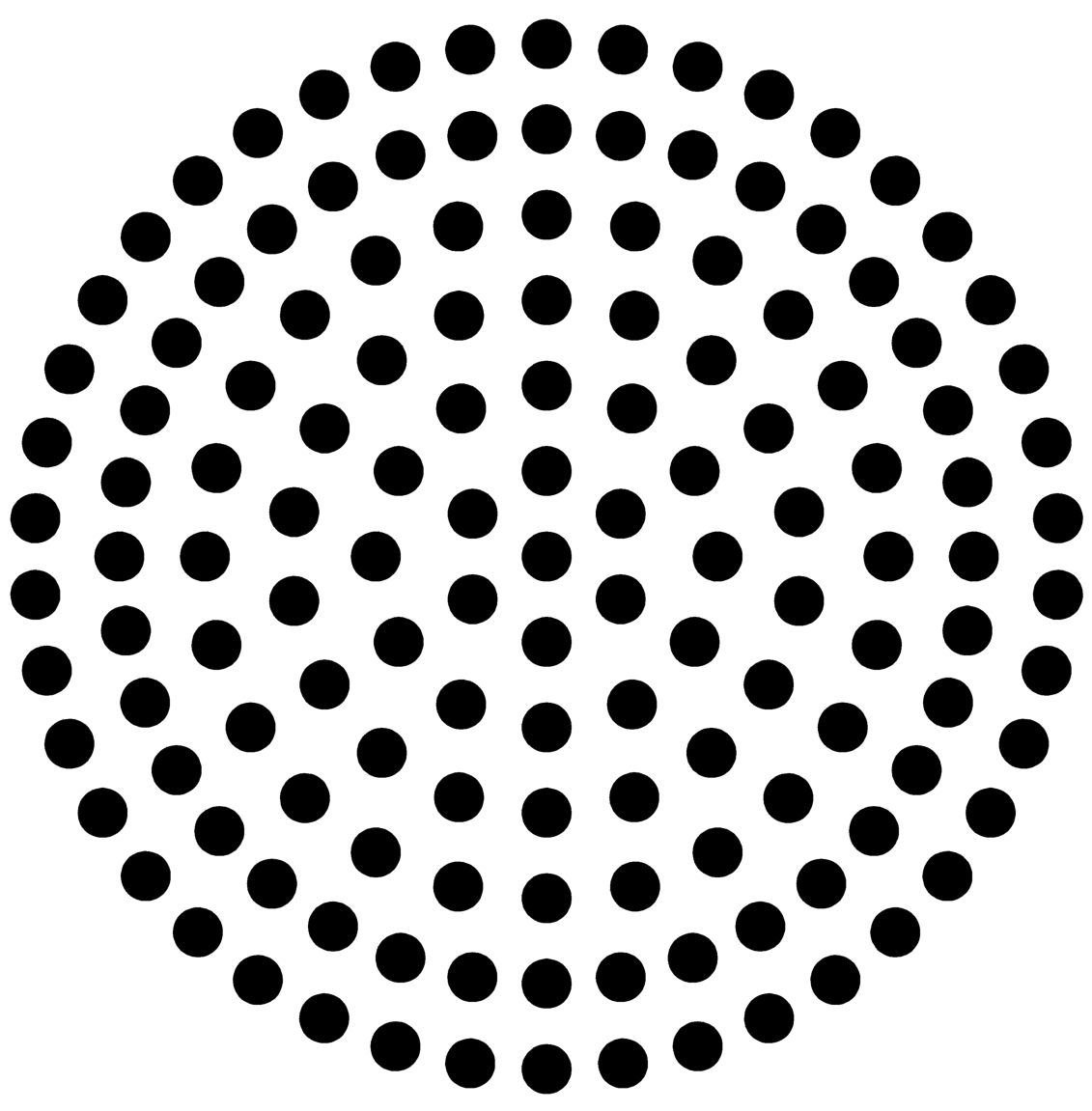}}}\hfill	
		\subfigure[Solid body, $\phi = 1$ \label{fig:51}{}]{
			\resizebox*{3.cm}{!}{\includegraphics{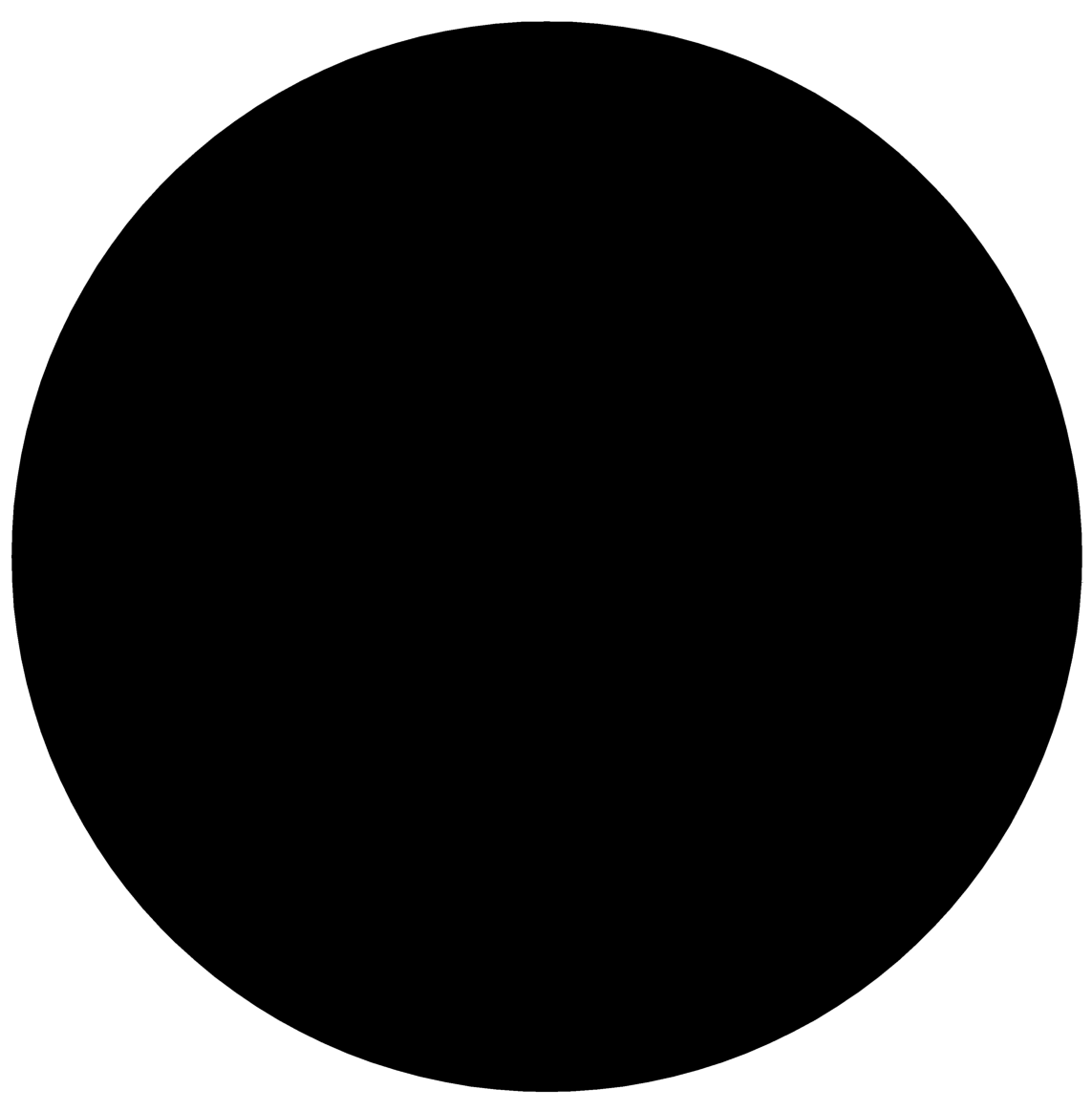}}}\hfill
		\caption{\label{fig:geoArray} Geometric configurations of the cylinder arrays. The solid fraction $\phi$ increases from (a) $N_c=7$, $\phi=0.016$ to (f) $N_c=139$, $\phi=0.315$. The limit $\phi =1$ corresponds to a solid obstruction (g). The arrays are constructed with six-fold rotational symmetry about a central cylinder at the origin.}
	\end{center}
\end{figure}

The physical model employed in this study is illustrated in figure~\ref{fig:geoArray} .
Overall, we followed the approach of \cite{NICOLLE2011} to construct the cylinder array, including the concentric ring structure and the number of rows for each array.
However, we further adjusted the arrangement of the cylinders to ensure that the geometry of each cylinder array is symmetric with respect to $y = 0$ (where $x$ is the inflow direction). Consequently, the inter-row distances and the number of cylinders per row were also modified.
Unlike the construction method in \cite{NICOLLE2011}, we adopted a ``filling outwards'' strategy with an additional cylinder placed at the center with its centroid at $(0,0)$. When there is only one row of cylinders (i.e., $N_c = 7$), the cylinder array is identical to the structure in \cite{NICOLLE2011}.
When there are two rows of cylinders, the configuration is constructed as follows:
(a) Three cylinders are uniformly placed along $x = 0$, $y \in [0,D/2]$;
(b) The concentric rings are defined based on the center of each cylinder;
(c) The concentric rings are successively divided into equal segments. 
This procedure is generalized in a similar manner for the other cylinder arrays with more than two rows.
The number of cylinders in each row is a multiple of $6$, resulting a cylinder array exhibiting six-fold rotational symmetry.
The separation between the concentric rings and between the cylinders on each ring differs slightly but remains very close.

We consider circular arrays with solid fractions spanning $\phi = 0.016$ ($N_c = 7$) to $0.315$ ($N_c = 139$), representing configurations from widely spaced to closely packed cylinders. 
For context, the solid fractions in the study of~\cite{NICOLLE2011} range from $\phi = 0.0023$ ($N_c = 1$) to $0.3016$ ($N_c = 133$).
Additionally, the case of a solid body ($\phi = 1$) is included as a reference. All geometric parameters are listed in Table~\ref{table1:geometry}. The overall diameter of the cylindrical array is $D = 1$~m, while each constituent cylinder has a diameter $d = 4.762 \times 10^{-2}$~m, making $D/d = 21$.

\begin{table}
	\centering	
\begin{tabular}{lccccccc}
    Case & $C_{7}$ & $C_{19}$ & $C_{37}$ & $C_{61}$ & $C_{97}$ & $C_{139}$ & $C_{\text{Solid}}$ \\
    $N_c$ & 7 & 19 & 37 & 61 & 97 & 139 & 1 \\
    $\phi$ & 0.016 & 0.043 & 0.084 & 0.138 & 0.220 & 0.315 & 1 \\
    Rows & 1 & 2 & 3 & 4 & 5 & 6 & -- \\
	\end{tabular}
	\caption{Summary of the cylinder array configurations. The solid fraction $\phi = (d/D)^2 N_c$ increases with the number of cylinders $N_c$. The array diameter $D = 1$~m and individual cylinder diameter $d = 4.762 \times 10^{-2}$~m.}    
	\label{table1:geometry}
\end{table}

\subsection{Flow configuration and governing equations}
\begin{figure}
	\centering
	\includegraphics[width=0.75\linewidth]{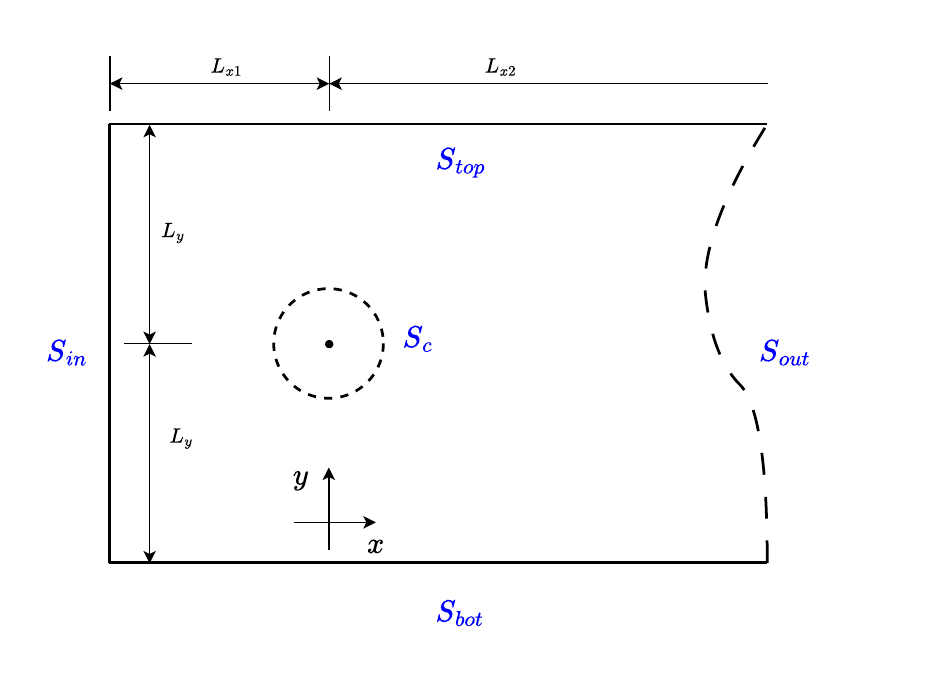}
	\caption{Schematic illustration of the problem setup for stability analysis of flow through a cylinder array, showing the computational domain and boundary conditions. The coordinate origin is at the array centre, which is denoted by the black dot.}
	\label{fig:cylinder-region}
\end{figure}

 The computational domain, boundary conditions, and geometry are shown schematically in figure~\ref{fig:cylinder-region}. The origin of the Cartesian coordinate system is located at the centre of the cylinder array.  

The governing equations of mass and momentum equation are given by 
\begin{equation}\label{eqn:govern}
	\begin{gathered}
		\nabla \cdot \boldsymbol{U}=0 \; , \\
		\frac{\partial \boldsymbol U}{\partial t}+(\boldsymbol{U} \cdot \nabla) \boldsymbol{U}=-\nabla P+\frac{1}{R e} \nabla^2 \boldsymbol{U} \; , 
	\end{gathered}	 
\end{equation}
where $\boldsymbol{U} = (U_x, U_y)$ is the velocity vector with components in the Cartesian directions, $P$ is the pressure, $t$ denotes time, and $\nabla$ represents the gradient operator in the spatial domain.

In Eqn.~\eqref{eqn:govern}, the array diameter $D$ is used as the reference length, the velocity $U_\infty$ of the uniform incoming flow as the reference velocity. The Reynolds number, Strouhal number, drag coefficient $C_d$, lift coefficient $C_l$ are
defined respectively as
\begin{equation}\label{eqn:cd}
	R e=\frac{U_{\infty} D}{\nu}, \quad S t=\frac{f D}{U_{\infty}}, \quad C_d=\frac{F_d}{\frac{1}{2} \rho U_{\infty}^2 A}, \quad C_l=\frac{F_l}{\frac{1}{2} \rho U_{\infty}^2 A}, 
\end{equation}
where $f$ is the frequency of vortex shedding, $\nu$ is the kinematic viscosity of the fluid and $\rho$ is the density. Here $F_d$ is the drag force on the surface of the cylinder, whose direction is the same as the streamwise direction. $F_l$ is the cross-stream lift force acting in either the $y$ direction.

The computational domain and boundary surfaces are shown schematically in figure~\ref{fig:cylinder-region}. The rectangular domain is bounded by an inlet surface $S_{{in}}$ and an outlet surface $S_{{out}}$, both with unit normal oriented along the $x$-direction. The surface of the cylinder array is denoted by $S_c$. The lateral boundaries are defined by two planar surfaces: $S_{top}$ and $S_{bot}$ are the top and bottom walls parallel to the $xy$-plane. 
No-slip boundary conditions are applied to the rigid surfaces of the bodies. The shear-free conditions are applied to the top and the bottom channel walls. A uniform velocity $\boldsymbol{U}=(U_\infty,0)$ is prescribed at the inlet. 
At the outlet, a zero-pressure condition ($P = 0$ as gauge pressure) is applied along with zero normal gradient for the tangential velocity components: 
\begin{equation}
    P = 0, \quad \frac{\partial \boldsymbol{U} }{\partial x} = 0 \; .
\end{equation}

\section{Linearisation}
For the analysis of the global LSA of the flow past the cylinder array, the total velocity and pressure  can be separated into a steady base state and infinitesimal perturbations: $\boldsymbol{U} = \boldsymbol{U}_b + \boldsymbol{u}$, $P = P_b + p$. Substituting this decomposition into the nonlinear governing equations (2.1) and subtracting the base-flow terms (which satisfy the steady Navier–Stokes equations) along with all nonlinear perturbation terms yields the linearized equations governing the evolution of the perturbations ($\boldsymbol{u}$, $p$) on the base state, i.e.

\begin{equation}\label{eqn:govern2}
	\begin{gathered}
		\nabla \cdot \boldsymbol{u}=0 \; , \\
		\frac{\partial \boldsymbol u}{\partial t}+\left(\boldsymbol{U}_b \cdot \nabla\right) \boldsymbol{u}+(\boldsymbol{u} \cdot \nabla) \boldsymbol{U}_b=-\nabla p+\frac{1}{R e} \nabla^2 \boldsymbol{u} \; , 
	\end{gathered}	
\end{equation}
where $\boldsymbol{u}$ is the 2-D perturbation velocity vector $\boldsymbol{u} = (u_x, u_y)$ and $p$ is the perturbation pressure. Homogeneous boundary conditions are applied for the perturbed variables

\begin{equation}
    \begin{gathered}
        \boldsymbol{u} = 0 \quad \text{on } S_{\text{c}}, \\
        u_x = U_\infty, \quad u_y = 0 \quad \text{on } S_{\text{in}}, \\
        \frac{\partial \boldsymbol{u}}{\partial x} = 0, \quad p = 0 \quad \text{on } S_{\text{out}}, \\
        \frac{\partial u_x}{\partial y} = u_y = \frac{\partial p}{\partial y} = 0 \quad \text{on } S_{\text{top}}, S_{\text{bot}}.
    \end{gathered}
\end{equation}

Here, $S_c$, $S_{in}$, and $S_{out}$ denote the cylinder surface, inflow, and outflow boundaries, respectively; $S_{top}$ and $S_{bot}$ represent the lateral boundaries where symmetry or free-slip conditions are imposed.

LSA is performed about two different reference states $\boldsymbol{Q}_b = ( \boldsymbol{U}_b, P_b)^T$: the steady base flow, obtained as the converged stationary solution of the Navier–Stokes equations, and the time-averaged mean flow computed from fully nonlinear unsteady simulations. Base-flow LSA characterizes the linear stability of the underlying steady solution and typically yields accurate growth rates of unstable modes; however, for self-excited oscillatory wakes the base-flow prediction of frequency is often inaccurate. By contrast, time-mean-flow LSA—linearization about the Reynolds-averaged mean—has been shown to capture the saturated oscillation frequency of nonlinear limit cycles, though it can underestimate linear growth rates. To exploit these complementary strengths we compute eigenvalues and leading global modes for both reference states using the same discretization and eigenvalue solver (specify solver, numerical resolution, and boundary conditions here). We then compare growth rates and frequencies, inspect mode shapes, and compute sensitivity/wavemaker maps to identify regions most receptive to base-flow modification for control. This combined analysis allows us to (i) identify the instability mechanism, (ii) reconcile linear predictions with nonlinear saturated dynamics, and (iii) propose targeted control strategies based on sensitivity results.

In addition, to obtain unstable steady base flows, we employ the selective frequency damping (SFD) method proposed by \cite{Akervik2006}. This technique adds a damping term to the Navier–Stokes equations that selectively suppresses temporal frequencies above a chosen cutoff, allowing convergence to an unstable steady solution.





\subsection{Sensitivity analysis}

The structural sensitivity of the ﬂow can be analysed using the adjoint equations~\cite{Luchini2014a}, which reveals core of an instability mechanism.  
Flow control such as destabilizing or stabilizing the global mode can be effectively achieved by perturbing the base flow within this region, as down experimentally by~\cite{Strykowski_Sreenivasan_1990}

Following~\cite{GIANNETTI_LUCHINI_2007}, the adjoint equations of the linearised NS equations

\begin{equation}\label{eqn:adj}
\begin{gathered}
\nabla\cdot\boldsymbol{u}^{+}= 0  \quad , \\[6pt]
-\frac{\partial\boldsymbol{u}^{+}}{\partial t}-\boldsymbol{U}_{b}\cdot(\nabla\boldsymbol{u}^{+})+(\nabla\boldsymbol{U}_{b})\cdot\boldsymbol{u}^{+}=-\nabla p^{+}+\frac{1}{Re}\nabla^{2}\boldsymbol{u}^{+} \quad , 
\end{gathered}
\end{equation}
where $\boldsymbol{u}^{+}$ and $p^{+}$ are the adjoint vector of perturbation field $\boldsymbol{u}$ and $p$, respectively.

The boundary conditions for the adjoint operator is obtained in process of deriving the adjoint operator from the direct linear operator. The bilinear concomitant--the term contains all the boundary contributions--must vanish to ensure that the adjoint operator is correct.~\cite{GIANNETTI_LUCHINI_2007}. As pointed out by \cite{Liu2016}, and \cite{Barkley2008}, Dirichlet for velocity and Neumann for pressure at the outlet is not formally correct, but acceptable for which is sufficiently far enough for the adjoint mode to be vanished at the far-field. This is true in this work considering the outlet boundary locates at $240D$ far downstream of the body. Following~\cite{Liu2016}, the boundary conditions of the adjoint equations are set as

\begin{equation}\label{eqn:directbc}
\begin{aligned}
\boldsymbol{u} = 0 
    &\quad \mathrm{on} \ S_{c} \ \mathrm{and} \ S_{in}, \\
(p\mathbf{I} - Re^{-1}\nabla\boldsymbol{u}) \cdot \mathbf{n} = 0 
    &\quad \mathrm{on} \ S_{out}, \\
\frac{\partial u_{x}}{\partial y} = u_{y} = \frac{\partial u_{z}}{\partial y} = \frac{\partial p}{\partial y} = 0 
    &\quad \mathrm{on} \ S_{xz,f}, S_{xz,b}, \\
\frac{\partial u_{x}}{\partial z} = \frac{\partial u_{y}}{\partial z} = u_{z} = \frac{\partial p}{\partial z} = 0 
    &\quad \mathrm{on} \ S_{xy,t}, S_{xy,b}.
\end{aligned}
\end{equation}

Combination of the direct and adjoint eigenfunctions obtains a spatial map of the eigenvalue sensitivity wavemaker, $\zeta$, given by 

\begin{equation}\label{eqn:wavemaker}
    \zeta = \frac{|\boldsymbol{u}||\boldsymbol{u}^{+}|}{\langle\boldsymbol{u}, \boldsymbol{u}^{+}\rangle}.
\end{equation}

In case of the LNSE, the boundary conditions are derived straight forward from the linearization of the ones used for the base flow, which are zero inflow, no-slip at the cylinder surface and outflow at the rear part. This reads:

\begin{equation}
\begin{gathered}\label{eqn:LNSE-direct}
\boldsymbol{u}(x,y) = (0,0), \quad \frac{\partial p(x,y)}{\partial \mathbf{n}} = 0 \quad \text{on } S_{\text{in}}, \\
\boldsymbol{u}(x,y) = (0,0), \quad \frac{\partial p(x,y)}{\partial \mathbf{n}} = 0 \quad \text{on } S_{\text{c}}, \\
\frac{\partial \boldsymbol{u}}{\partial \mathbf{n}} = 0, \quad p(x,y) = 0 \quad \text{on } S_{\text{out}}.
\end{gathered}
\end{equation}

Regarding the adjoint equations, the most suitable boundary conditions are:
\begin{equation}
\begin{gathered}\label{eqn:LNSE-adj}
\boldsymbol{v}'(x,y) = (0,0), \quad \frac{\partial m'(x,y)}{\partial \mathbf{n}} = 0 \quad \text{on } S_{\text{in}}, \\
\boldsymbol{v}'(x,y) = (0,0), \quad \frac{\partial m'(x,y)}{\partial \mathbf{n}} = 0 \quad \text{on } S_{\text{c}}, \\
\boldsymbol{v}'(x,y) = (0,0), \quad m'(x,y) = 0 \quad \text{on } S_{\text{out}}.
\end{gathered}
\end{equation}


\newpage

\section{Linearisation}
For the analysis of the global LSA of the flow past the cylinder array, the total velocity and pressure  can be separated into a steady base state and infinitesimal perturbations: $\boldsymbol{U} = \boldsymbol{U}_b + \boldsymbol{u}$, $P = P_b + p$. Substituting this decomposition into the nonlinear governing equations (2.1) and subtracting the base-flow terms (which satisfy the steady Navier–Stokes equations) along with all nonlinear perturbation terms yields the linearized equations governing the evolution of the perturbations ($\boldsymbol{u}$, $p$) on the base state, i.e.

\begin{equation}\label{eqn:govern2}
    \begin{gathered}
        \nabla \cdot \boldsymbol{u}=0 \; , \\
        \frac{\partial \boldsymbol u}{\partial t}+\left(\boldsymbol{U}_b \cdot \nabla\right) \boldsymbol{u}+(\boldsymbol{u} \cdot \nabla) \boldsymbol{U}_b=-\nabla p+\frac{1}{R e} \nabla^2 \boldsymbol{u} \; , 
    \end{gathered}	
\end{equation}
where $\boldsymbol{u}$ is the 2-D perturbation velocity vector $\boldsymbol{u} = (u_x, u_y)$ and $p$ is the perturbation pressure. Homogeneous boundary conditions are applied for the perturbed variables

\begin{equation}
    \begin{gathered}
        \boldsymbol{u} = 0 \quad \text{on } S_{\text{c}}, \\
        u_x = U_\infty, \quad u_y = 0 \quad \text{on } S_{\text{in}}, \\
        \frac{\partial \boldsymbol{u}}{\partial x} = 0, \quad p = 0 \quad \text{on } S_{\text{out}}, \\
        \frac{\partial u_x}{\partial y} = u_y = \frac{\partial p}{\partial y} = 0 \quad \text{on } S_{\text{top}}, S_{\text{bot}}.
    \end{gathered}
\end{equation}


\subsection{Sensitivity analysis}

The structural sensitivity of the ﬂow can be analysed using the adjoint equations~\cite{Luchini2014a}, which reveals core of an instability mechanism.  
Flow control such as destabilizing or stabilizing the global mode can be effectively achieved by perturbing the base flow within this region, as down experimentally by~\cite{Strykowski_Sreenivasan_1990}

Following~\cite{GIANNETTI_LUCHINI_2007}, the adjoint equations of the linearised NS equations

\begin{equation}\label{eqn:adj}
\begin{gathered}
\nabla\cdot\boldsymbol{u}^{+}= 0  \quad , \\[6pt]
-\frac{\partial\boldsymbol{u}^{+}}{\partial t}-\boldsymbol{U}_{b}\cdot(\nabla\boldsymbol{u}^{+})+(\nabla\boldsymbol{U}_{b})\cdot\boldsymbol{u}^{+}=-\nabla p^{+}+\frac{1}{Re}\nabla^{2}\boldsymbol{u}^{+} \quad , 
\end{gathered}
\end{equation}
where $\boldsymbol{u}^{+}$ and $p^{+}$ are the adjoint vector of perturbation field $\boldsymbol{u}$ and $p$, respectively.

The boundary conditions for the adjoint operator are obtained by deriving the adjoint from the direct linear operator, ensuring the bilinear concomitant vanishes~\cite{GIANNETTI_LUCHINI_2007}. For the two-dimensional flow considered here, the boundary conditions are set as follows:

\begin{equation}\label{eqn:directbc}
\begin{aligned}
\boldsymbol{u} = 0 
    &\quad \mathrm{on} \ S_{\text{c}} \ \mathrm{and} \ S_{\text{in}}, \\
(p\mathbf{I} - Re^{-1}\nabla\boldsymbol{u}) \cdot \mathbf{n} = 0 
    &\quad \mathrm{on} \ S_{\text{out}}, \\
\frac{\partial u_{x}}{\partial y} = u_{y} = \frac{\partial p}{\partial y} = 0 
    &\quad \mathrm{on} \ S_{\text{top}}, S_{\text{bot}}.
\end{aligned}
\end{equation}

\section{Numerical methods and validation}\label{Numerical}

\subsection{Numerical methods: DNS }

In order to obtain the accurate wake pattern and the base states of the flow past and through the cylinder array at low Reynolds numbers, we adopt the parallelised open-source code OpenFOAM (\cite{openfoam_website}) for the numerical simulations. 
Equations~\eqref{eqn:govern} are solved with the finite-volume approach and the PISO (pressure implicit with splitting of operators) algorithm (\cite{ISSA198640}). 
The convection term is discretized using the fourth-order cubic scheme while the diffusion term is discretized using a second-order linear scheme. A blended scheme consisting of the second-order Crank–Nicolson scheme and a first-order Euler implicit scheme is used to integrate the equations in time.

The grid generation is performed by following the approach used in ~\cite{Jiang2016,Jiang_Cheng_2017}.
In particular, the reference mesh in the present study uses the same parameters of the standard mesh in ~\cite{Jiang2016}. The computational mesh is then refined and domain size is increased to fulfill the requirement of linear stabilities. A thorough parameter dependence check is reported separately in appendix A.

\begin{figure}
	\begin{center}
		\subfigure[$C_7$ \label{fig:C7-subfig1}]{
			\resizebox*{3.55cm}{!}{\includegraphics{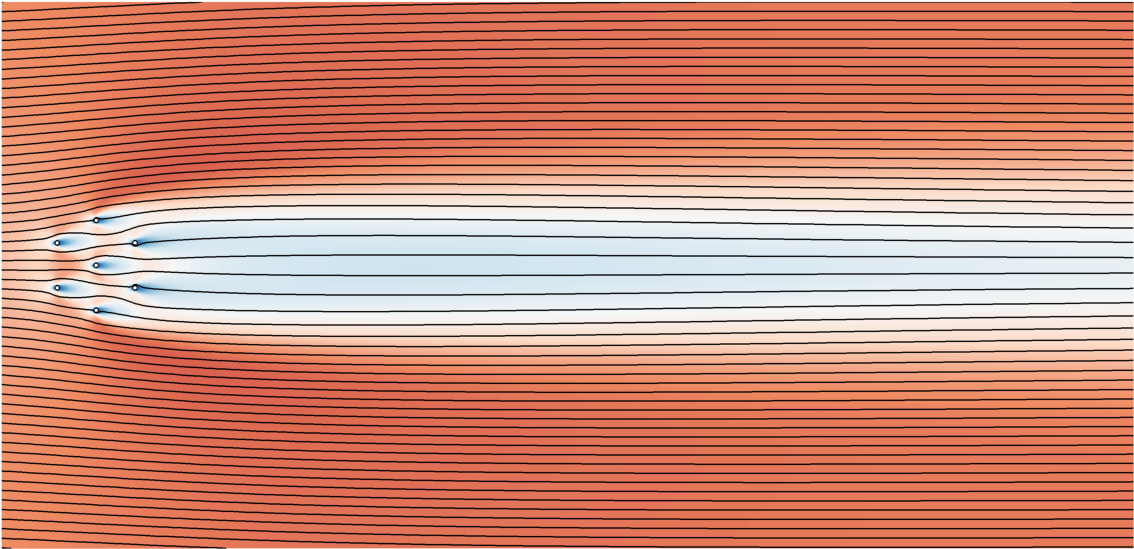}}}\hfill
		\subfigure[$C_7$ \label{fig:C7-subfig2}]{
			\resizebox*{3.55cm}{!}{\includegraphics{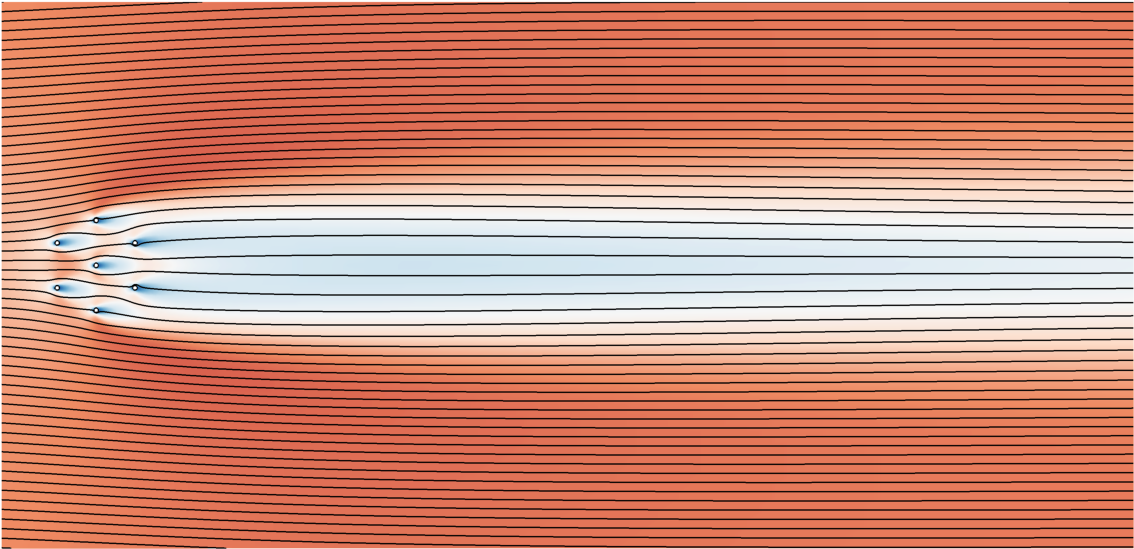}}}\hfill
		\subfigure[$C_7$ \label{fig:C7-subfig3}]{
			\resizebox*{3.9cm}{!}{\includegraphics{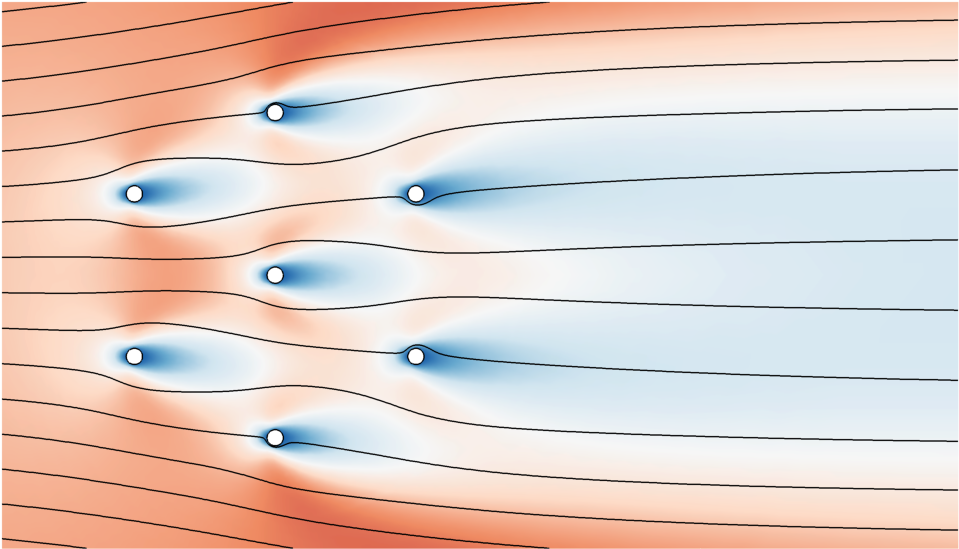}}}\\
			
		\subfigure[$C_{19}$ \label{fig:C19-subfig1}]{
			\resizebox*{3.55cm}{!}{\includegraphics{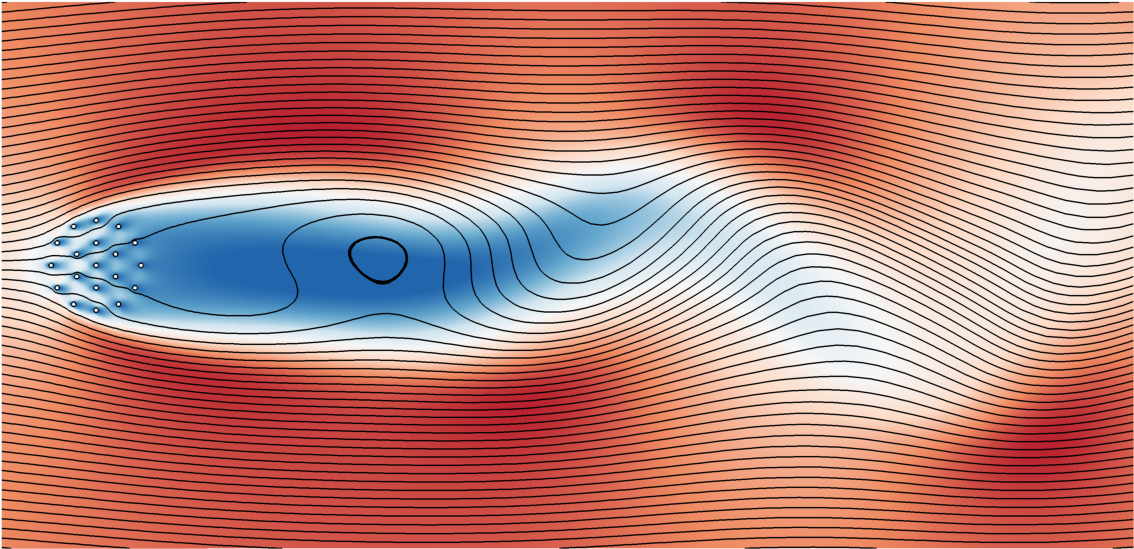}}}\hfill
		\subfigure[$C_{19}$ \label{fig:C19-subfig2}]{
			\resizebox*{3.55cm}{!}{\includegraphics{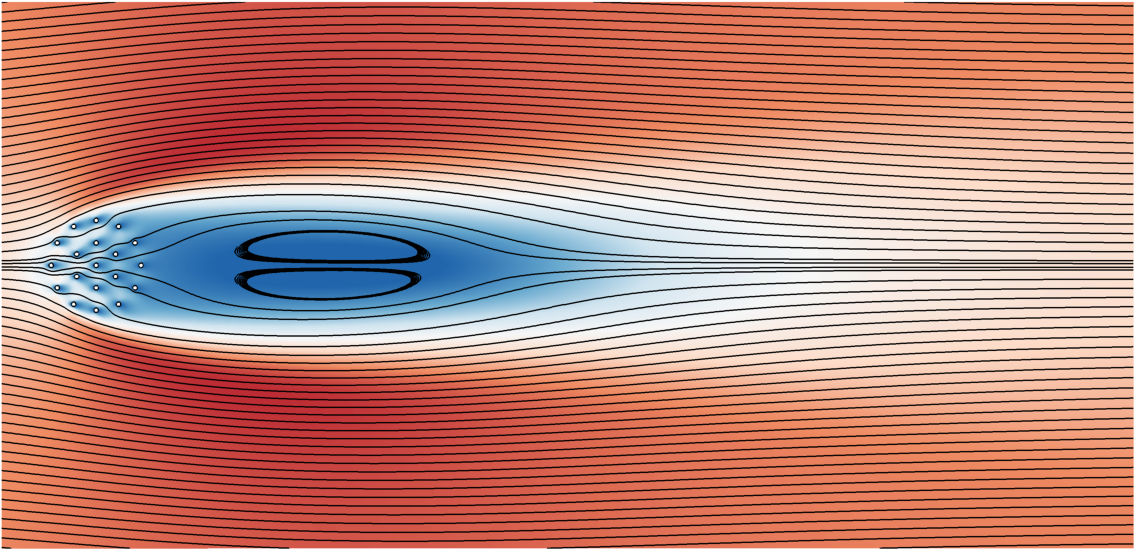}}}\hfill
		\subfigure[$C_{19}$ \label{fig:C19-subfig3}]{
			\resizebox*{3.9cm}{!}{\includegraphics{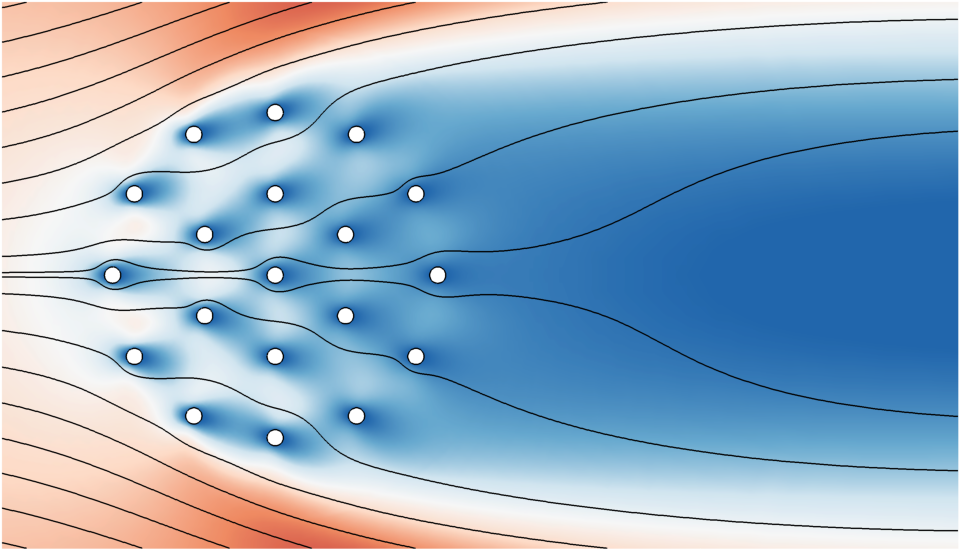}}}\\
			
		\subfigure[$C_{37}$ \label{fig:C37-subfig1}]{
			\resizebox*{3.55cm}{!}{\includegraphics{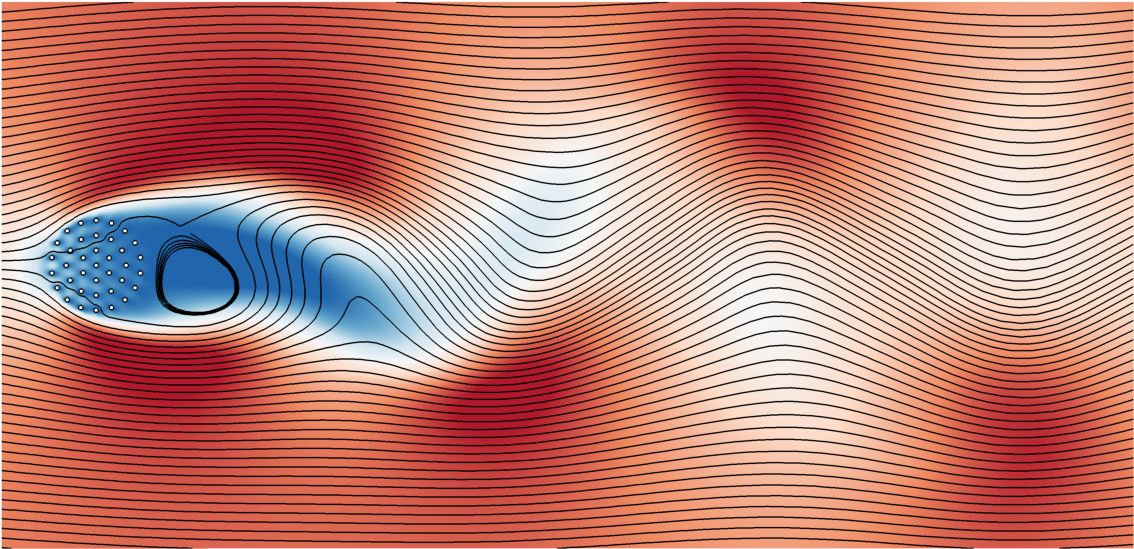}}}\hfill
		\subfigure[$C_{37}$ \label{fig:C37-subfig2}]{
			\resizebox*{3.55cm}{!}{\includegraphics{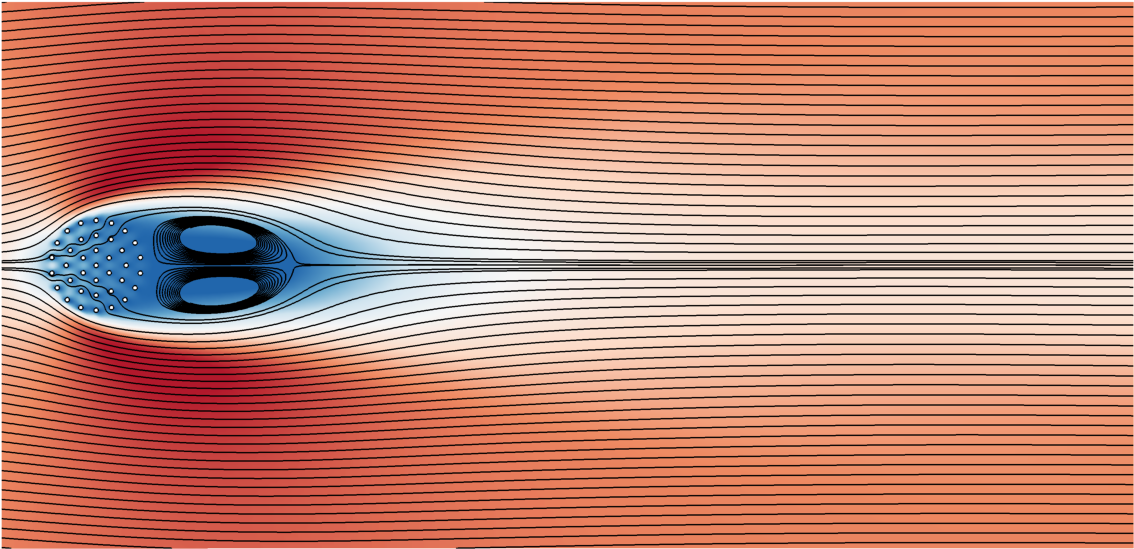}}}\hfill
		\subfigure[$C_{37}$ \label{fig:C37-subfig3}]{
			\resizebox*{3.9cm}{!}{\includegraphics{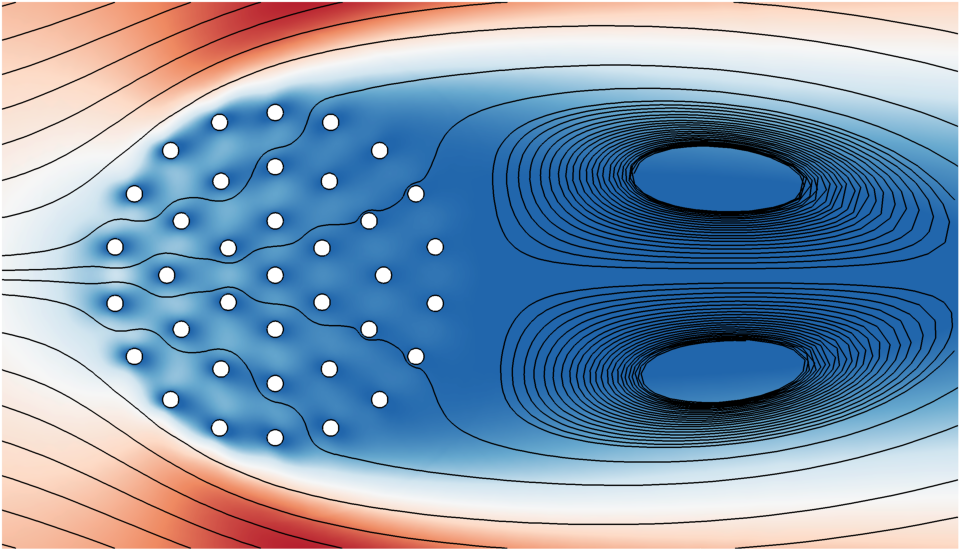}}}\\
			
		\subfigure[$C_{61}$ \label{fig:C61-subfig1}]{
			\resizebox*{3.55cm}{!}{\includegraphics{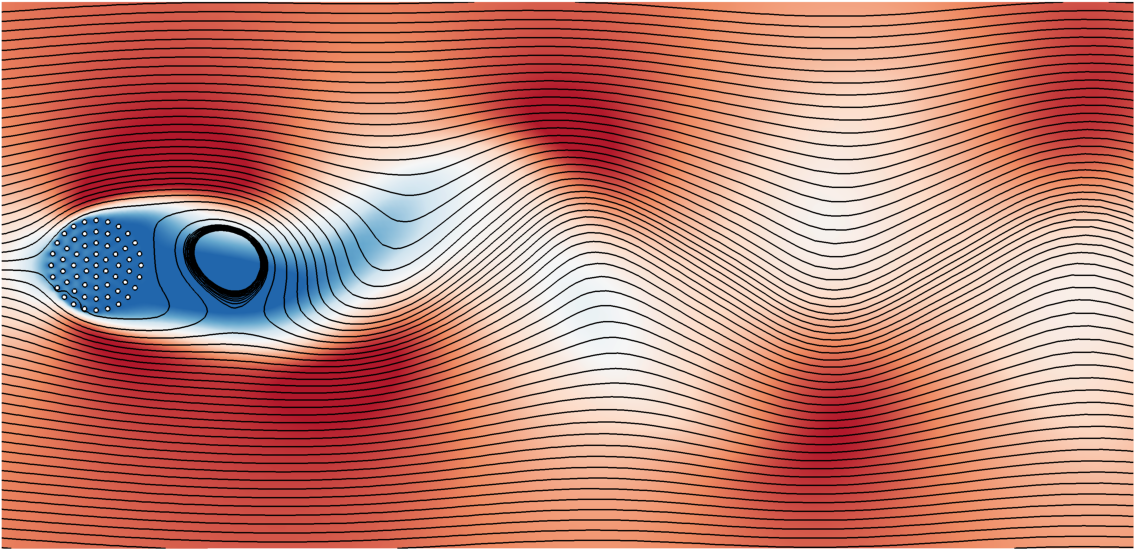}}}\hfill
		\subfigure[$C_{61}$ \label{fig:C61-subfig2}]{
			\resizebox*{3.55cm}{!}{\includegraphics{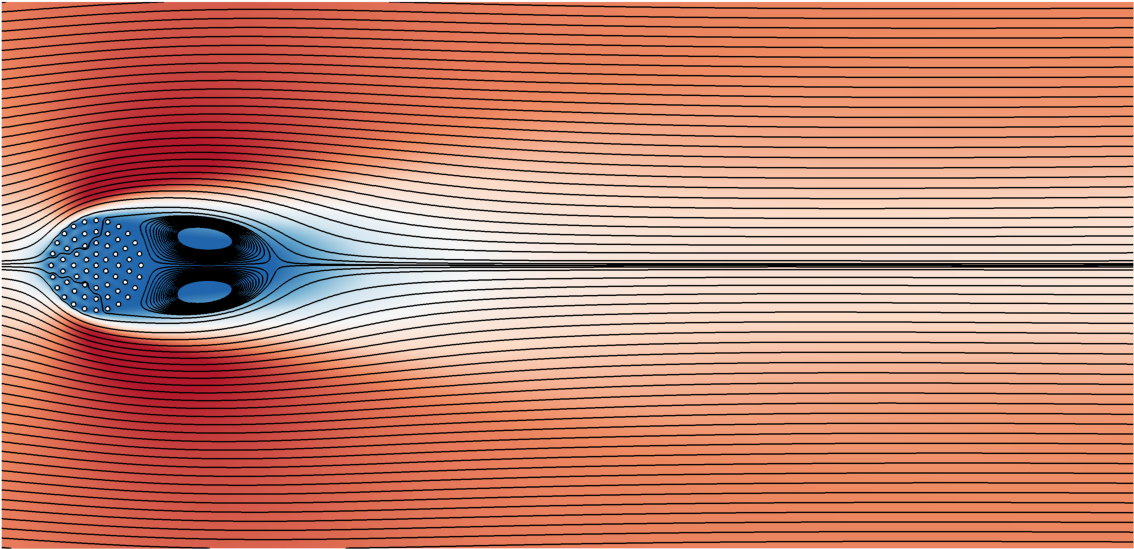}}}\hfill
		\subfigure[$C_{61}$ \label{fig:C61-subfig3}]{
			\resizebox*{3.9cm}{!}{\includegraphics{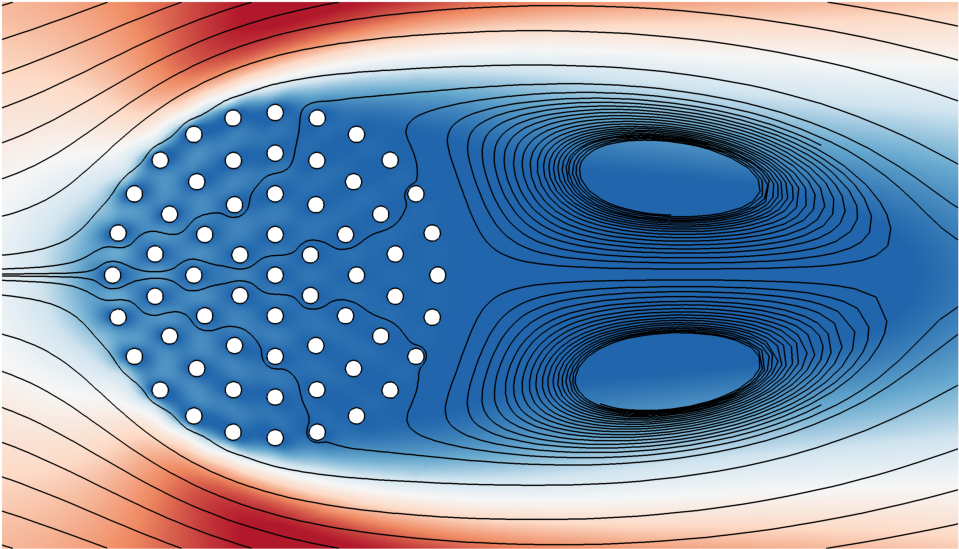}}}\\
			
		\subfigure[$C_{97}$ \label{fig:C97-subfig1}]{
			\resizebox*{3.55cm}{!}{\includegraphics{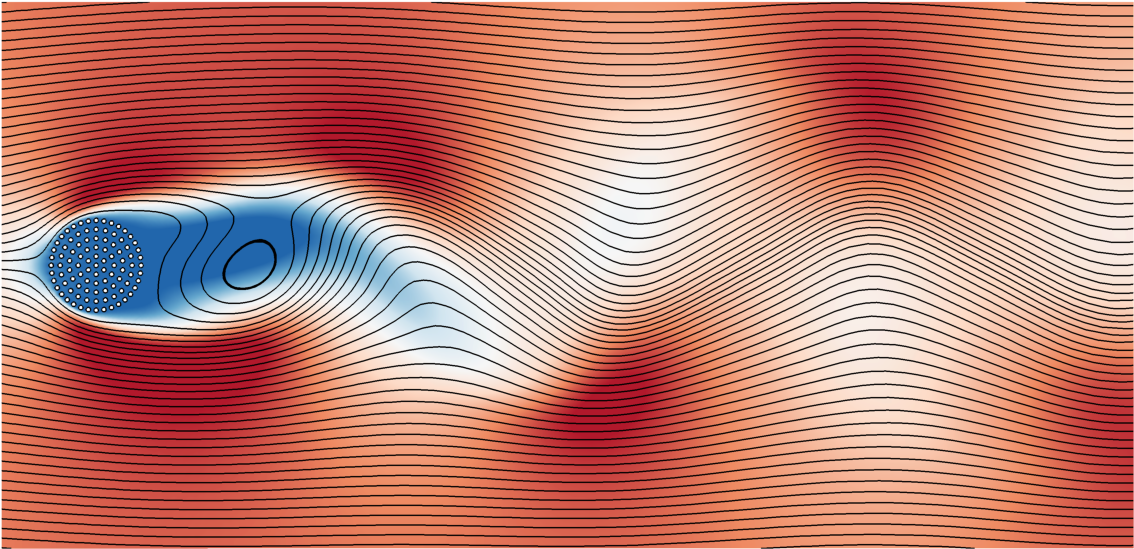}}}\hfill
		\subfigure[$C_{97}$ \label{fig:C97-subfig2}]{
			\resizebox*{3.55cm}{!}{\includegraphics{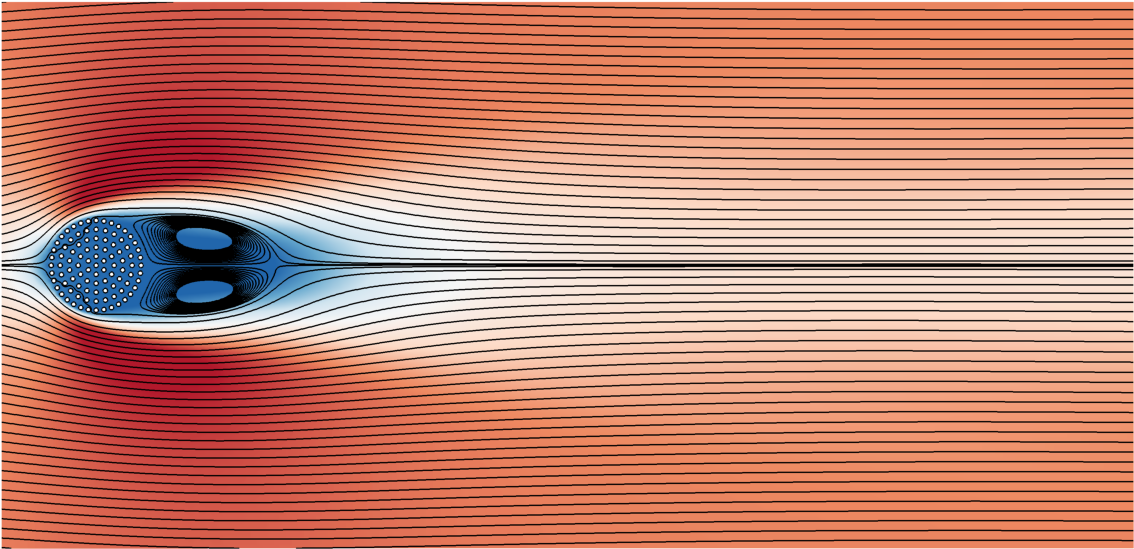}}}\hfill
		\subfigure[$C_{97}$ \label{fig:C97-subfig3}]{
			\resizebox*{3.9cm}{!}{\includegraphics{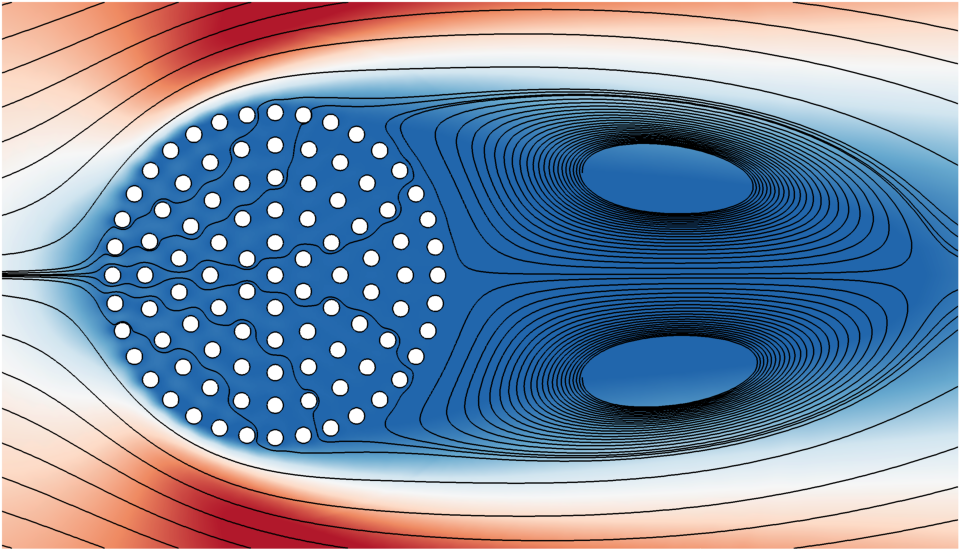}}}\\
			
		\subfigure[$C_{139}$ \label{fig:C139-subfig1}]{
			\resizebox*{3.55cm}{!}{\includegraphics{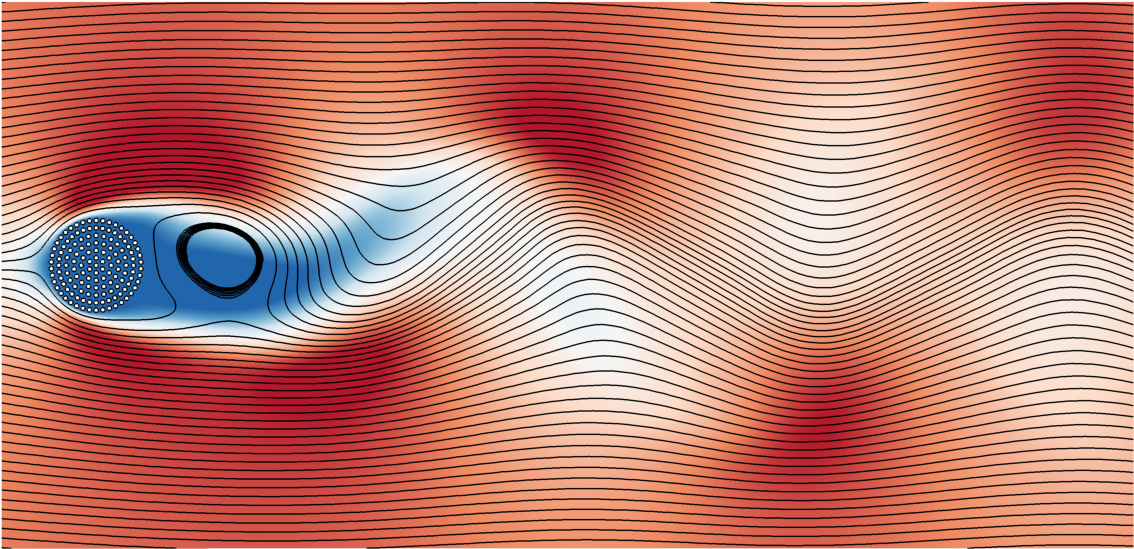}}}\hfill
		\subfigure[$C_{139}$ \label{fig:C139-subfig2}]{
			\resizebox*{3.55cm}{!}{\includegraphics{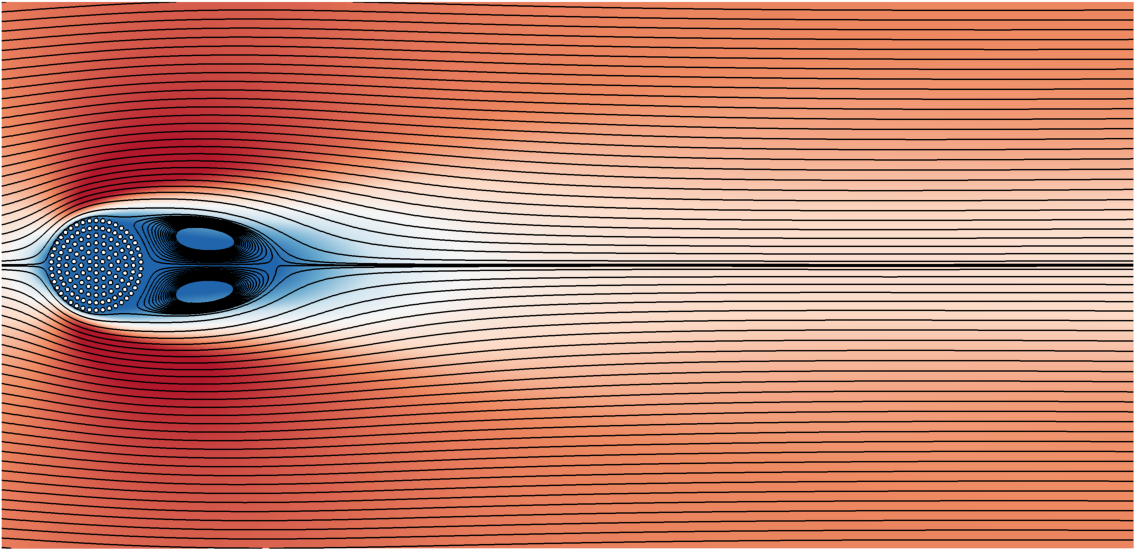}}}\hfill
		\subfigure[$C_{139}$ \label{fig:C139-subfig3}]{
			\resizebox*{3.9cm}{!}{\includegraphics{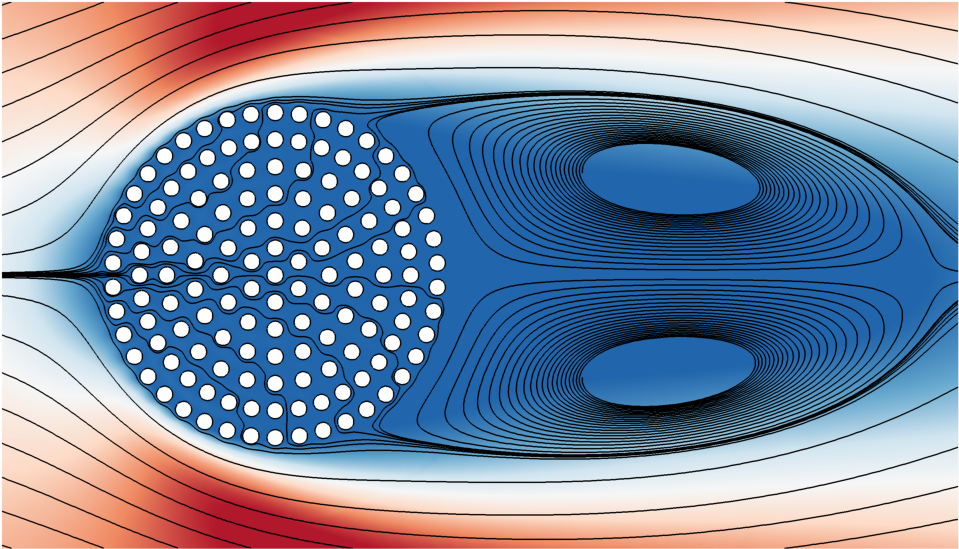}}}\\
			
		\subfigure[$C_{\text{solid}}$ \label{fig:solid-subfig1}]{
			\resizebox*{3.55cm}{!}{\includegraphics{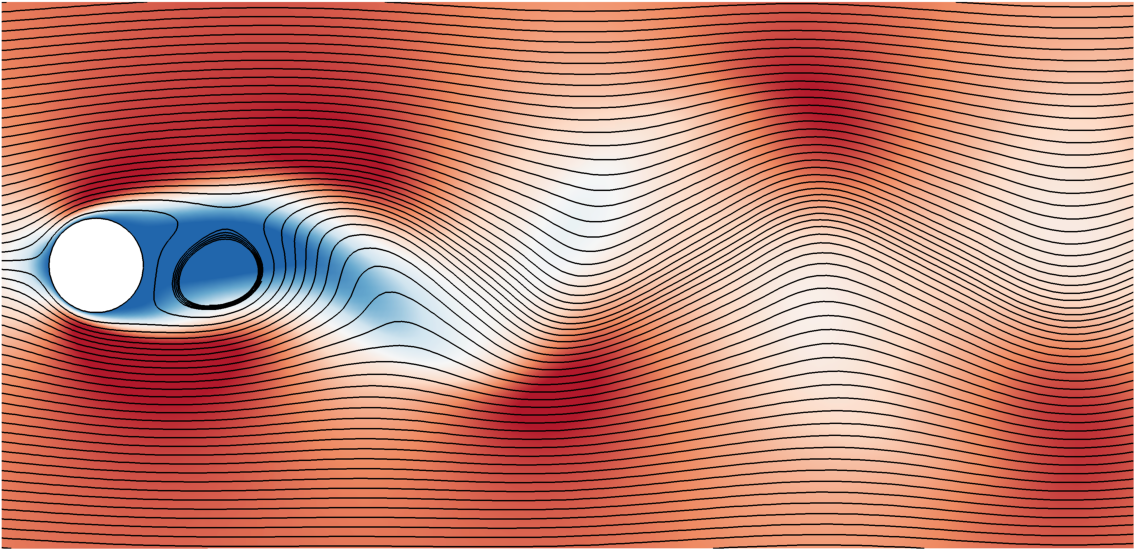}}}\hfill
		\subfigure[$C_{\text{solid}}$ \label{fig:solid-subfig2}]{
			\resizebox*{3.55cm}{!}{\includegraphics{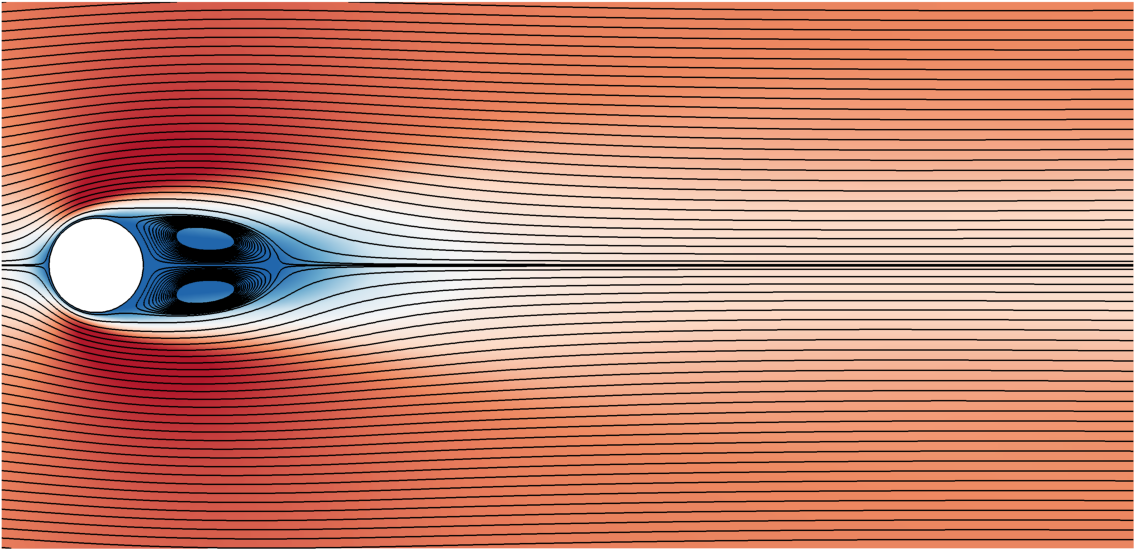}}}\hfill
		\subfigure[$C_{\text{solid}}$ \label{fig:solid-subfig3}]{
			\resizebox*{3.9cm}{!}{\includegraphics{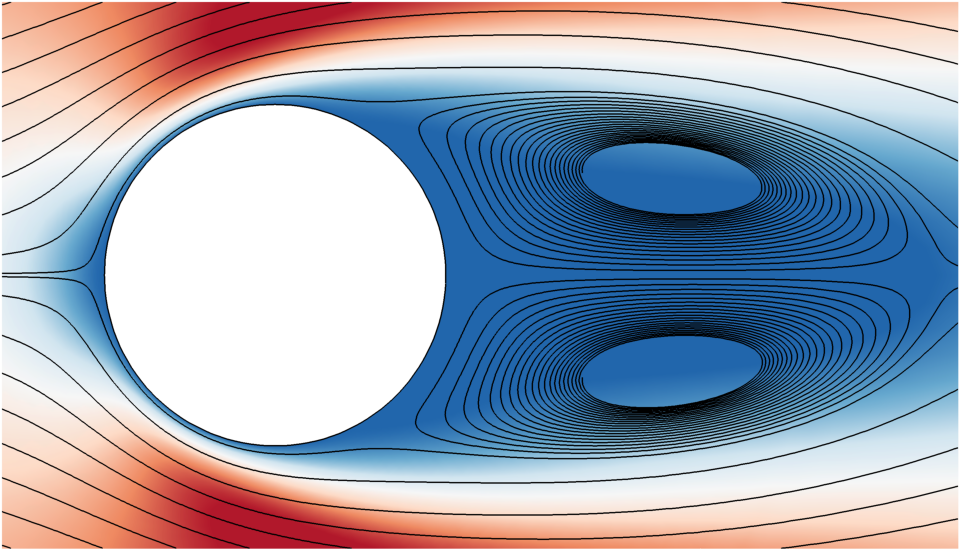}}}
		
		\caption{\label{fig:Streamlines} Instantaneous, time-averaged, and near-wake flow fields for different array configurations at $Re = 100$. Contours of the $u$-component velocity (colored) and streamlines for: the instantaneous flow field (left column), the time-averaged flow field (middle column), and the near-wake region (right column). Configurations shown are $(a)$–$(c)$ $C_{7}$, $(d)$–$(f)$ $C_{19}$, $(g)$–$(i)$ $C_{37}$, $(j)$–$(l)$ $C_{61}$, $(m)$–$(o)$ $C_{97}$, $(p)$–$(r)$ $C_{139}$, and $(s)$–$(u)$ $C_{\text{solid}}$. The flow direction is from left to right, with velocity contours indicating speed from low (blue) to high (red). Contour levels are uniformly spaced between $0$ and $1.2$ in increments of $0.1$.}
	\end{center}
\end{figure}

Instantaneous, time-averaged, and near-wake flow fields for seven different configurations at $Re = 100$ are illustrate in figure~\ref{fig:Streamlines}. 

In contrast to the wake of the other configurations, behind the porous patch $C_7$ the flow is stable. There is a steady wake region. The shear layers at the two wake edges grows but does not never meet, and there is no vortex street.

As the porosity $\phi$ is increased, e.g., $C_{19}$, the steady wake region still exits, and a vortex street is formed. This phenomena has been investigated by \cite{Zong_Nepf_2012}. They pointed out that the steady wake extends from the end of the patch to a distance downstream, characterized with constant velocity value. After the steady wake the velocity along the centreline increases, recovering toward the free-stream velocity.  The two shear layers at the two wake edges grow until they meet at the middle of the wake, after which a single vortex street can form. They also suggested that the length of the steady wake, can be estimated using previous descriptions of planar shear layer growth. 

As the porosity $\phi$ is further increased,  the wake length is decreased and the the vortex region move upstream (as indicated by the time-average flow field).

\subsection{Flow diagnostics}

To analyse the flow characteristics, a set of diagnostic tools was employed. The first technique was to analyse the distribution of the out-of-plane vorticity component $\omega=(\nabla \times \boldsymbol{u}) \cdot \hat{\boldsymbol{x}}_3$ within and downstream of the array. To distinguish between regions of the flow that were dominated by straining or rotation, we analysed the relative strength of the symmetric and non-symmetric components of the velocity gradient tensor defined by

$$
\boldsymbol{\Sigma}=\frac{1}{2}\left(\nabla \boldsymbol{u}+(\nabla \boldsymbol{u})^{\mathrm{T}}\right), \quad \boldsymbol{\Omega}=\frac{1}{2}\left(\nabla \boldsymbol{u}-(\nabla \boldsymbol{u})^{\mathrm{T}}\right) .
$$

A usual measure of the magnitude of these tensors is $\|\boldsymbol{\Sigma}\|^2=\Sigma_{i j} \Sigma_{i j}$ and $\|\boldsymbol{\Omega}\|^2=\Omega_{i j} \Omega_{i j}$. Straining and vortical regions within a flow can be distinguished by calculating the second invariant of the velocity gradient tensor (Hunt, Wray \& Moin 1988). We apply the dimensionless (second invariant of the velocity gradient tensor) measure,

$$
Q=\frac{\|\boldsymbol{\Sigma}\|^2-\|\boldsymbol{\Omega}\|^2}{\|\boldsymbol{\Sigma}\|^2+\|\boldsymbol{\Omega}\|^2},
$$

which is bounded between -1 and 1 (Davidson 2004) and takes the values of -1 , 0 and 1 for vortical, shearing and irrotational flows, respectively. Equation (3.7) is a much more sensitive and precise indicator of whether fluid is irrotational rather than thresholding vorticity. While this measure has been applied to many turbulent flow studies (Jeong \& Hussain 2006), it has not been widely applied to multi-body flows.

\begin{figure}
	\begin{center}
		\subfigure[$C_{7}$ \label{fig2:C7-subfig2}]{
			\resizebox*{9cm}{!}{\includegraphics{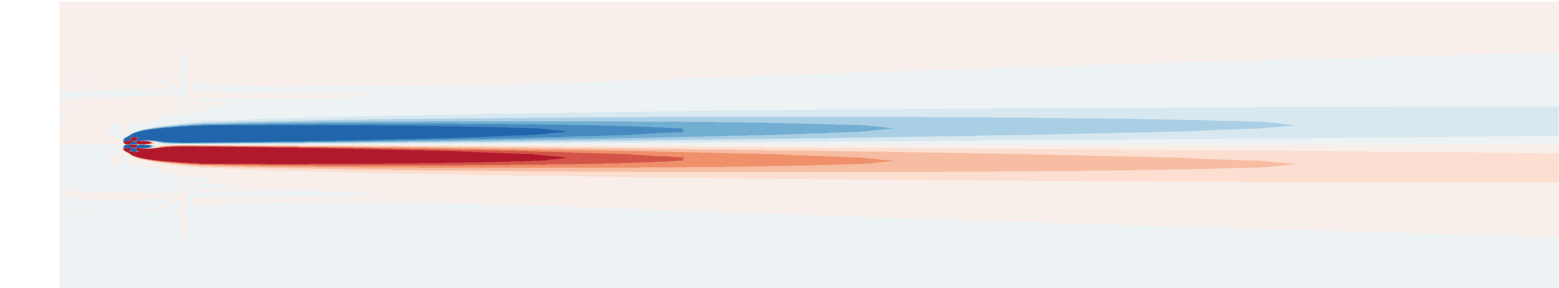}}}\hfill
		\subfigure[$C_{7}$ \label{fig2:C7-subfig3}]{
			\resizebox*{4cm}{!}{\includegraphics{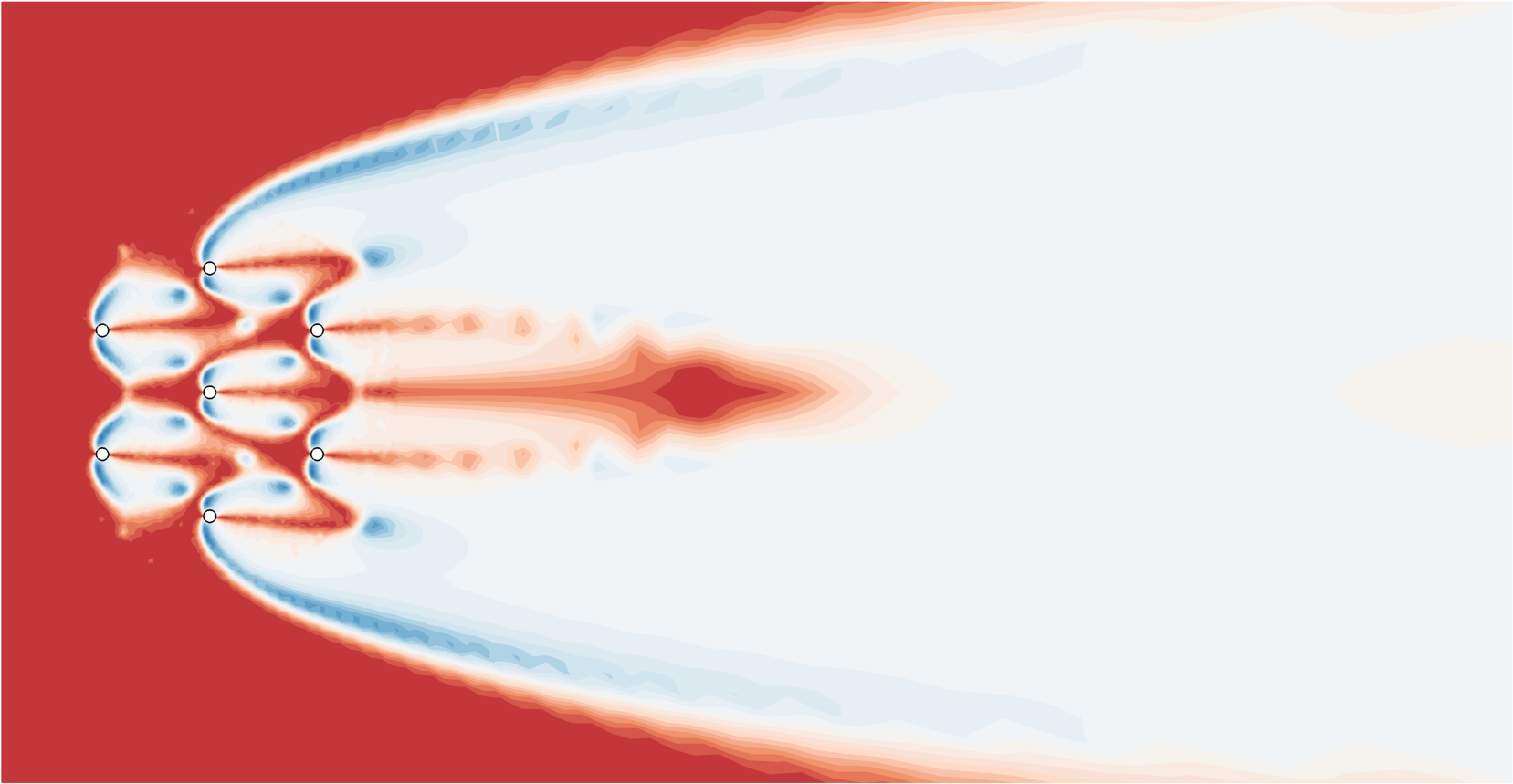}}}\\
			
		\subfigure[$C_{19}$ \label{fig2:C19-subfig2}]{
			\resizebox*{9cm}{!}{\includegraphics{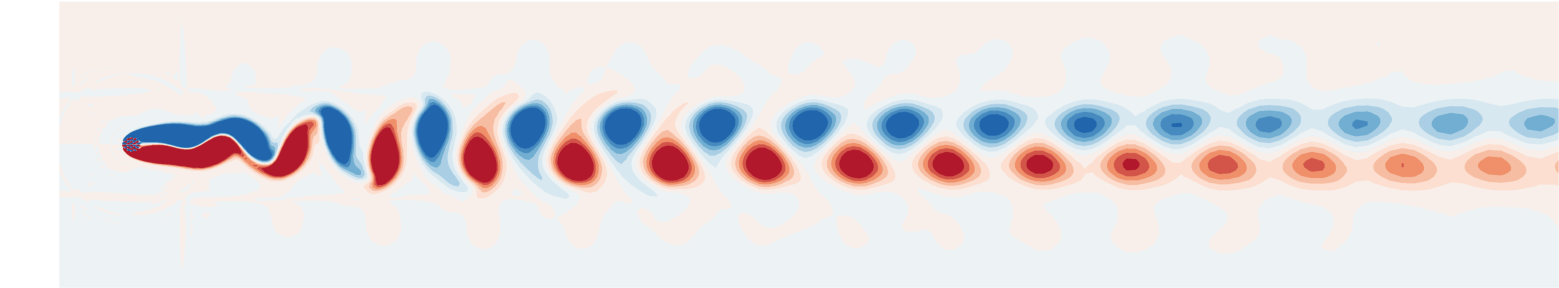}}}\hfill
		\subfigure[$C_{19}$ \label{fig2:C19-subfig3}]{
			\resizebox*{4cm}{!}{\includegraphics{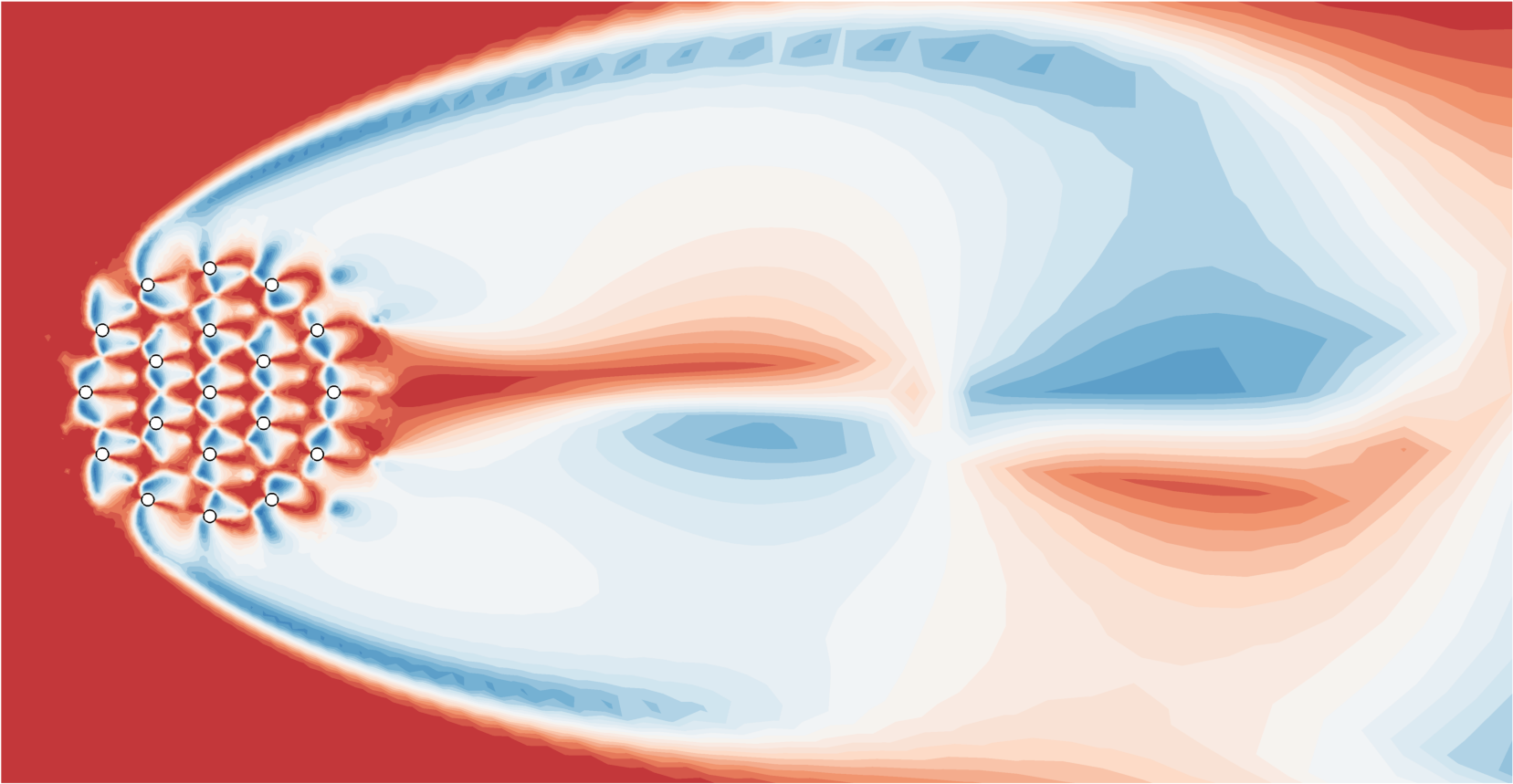}}}\\
			
		\subfigure[$C_{37}$ \label{fig2:C37-subfig2}]{
			\resizebox*{9cm}{!}{\includegraphics{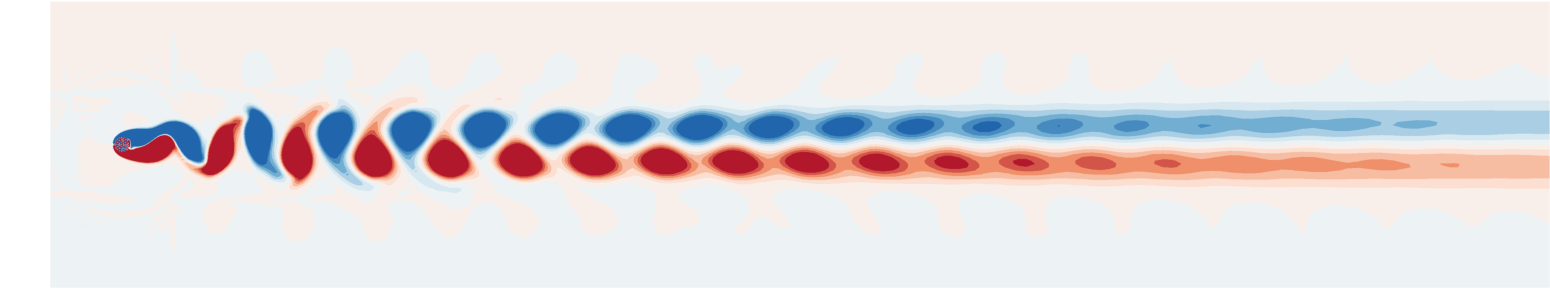}}}\hfill
		\subfigure[$C_{37}$ \label{fig2:C37-subfig3}]{
			\resizebox*{4cm}{!}{\includegraphics{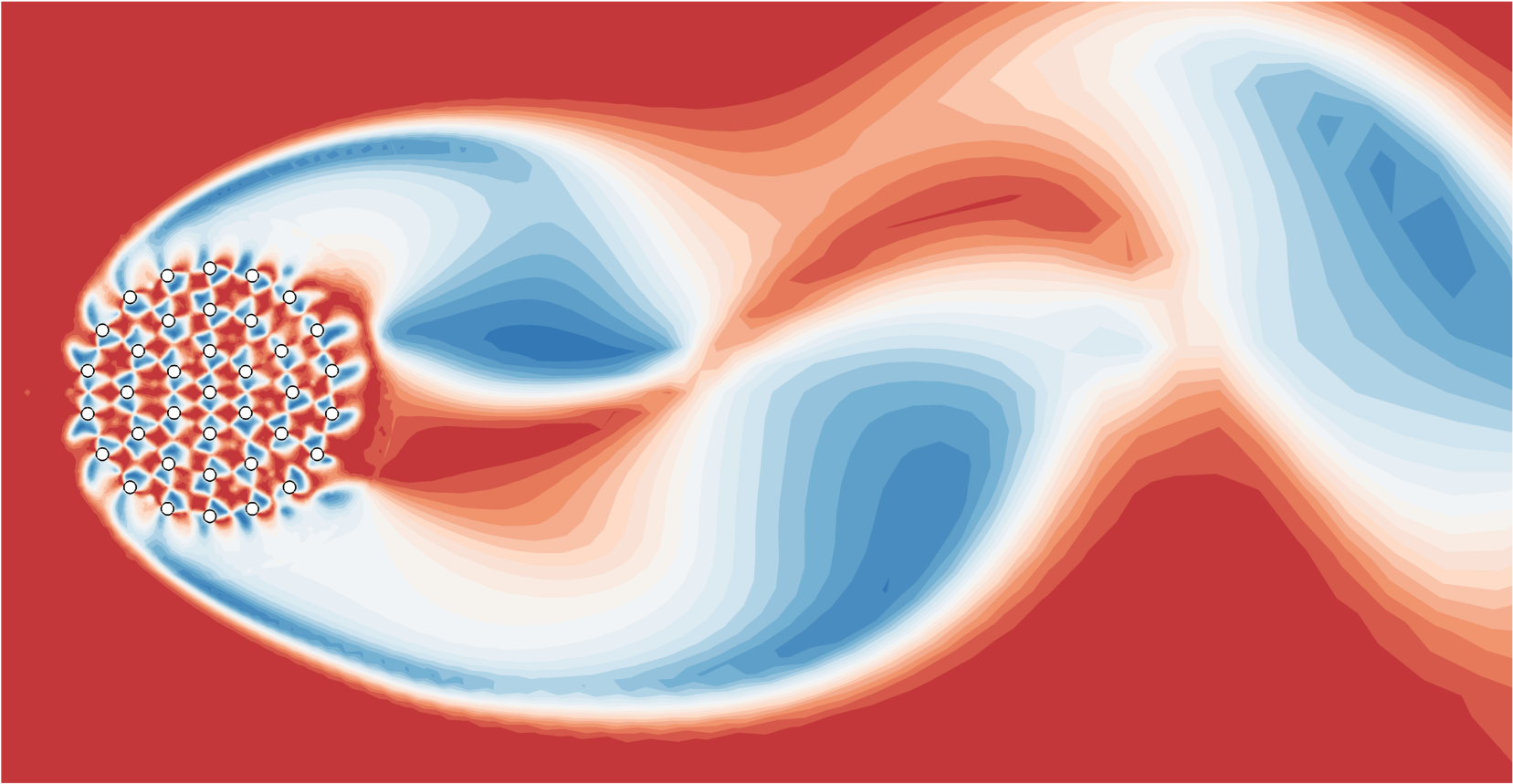}}}\\

		\subfigure[$C_{61}$ \label{fig2:C61-subfig2}]{
			\resizebox*{9cm}{!}{\includegraphics{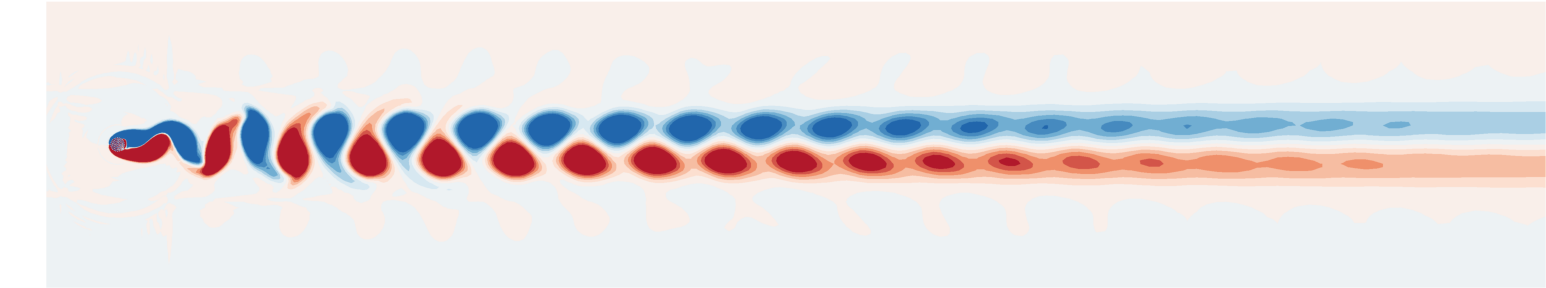}}}\hfill
		\subfigure[$C_{61}$ \label{fig2:C61-subfig3}]{
			\resizebox*{4cm}{!}{\includegraphics{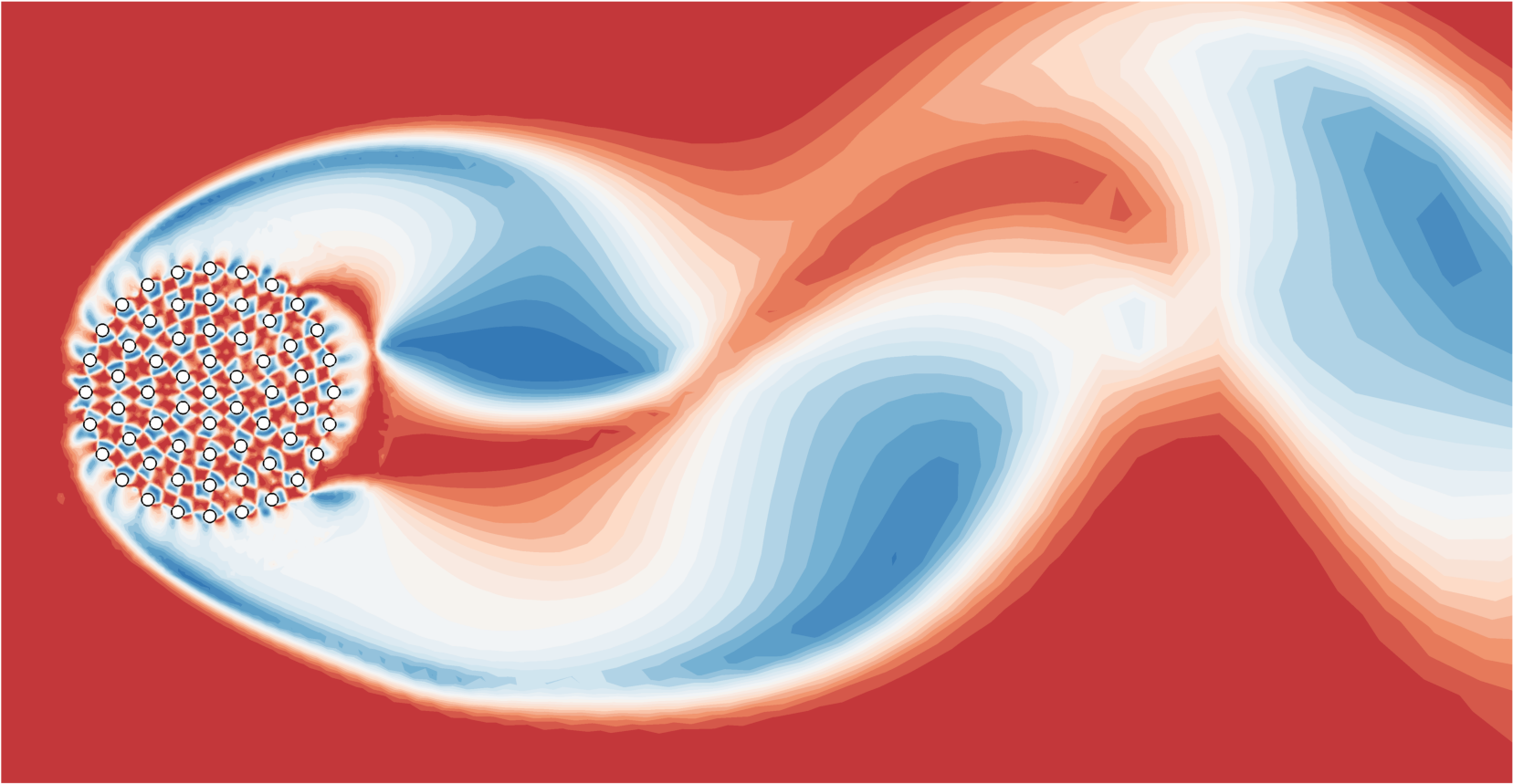}}}\\

		\subfigure[$C_{97}$ \label{fig2:C97-subfig2}]{
			\resizebox*{9cm}{!}{\includegraphics{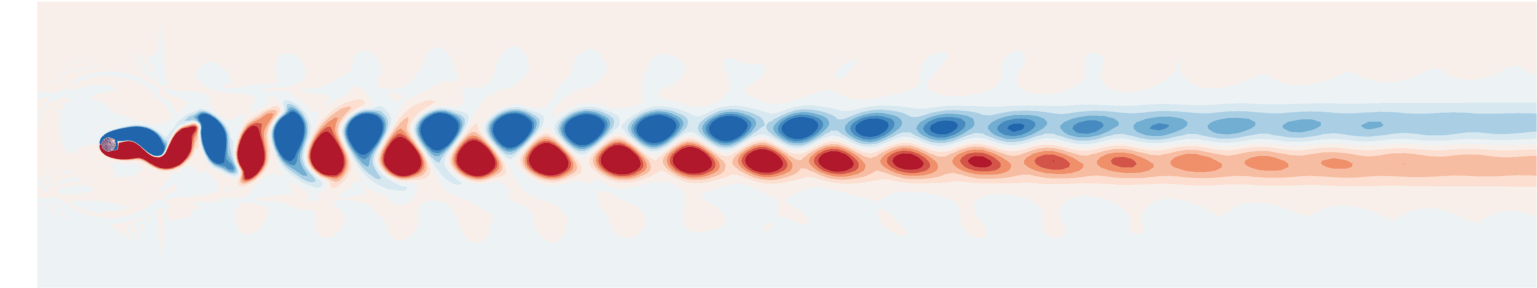}}}\hfill
		\subfigure[$C_{97}$ \label{fig2:C97-subfig3}]{
			\resizebox*{4cm}{!}{\includegraphics{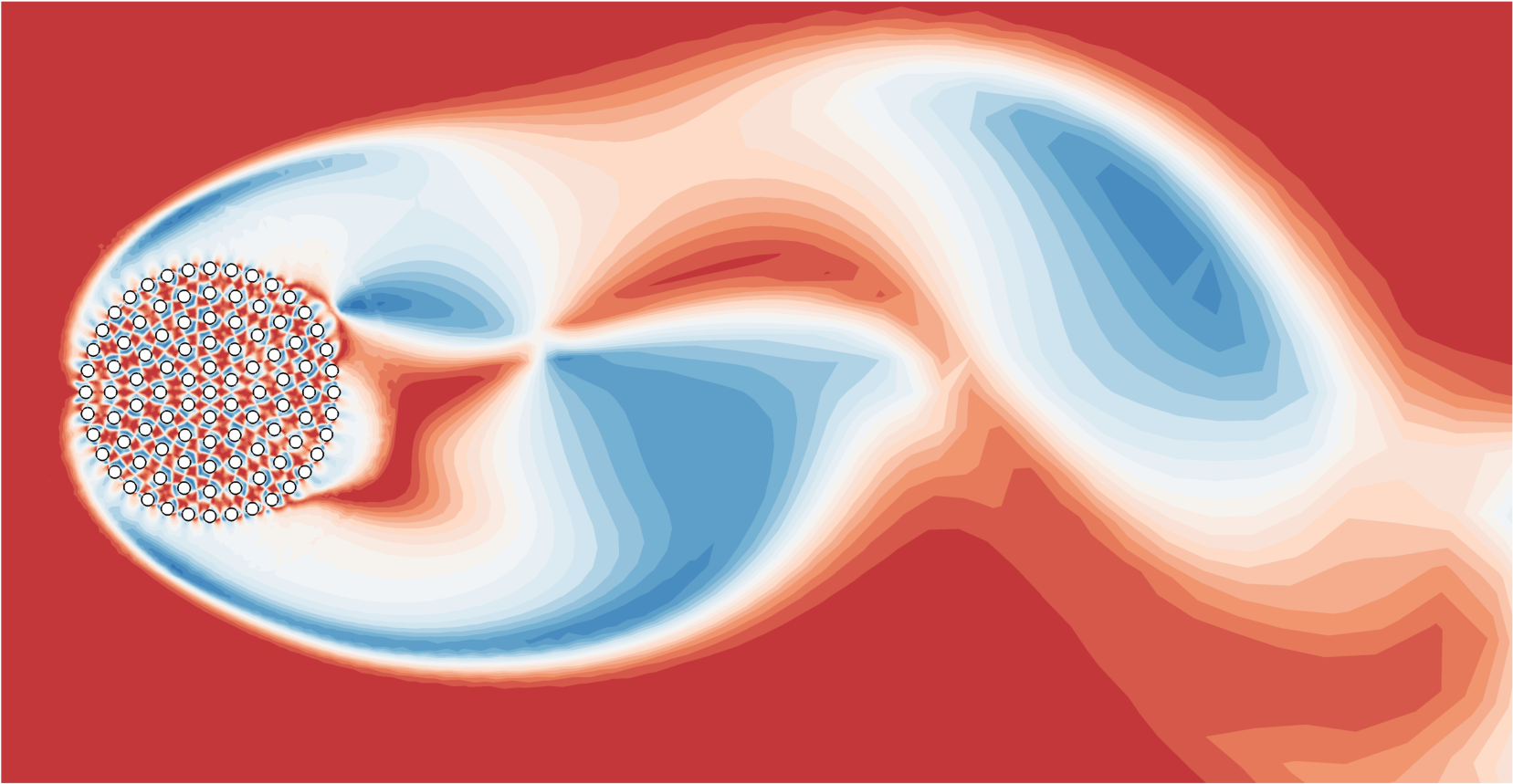}}}\\

		\subfigure[$C_{139}$ \label{fig2:C139-subfig2}]{
			\resizebox*{9cm}{!}{\includegraphics{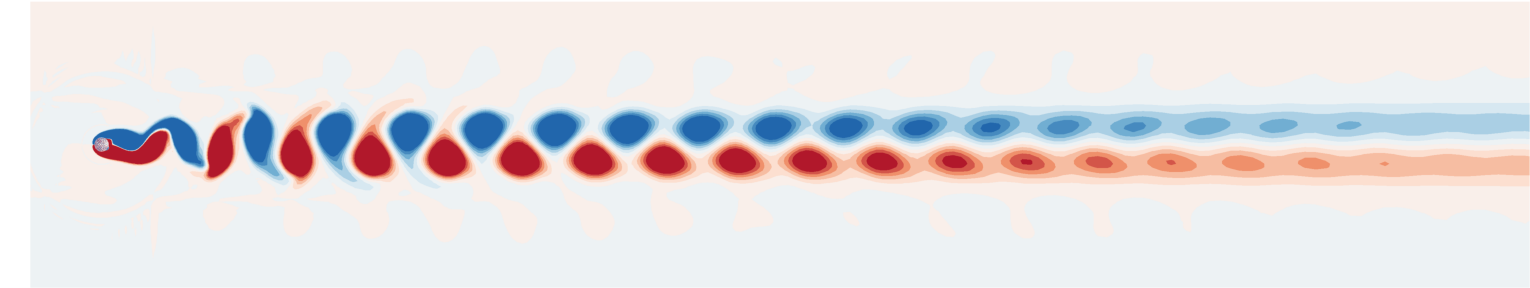}}}\hfill
		\subfigure[$C_{139}$ \label{fig2:C139-subfig3}]{
			\resizebox*{4cm}{!}{\includegraphics{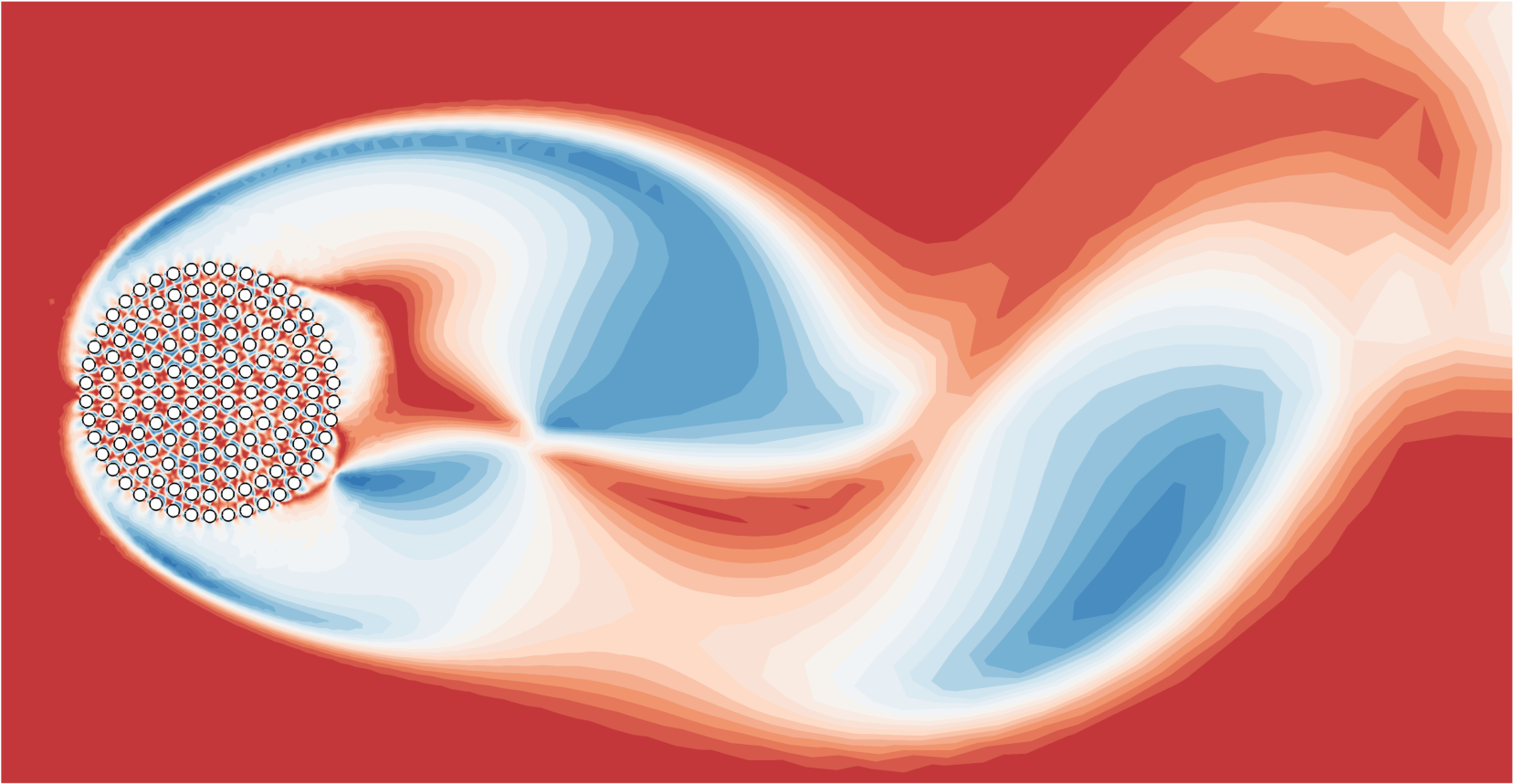}}}\\

		\subfigure[$C_{\text{solid}}$ \label{fig2:solid-subfig2}]{
			\resizebox*{9.cm}{!}{\includegraphics{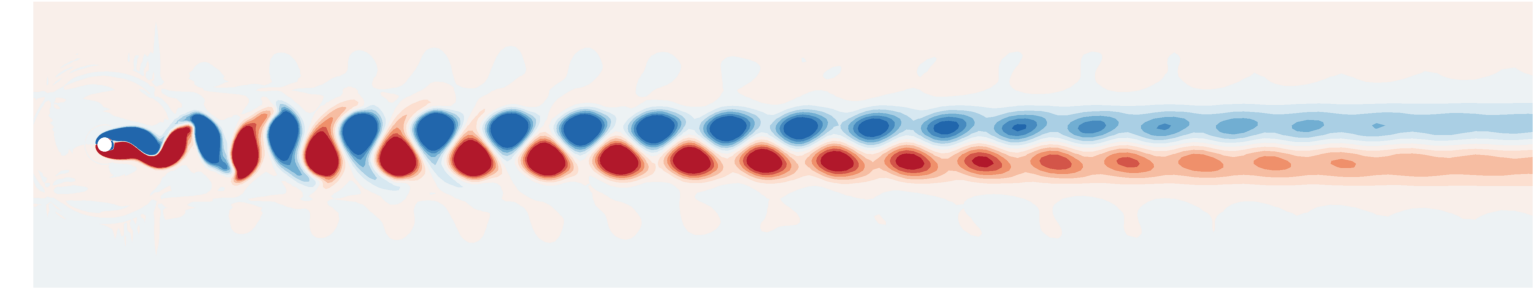}}}\hfill
		\subfigure[$C_{\text{solid}}$ \label{fig2:solid-subfig3}]{
			\resizebox*{4.cm}{!}{\includegraphics{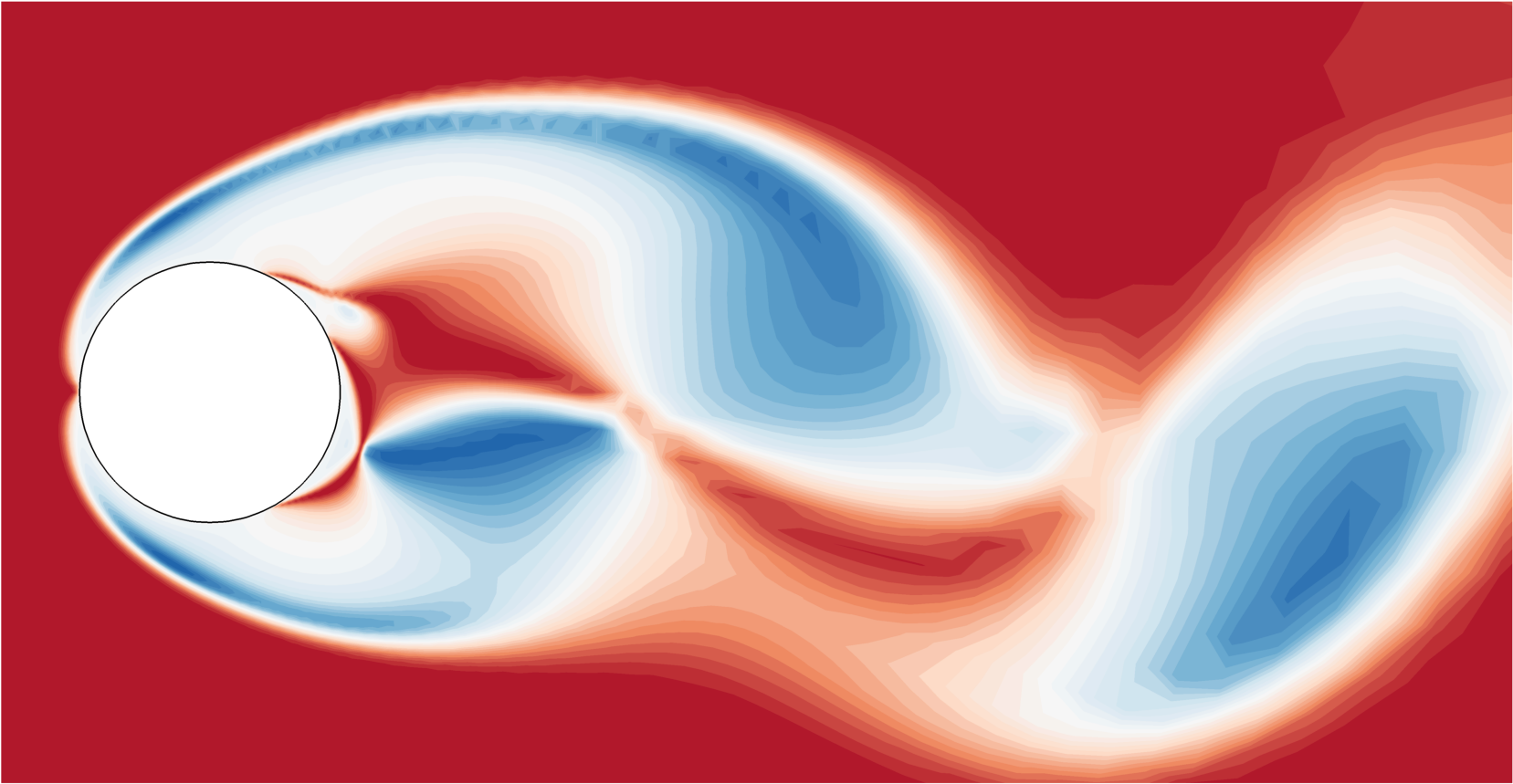}}}
		
    \caption{\label{fig2:omegaandQ2} Vorticity field, $\omega$ (left), and dimensionless second invariant of the velocity gradient tensor, $Q$ (right), for different array configurations. Configurations shown are: (a,b) $C_7$, (c,d) $C_{19}$, (e,f) $C_{37}$, (g,h) $C_{61}$, (i,j) $C_{97}$, (k,l) $C_{139}$, and (m,n) $C_{\text{solid}}$. Twenty contour levels are uniformly spaced between $-0.2$ and $0.2$ for $\omega$, and between $-1.0$ and $1.0$ for $Q$, indicating values from low (blue) to high (red). }
	\end{center}
\end{figure}

\subsection{Base flows and pressure by SFD method}

In this section we present the base state of the flow past the cylinder array.

\subsection{Forces acting on individual cylinders within the array}
\begin{figure}
	\begin{center}
		\subfigure[$C_7$ \label{fig:fig-ScatterForcesC7}]{
			\resizebox*{3.55cm}{!}{\includegraphics{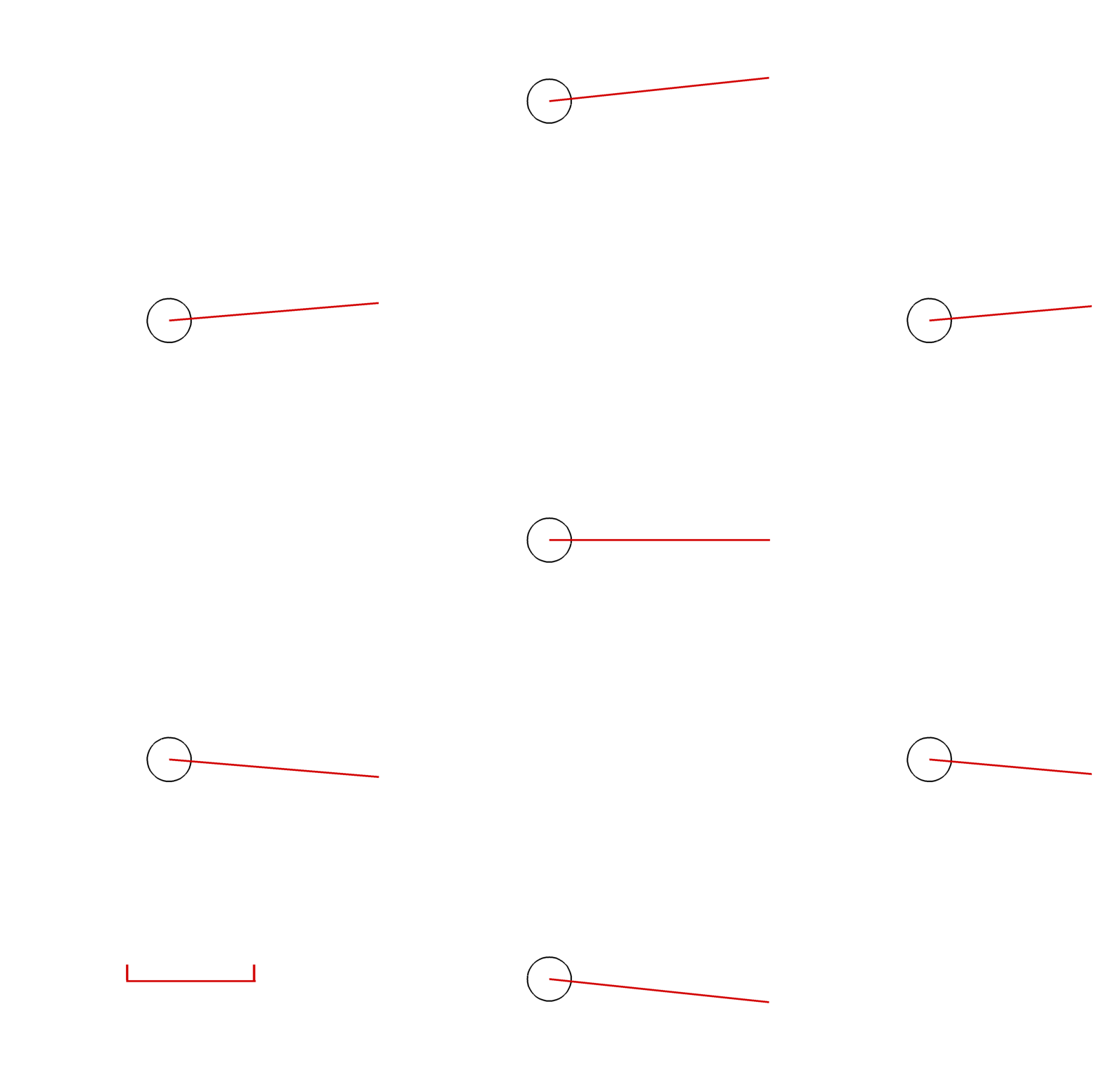}}}\hfill
		\subfigure[$C_{19}$ \label{fig:fig-ScatterForcesC19}]{
			\resizebox*{3.55cm}{!}{\includegraphics{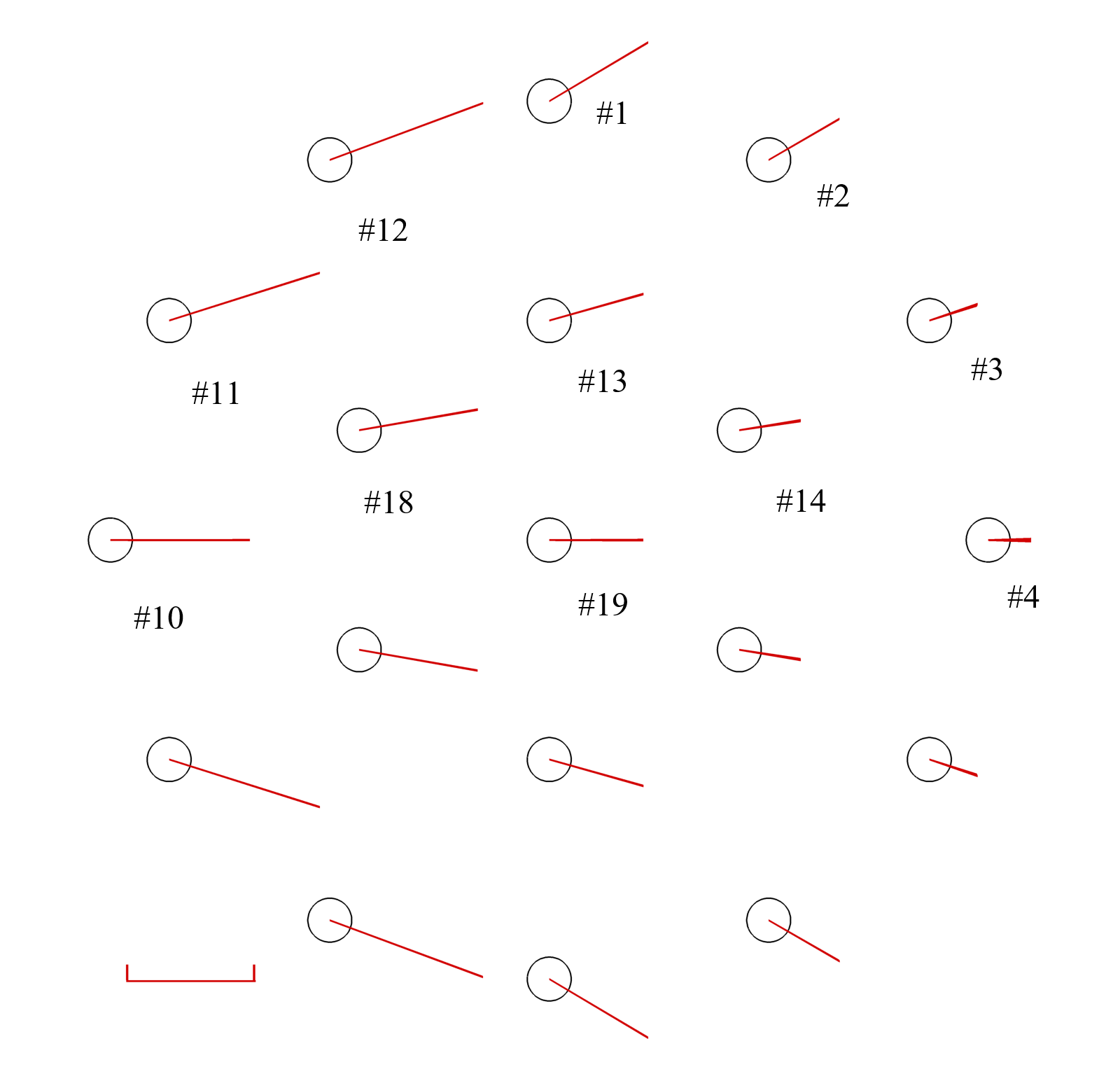}}}\hfill
		\subfigure[$C_{37}$ \label{fig:fig-ScatterForcesC37}]{
			\resizebox*{3.55cm}{!}{\includegraphics{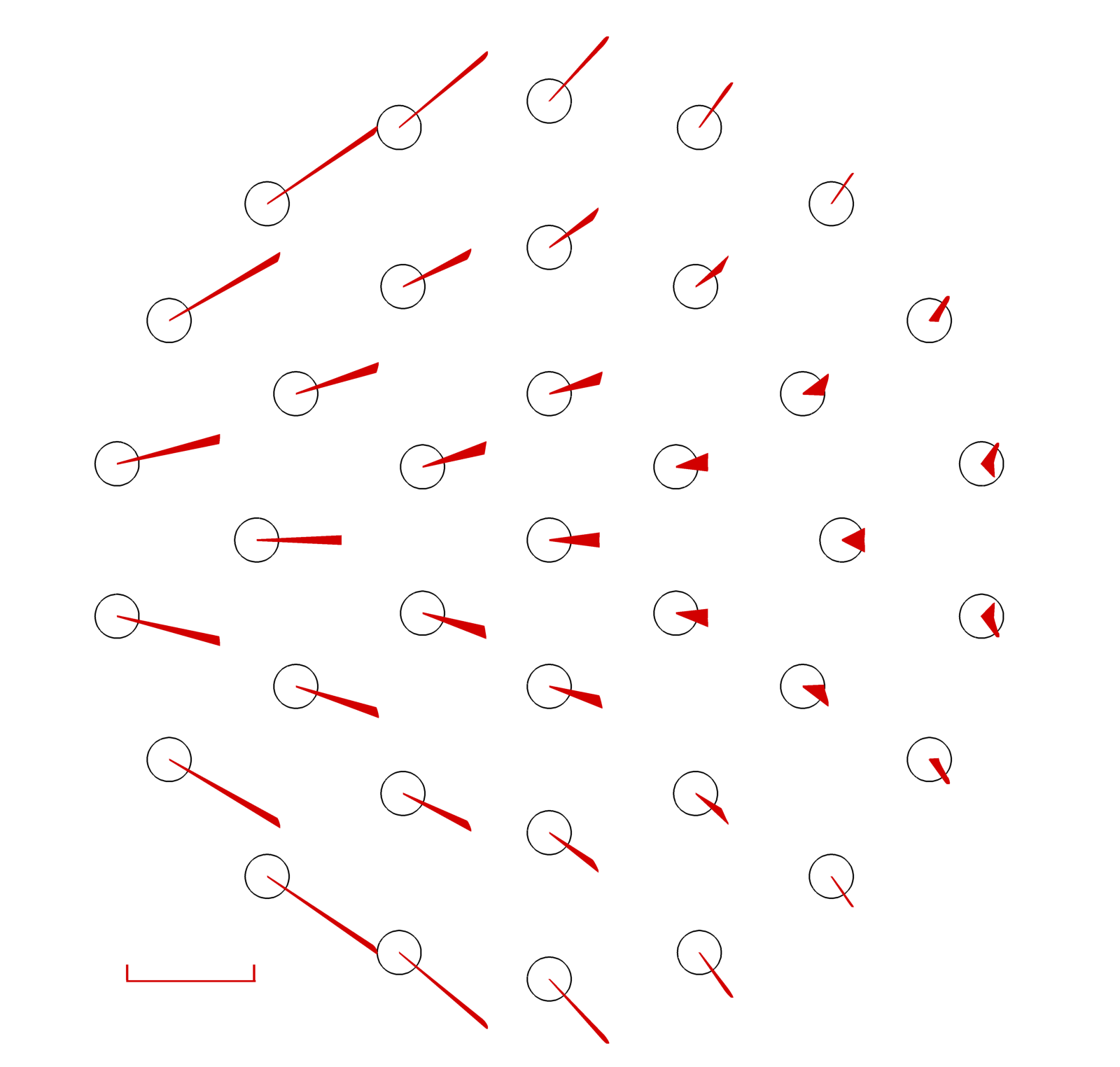}}}\\
			
		\subfigure[$C_{61}$ \label{fig:fig-ScatterForcesC61}]{
			\resizebox*{3.55cm}{!}{\includegraphics{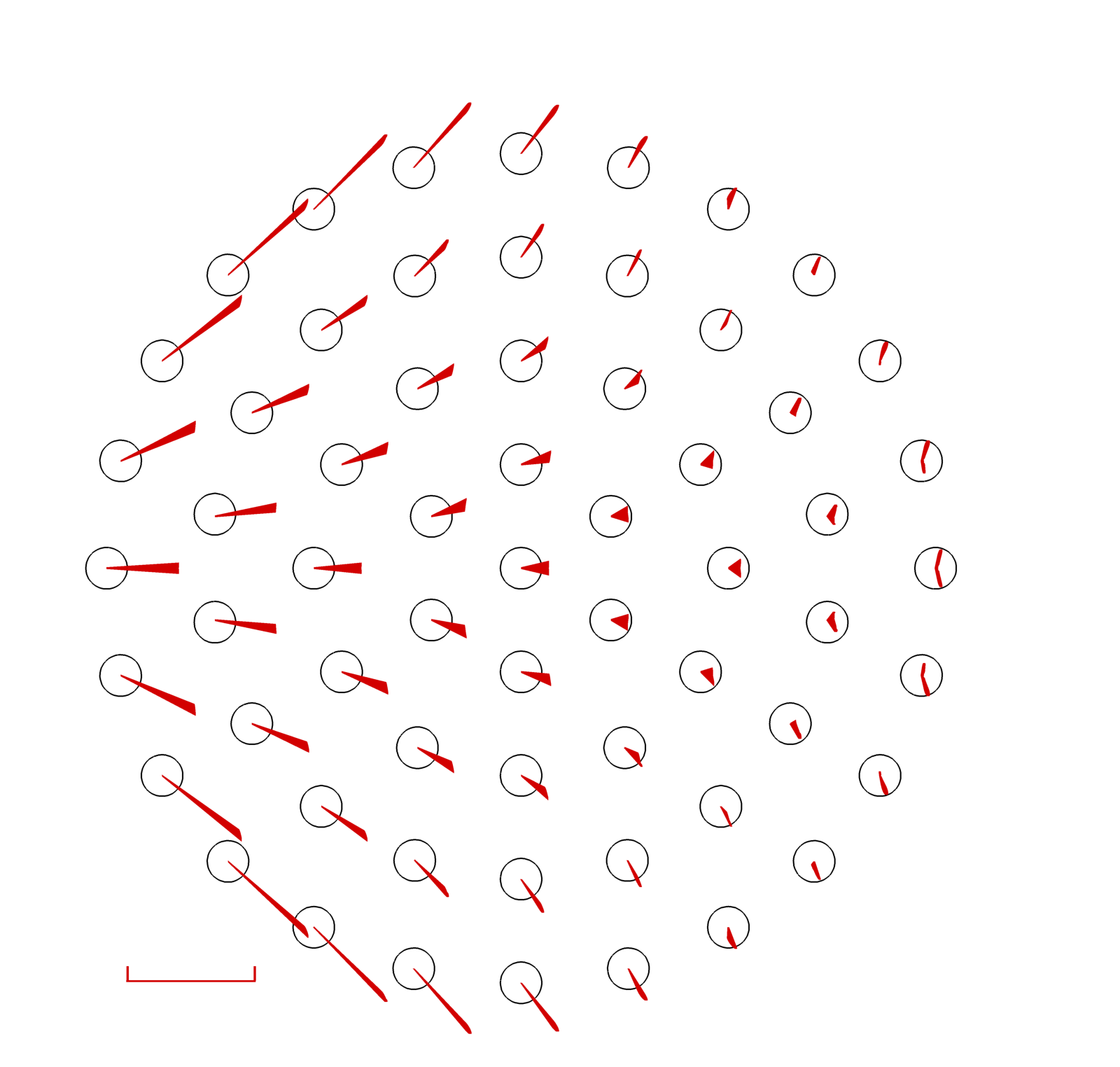}}}\hfill
		\subfigure[$C_{97}$ \label{fig:fig-ScatterForcesC97}]{
			\resizebox*{3.55cm}{!}{\includegraphics{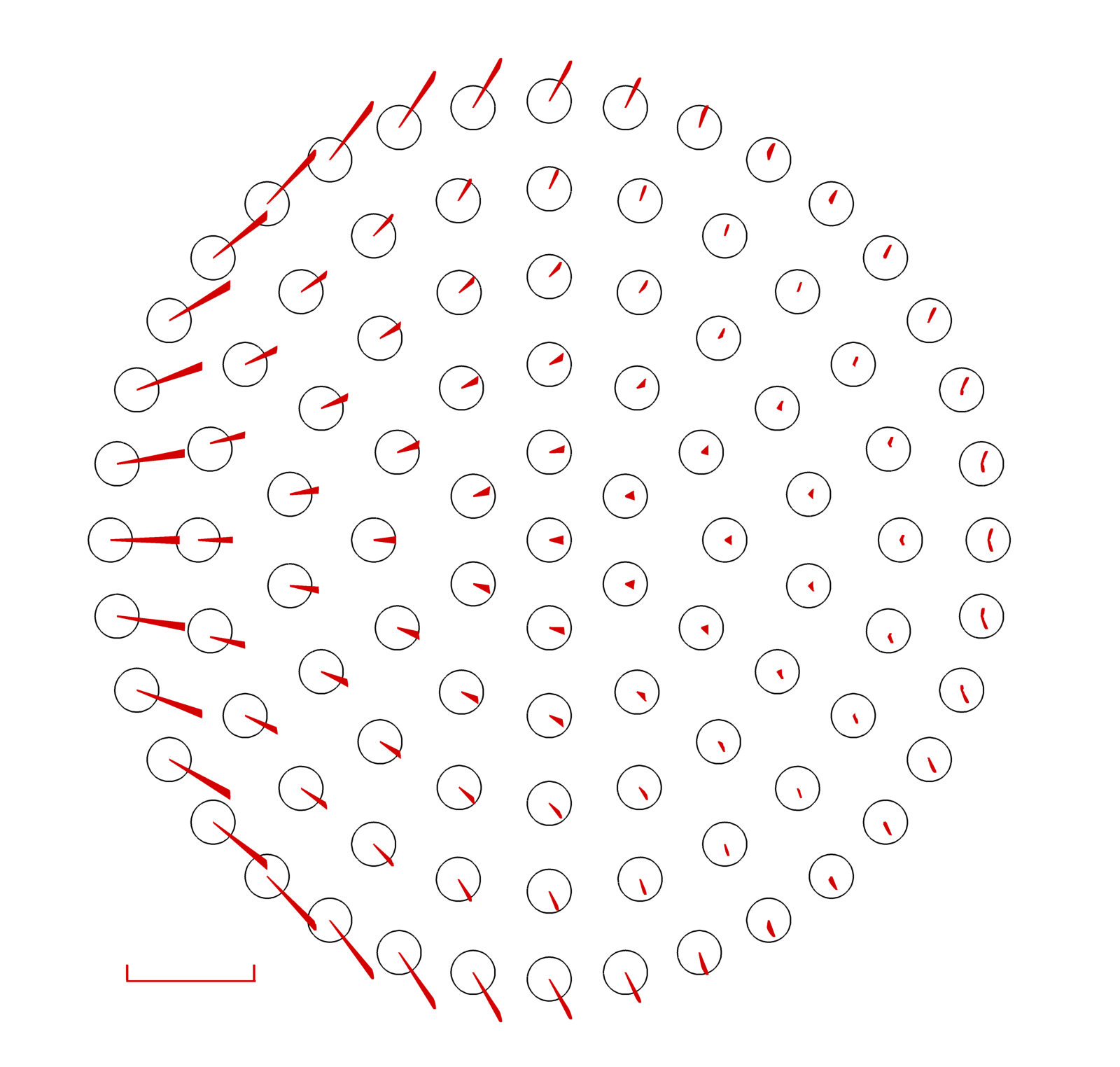}}}\hfill
		\subfigure[$C_{139}$ \label{fig:fig-ScatterForcesC139}]{
			\resizebox*{3.55cm}{!}{\includegraphics{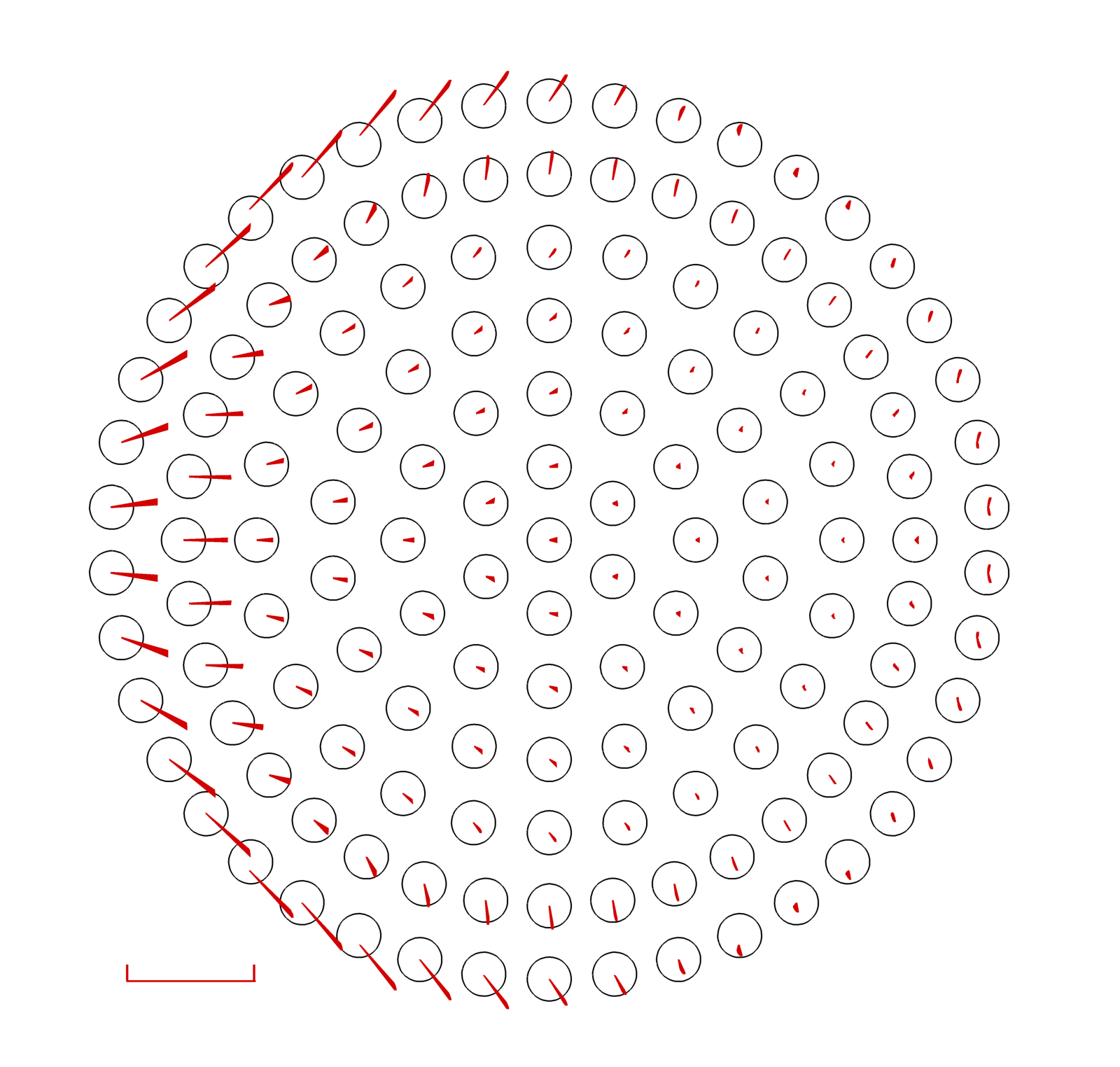}}}\\
			
		\subfigure[$C_{\text{solid}}$ \label{fig:fig-ScatterForcessolid}]{
			\resizebox*{3.55cm}{!}{\includegraphics{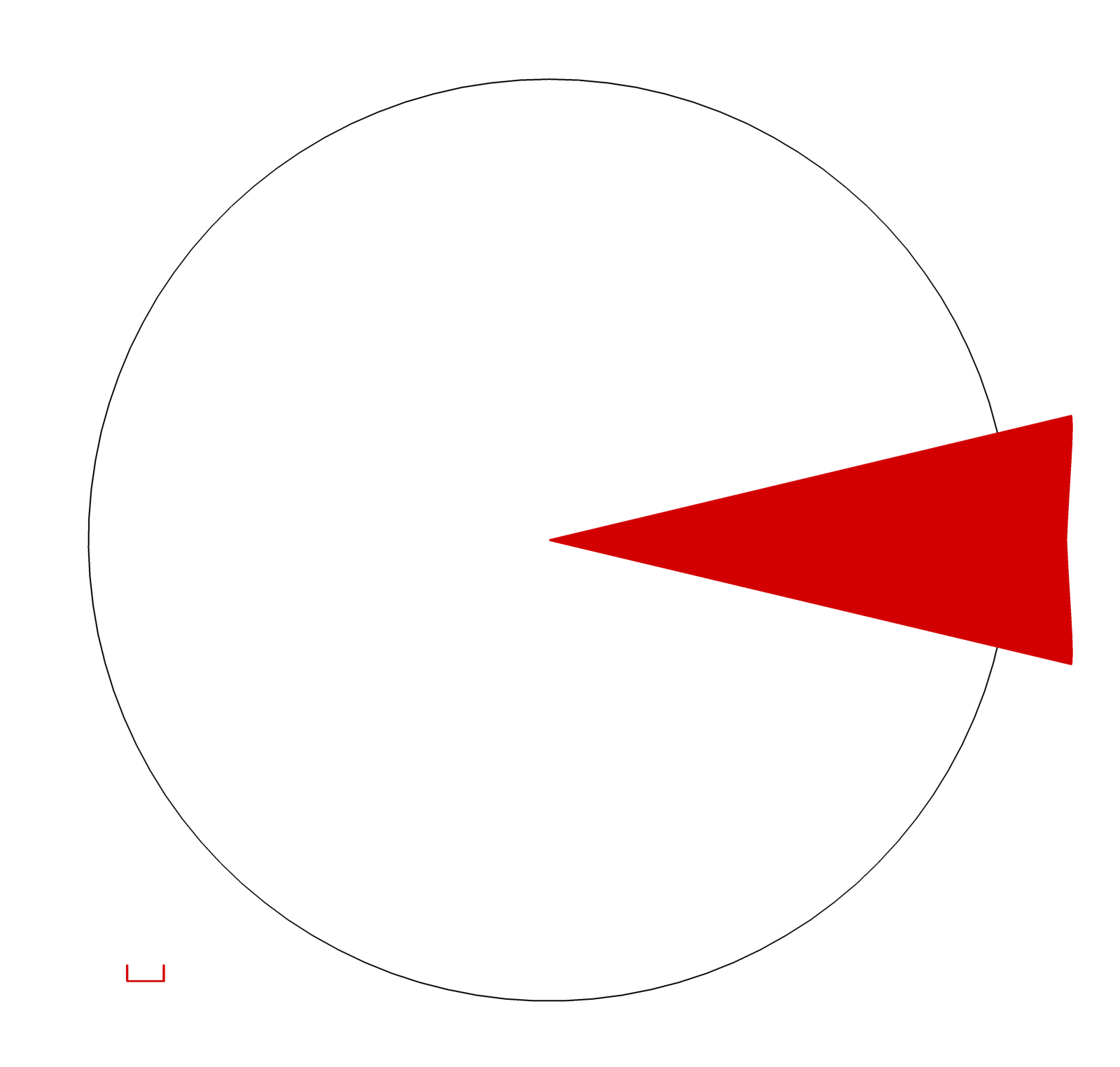}}}
		
		\caption{\label{fig:fig-ScatterForces} Fluid force scatter plots characterizing the instantaneous load distributions on cylinder arrays: (a) $C_7$, (b) $C_{19}$, (c) $C_{37}$, (d) $C_{61}$, (e) $C_{97}$, (f) $C_{139}$, and (g) $C_\text{solid}$ at $Re_D = 100$. A reference line indicating unit drag coefficient is provided beneath each configuration for quantitative comparison. Cylinder geometries are rendered to scale, with the mean flow oriented from left to right. The force scale for the solid cylinder case is adjusted to optimize visualization clarity. Several cylinders in configuration (b) are labelled for identification in subsequent analysis.}
	\end{center}
\end{figure}

Following~\cite{NICOLLE2011}, we examine the force scatter characteristics for each array configuration, which reveal both the steady (mean) and unsteady (fluctuating) components of the fluid forces acting on individual cylinders. Figure~\ref{fig:fig-ScatterForces} presents the force scatter plots at $Re_D = 100$. The resulting $Re_d = 4.76$ indicates that the flow around a small cylinder in unobstructed, uniform conditions would be steady.

In each panel, the outer boundary of every cylinder is shown in grey, while the scatter of dimensionless force vectors, originating from each cylinder's centre, is plotted in red. Each force vector comprises the streamwise drag coefficient and transverse lift coefficient. A reference drag coefficient of unity is included below each array for scale. For comparison, the solid cylinder exhibits an unsteady flow field corresponding to a fully developed vortex street. The unity drag coefficient length has been scaled for optimal visualization.

For array $C_7$ (figure~\ref{fig:fig-ScatterForcesC7}), the fluid forces acting on individual cylinders remain steady, with all forces exhibiting good symmetry about $y=0$. While the central cylinder experiences a force purely in the downstream direction, forces on the remaining cylinders are offset transversely, indicating flow divergence through the array. This behaviour arises from weak fluid dynamic interactions among the cylinder group. The two upstream cylinders experience slightly higher drag forces than their downstream counterparts, as expected given the flow obstruction created by upstream elements.

\begin{figure}
	\centering
	\includegraphics[width=0.75\linewidth]{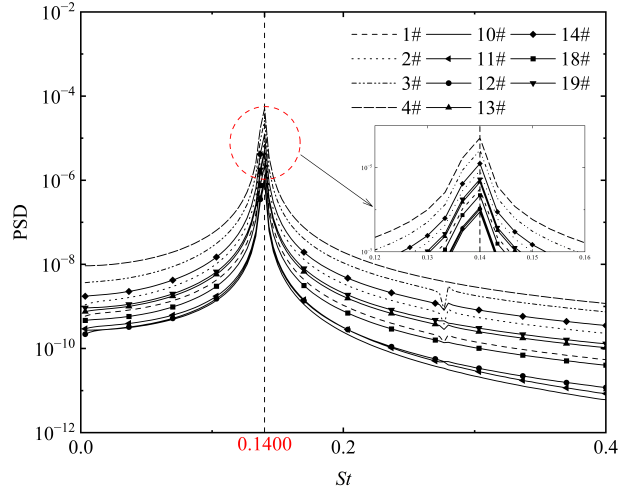}
	\caption{Power spectral density (PSD) of the lift coefficient ($C_l$) time histories for individual cylinders in the $C_{19}$ array at $Re_D = 100$. Owing to the symmetry of the configuration, only cylinders with $y \geq 0$ are shown. The numbering scheme for the cylinders is provided in figure~\ref{fig:fig-ScatterForcesC19}.}
\label{fig:PSD-c19}
\end{figure}

Array $C_{19}$ (figure~\ref{fig:fig-ScatterForcesC19}) demonstrates stronger inter-cylinder interactions due to closer spacing and increased blockage. All cylinders experience reduced drag forces, with more pronounced reduction occurring in the downstream portion of the array. The global flow field is unsteady, as evidenced by a recirculating wake that persists approximately $2D$ downstream of the array (figure~1d); nevertheless, the force scatters suggest a relatively weak influence of this wake on the downstream cylinders. Spectral analysis of the lift coefficient ($C_l$), presented in figure~\ref{fig:PSD-c19}, indicates that the unsteady forces on all cylinders are synchronized at a single frequency, $St = 0.1400$. This frequency is consistent with the vortex shedding from the array as a whole ( ). A spatial variation in the oscillation amplitude is observed: it is maximized for the downstream cylinders (\#4 and \#3) and minimized for the upstream cylinders (\#10 and \#11).

With further increase in solid fraction from intermediate to high values ($C_{37}$-$C_{139}$, figures~\ref{fig:fig-ScatterForcesC139}-~\ref{fig:fig-ScatterForcesC19}), the forces on individual cylinders become unsteady. Similar to the $C_{19}$ case, all cylinders within each configuration experience forces oscillating at the same frequency. In all cases, the mean forces on upstream cylinders indicate flow divergence at the array front. Downstream cylinders exhibit characteristic butterfly-wing-shaped force scatter patterns, resulting from strong influence by the array's recirculating wake.

The increased blockage generally reduces force magnitudes on all cylinders. The most upstream cylinders remain influenced by the incident flow, while the most downstream cylinders are dominated by wake effects. For array $C_{139}$, interior cylinders experience minimal fluid dynamic influence due to severely limited flow penetration, causing the array to behave essentially as a solid body.

\begin{figure}
	\begin{center}
		\subfigure[$C_7$ \label{fig:averageforcevector-C7}]{
			\resizebox*{3.55cm}{!}{\includegraphics{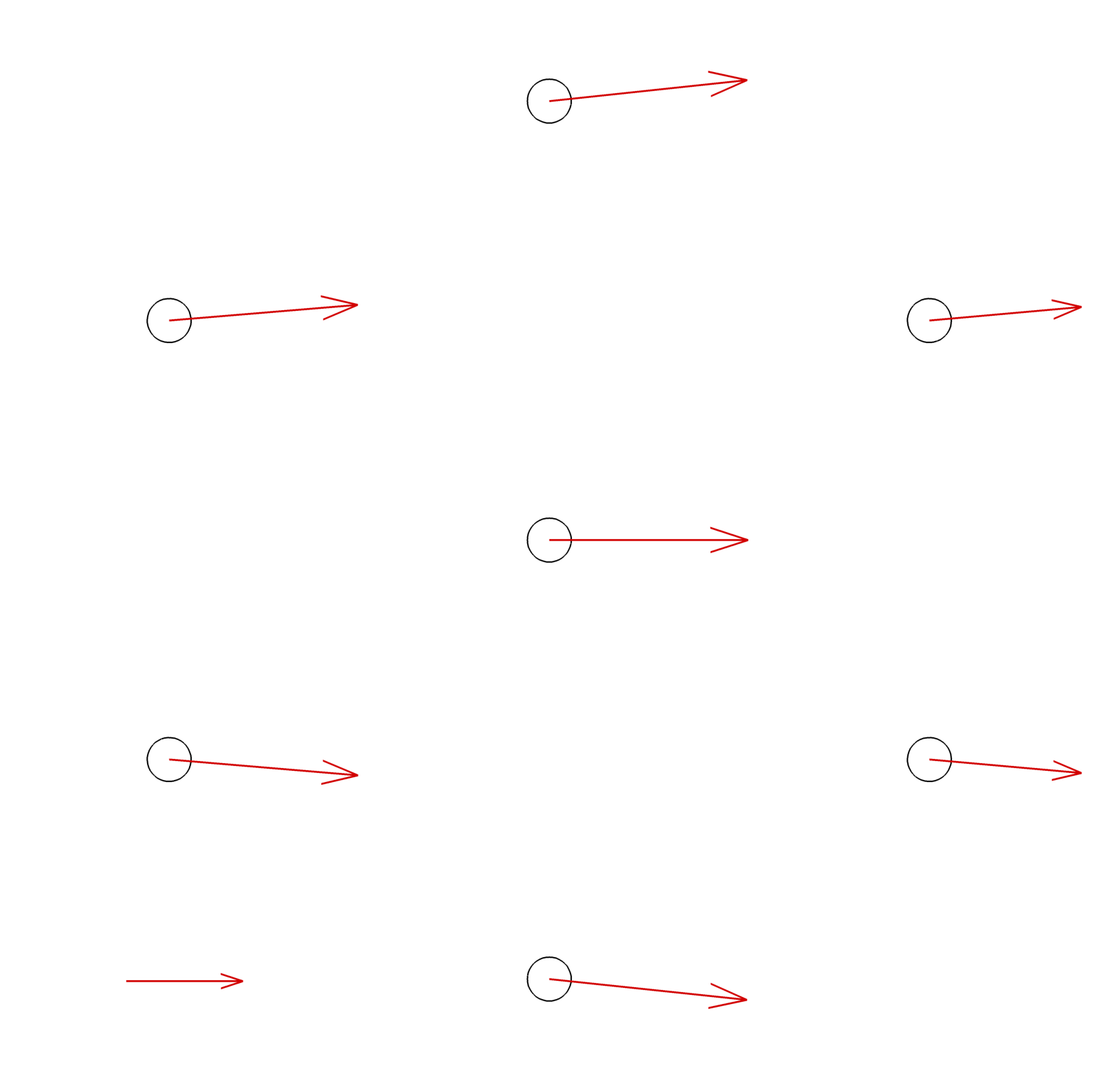}}}\hfill
		\subfigure[$C_{19}$ \label{fig:averageforcevector-C19}]{
			\resizebox*{3.55cm}{!}{\includegraphics{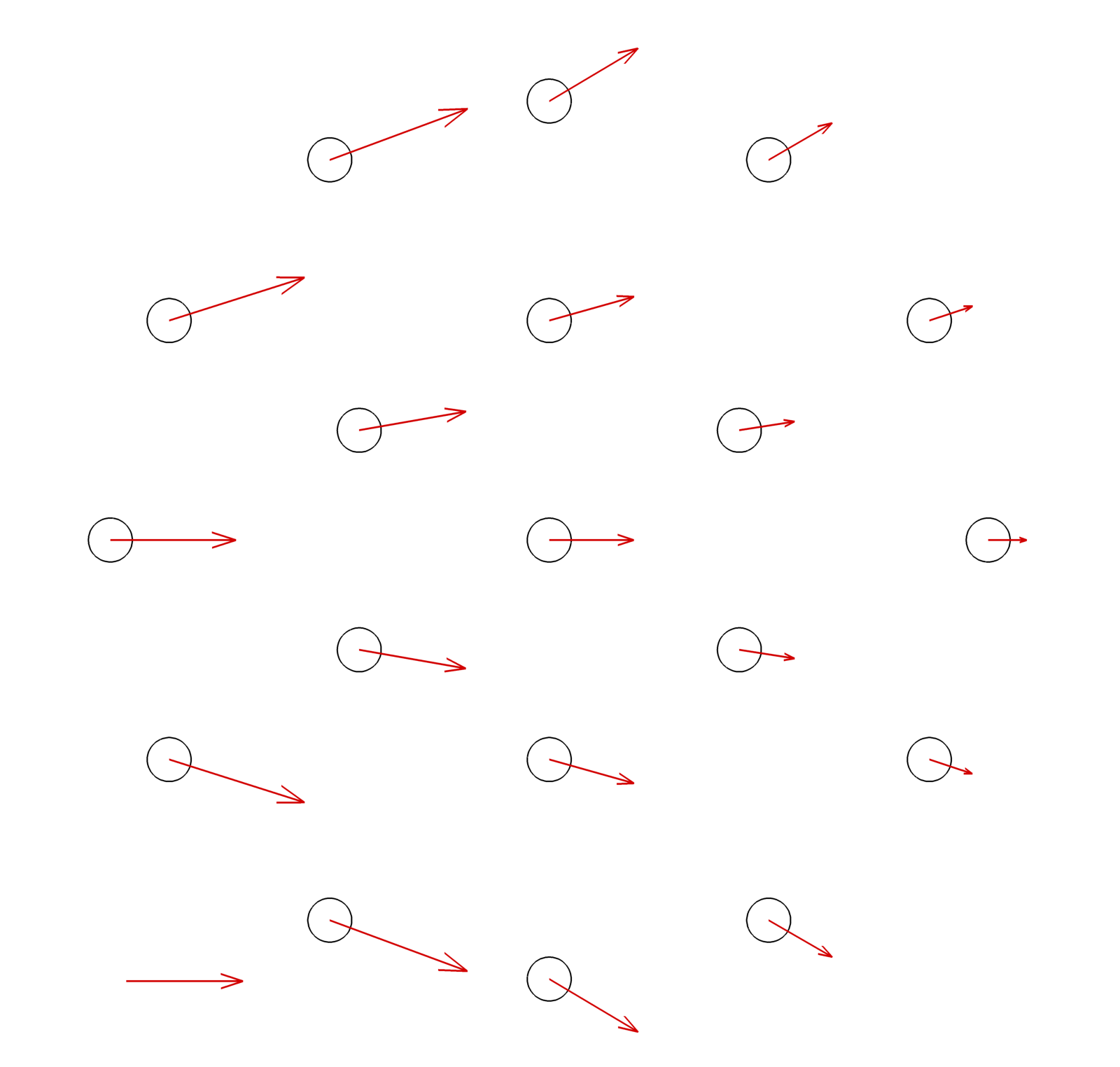}}}\hfill
		\subfigure[$C_{37}$ \label{fig:averageforcevector-C37}]{
			\resizebox*{3.55cm}{!}{\includegraphics{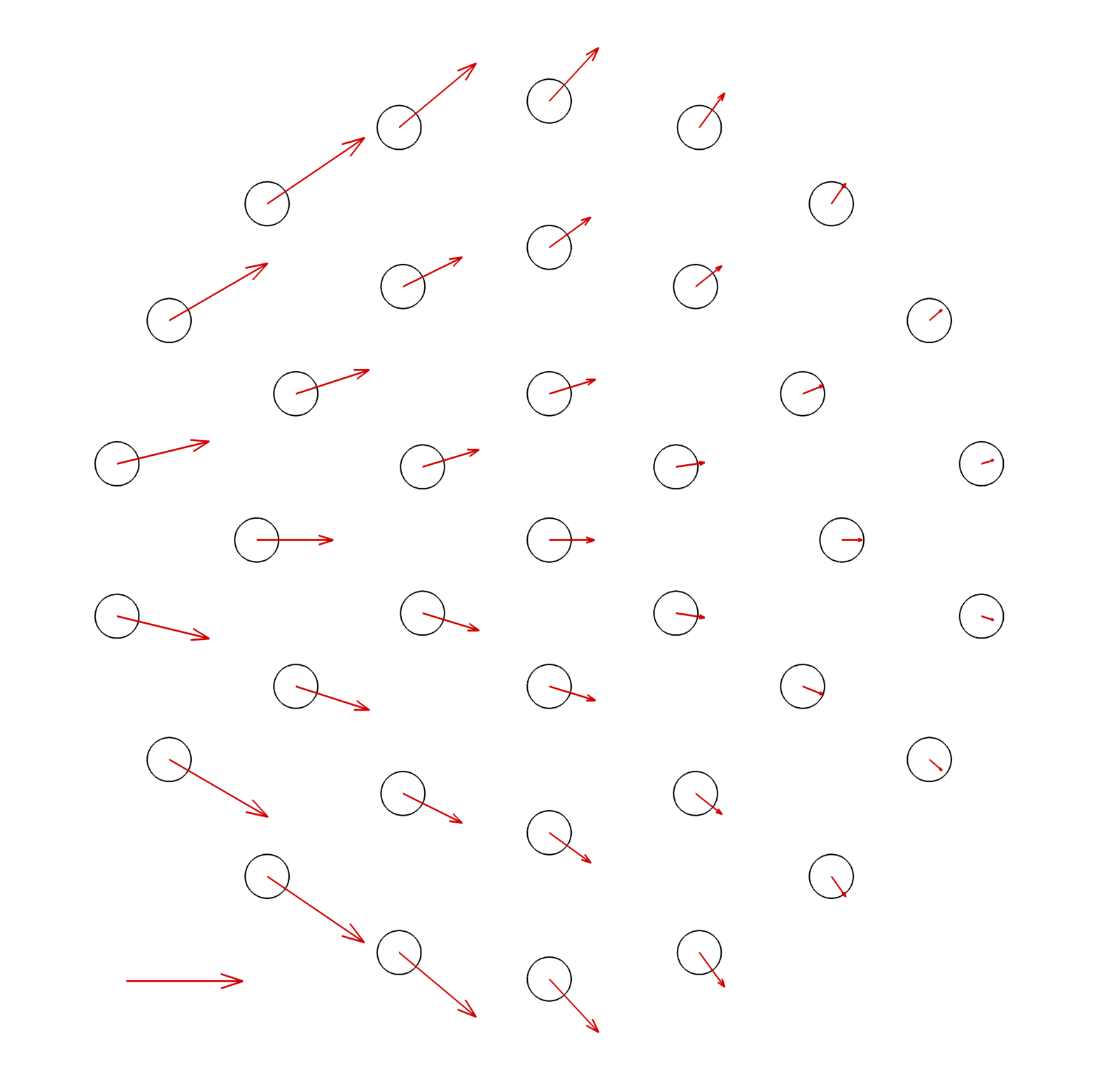}}}\\
			
		\subfigure[$C_{61}$ \label{fig:averageforcevector-C61}]{
			\resizebox*{3.55cm}{!}{\includegraphics{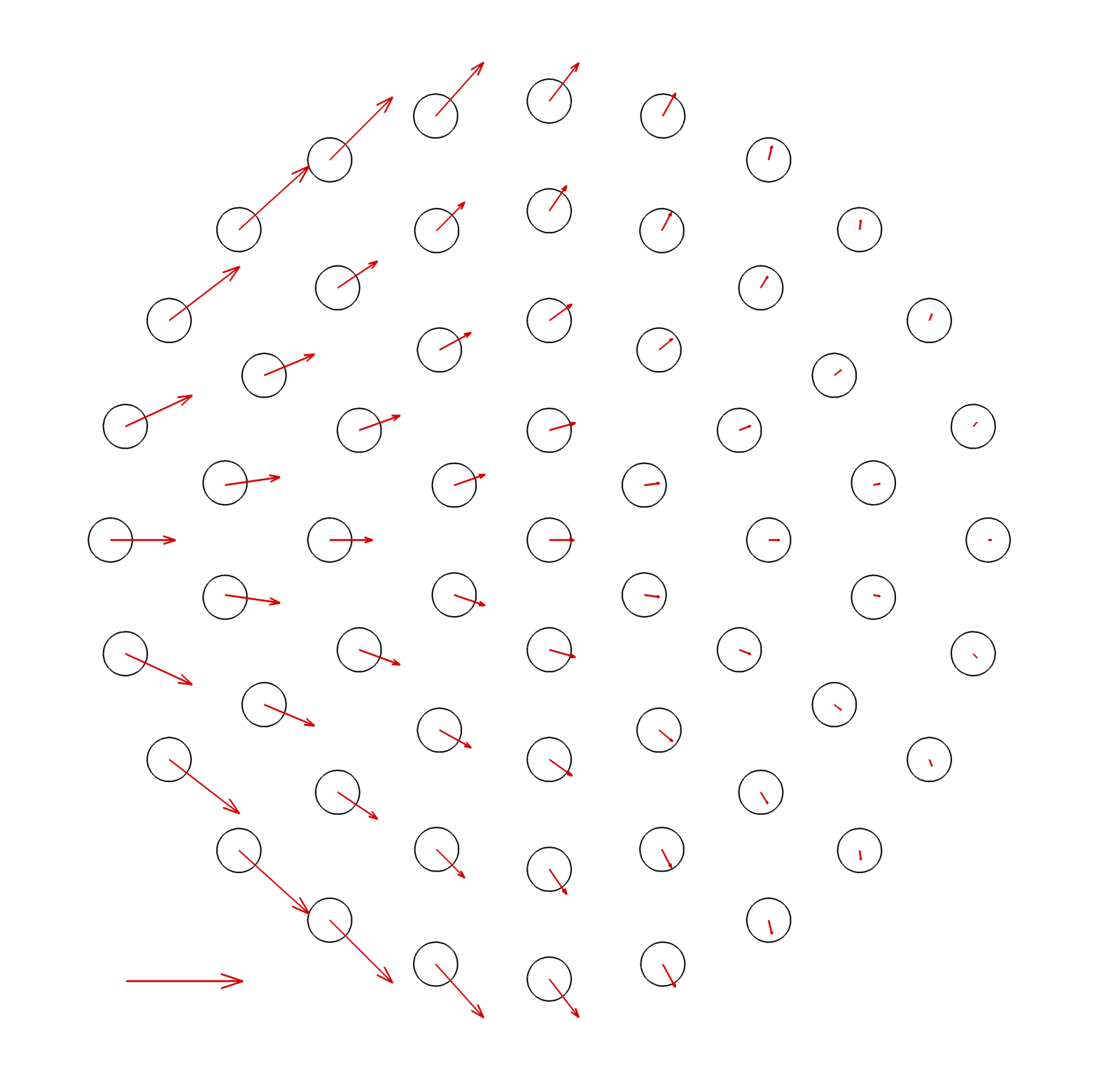}}}\hfill
		\subfigure[$C_{97}$ \label{fig:averageforcevector-C97}]{
			\resizebox*{3.55cm}{!}{\includegraphics{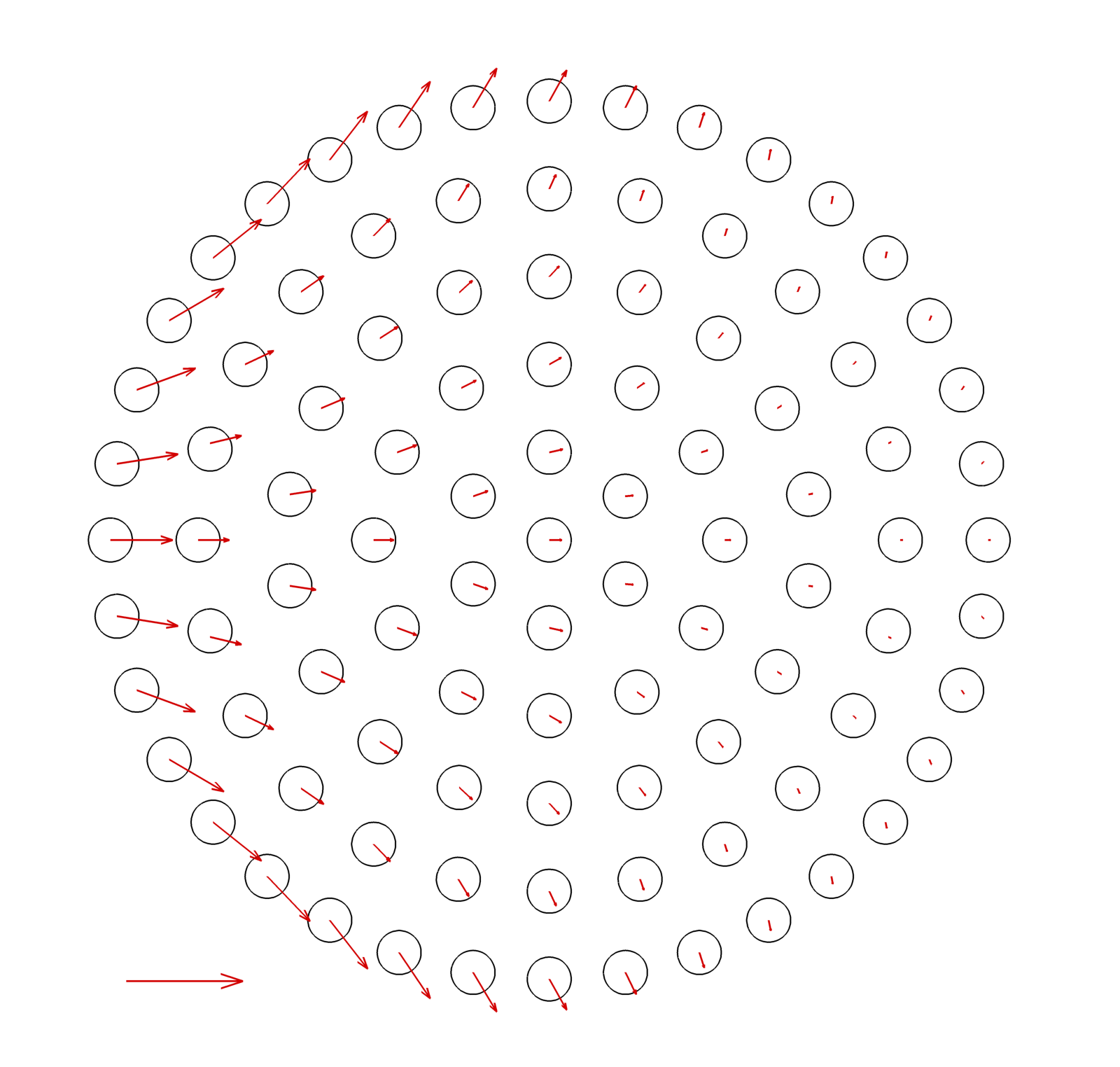}}}\hfill
		\subfigure[$C_{139}$ \label{fig:averageforcevector-C139}]{
			\resizebox*{3.55cm}{!}{\includegraphics{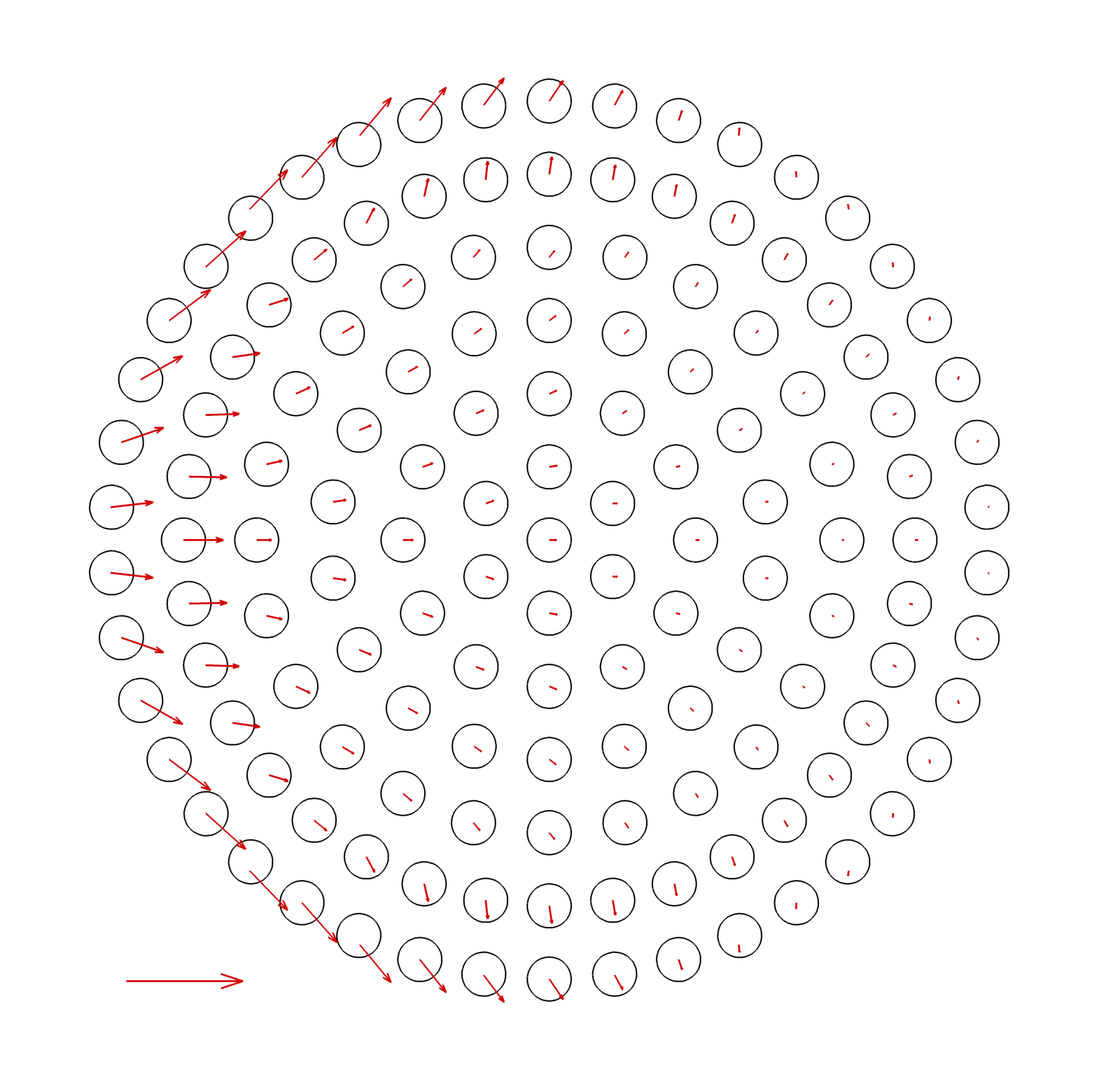}}}\\
			
		\subfigure[$C_{\text{solid}}$ \label{fig:averageforcevector-solid}]{
			\resizebox*{3.55cm}{!}{\includegraphics{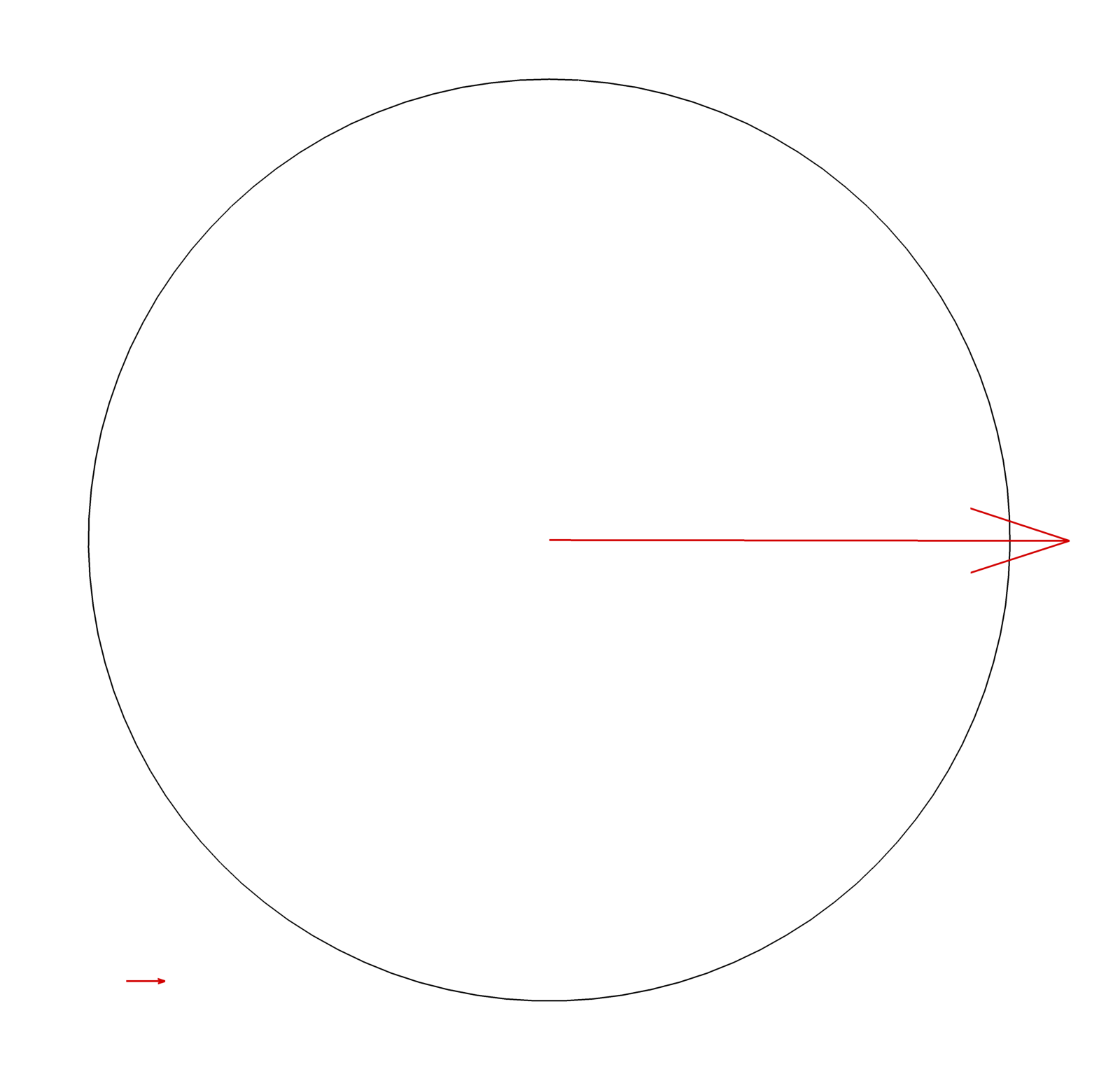}}}
		
		\caption{\label{fig:averageforcevector} Predicted mean fluid force distributions on the cylinder arrays: (a) $C_7$, (b) $C_{19}$, (c) $C_{37}$, (d) $C_{61}$, (e) $C_{97}$, (f) $C_{139}$, and (g) the solid cylinder. Results are for $Re_D = 100$.}
	\end{center}
\end{figure}

\subsection{Validation of the global stability analysis}

\begin{figure}
	\begin{center}
		\subfigure[\label{fig:11}{}]{
			\resizebox*{6.5cm}{!}{\includegraphics{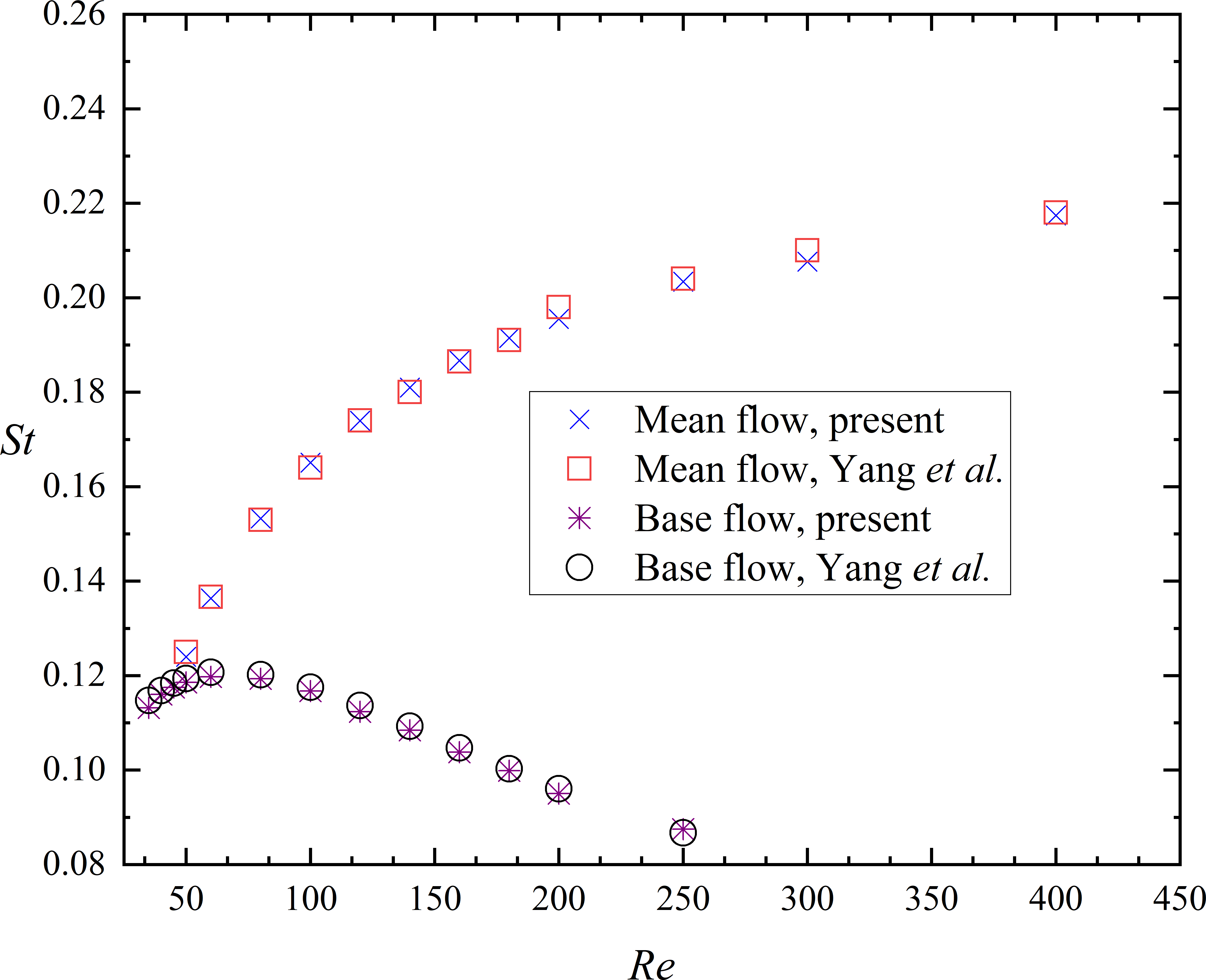}}}
		\subfigure[\label{fig:12}{}]{
			\resizebox*{6.5cm}{!}{\includegraphics{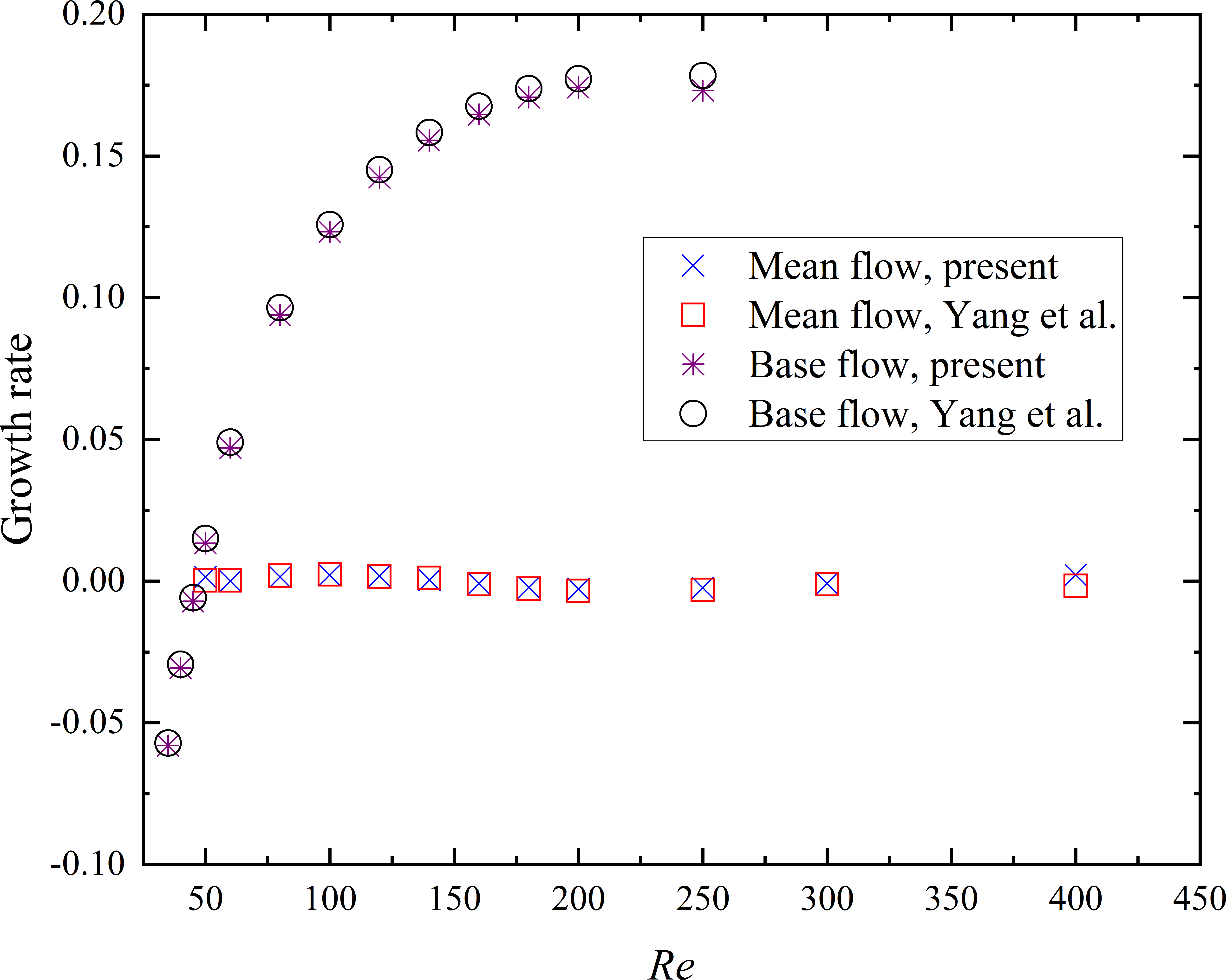}}}		
		\caption{\label{fig:g1fig} Comparison of $St$ and $Gr$ between present solutions and the results in the literature for a 2-D cylinder wake flow at Re = 100. }
	\end{center}
\end{figure}

\begin{figure}
	\begin{center}
		\subfigure[\label{fig2:11}{}]{
			\resizebox*{6.5cm}{!}{\includegraphics{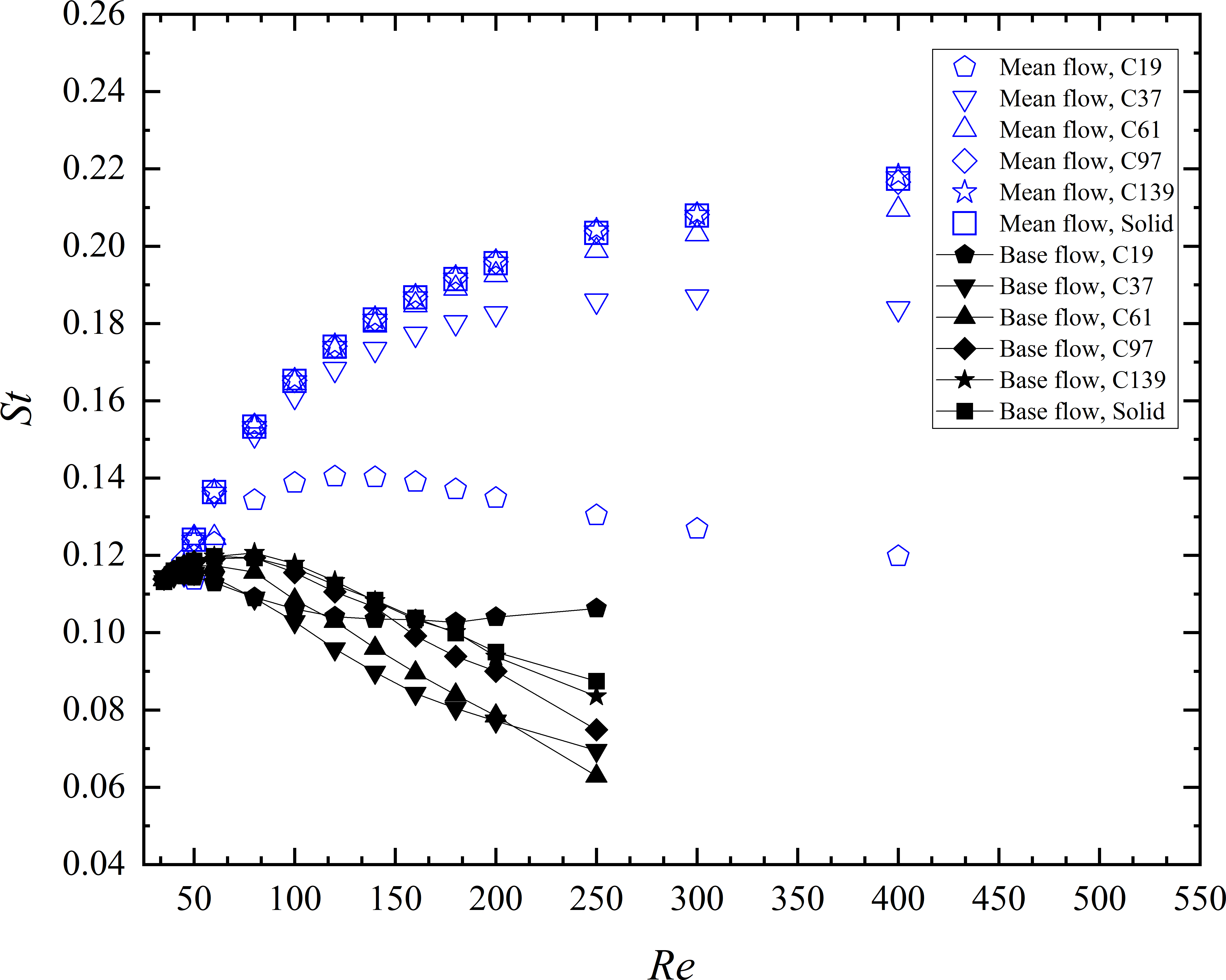}}}
		\subfigure[\label{fig2:12}{}]{
			\resizebox*{6.5cm}{!}{\includegraphics{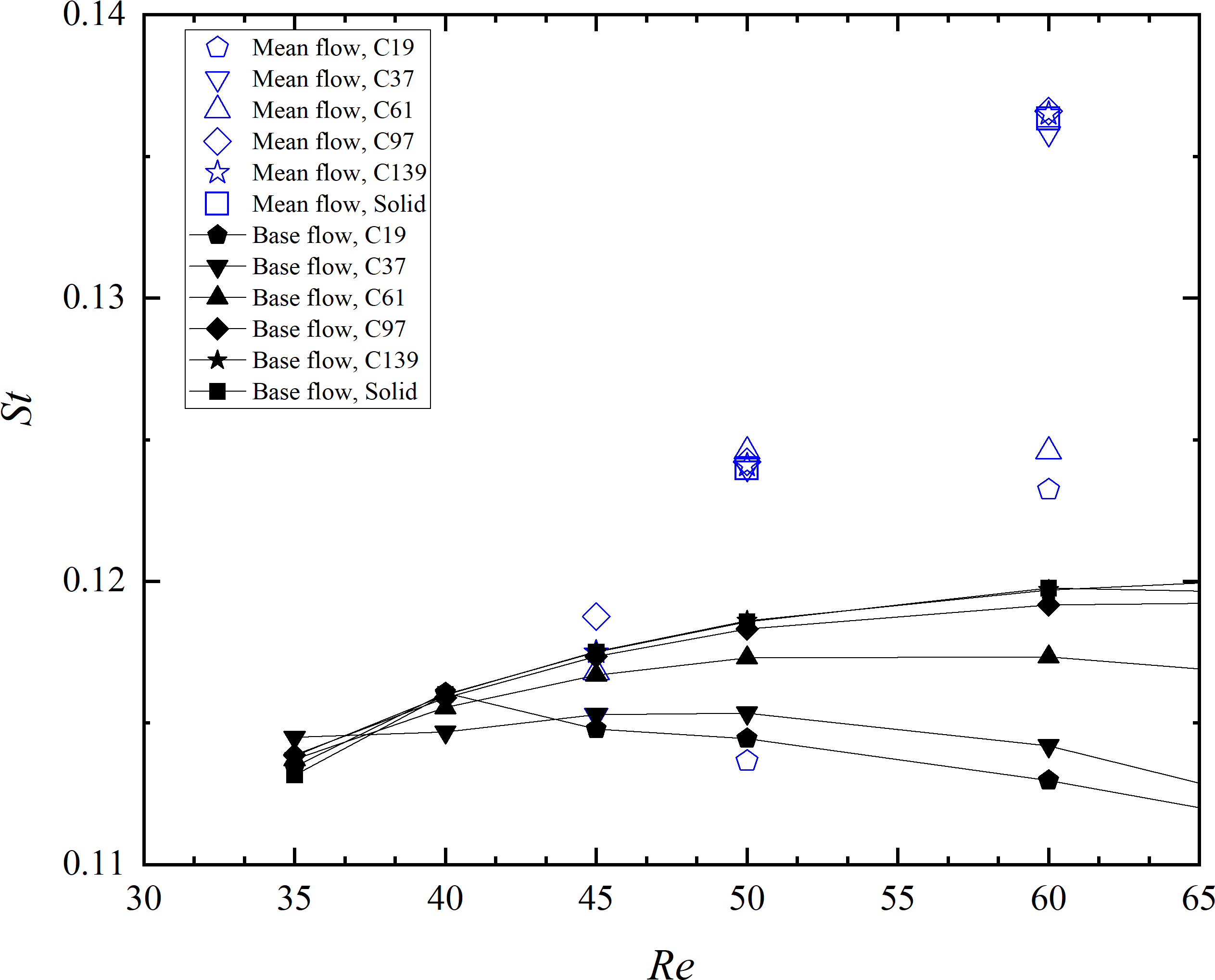}}}		
		\caption{\label{fig2:g1fig} Comparison of $St$ between present solutions and the results in the literature for the 2D flow through and around the cylinder array at Re = 100. }
	\end{center}
\end{figure}

\section{Results and Discussion} \label{results1}

Here, I \textbf{Quote} `` 
The spatial structure of the global mode changes considerably in the range of
Reynolds numbers investigated. The maxima of $\hat{u}$, $\hat{v}$ and $\hat{p}$, in fact, move gradually
upstream when Re becomes larger, while the size of the separation bubble tends to
increase linearly with it.  Goujon-Durand, Jenffer \& Wesfreid (1994) and Zielinska \&Wesfreid (1995) dete
''

\begin{figure}
	\begin{center}
		\subfigure[\label{fig3:11}{}]{
			\resizebox*{6.5cm}{!}{\includegraphics{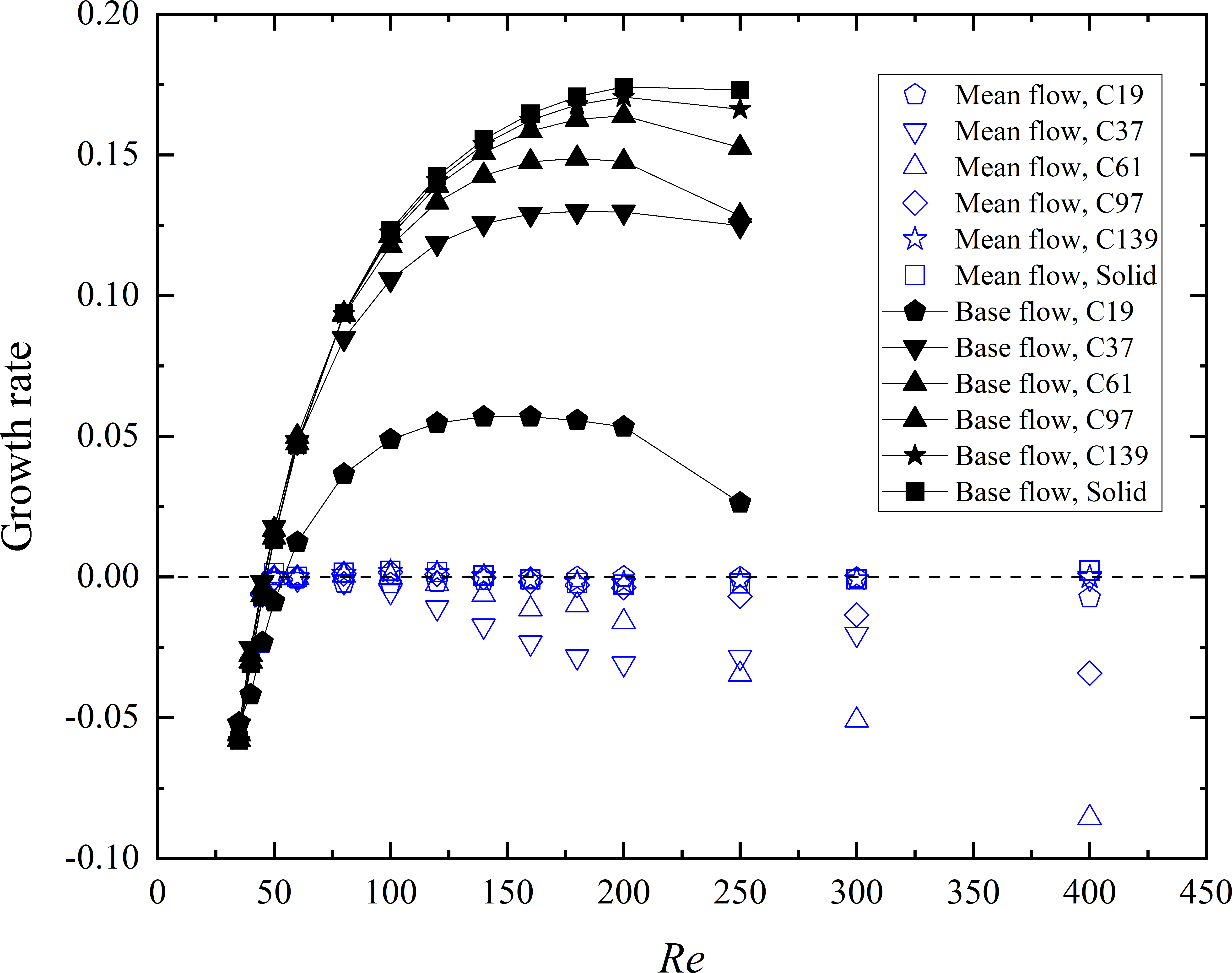}}}
		\subfigure[\label{fig3:12}{}]{
			\resizebox*{6.5cm}{!}{\includegraphics{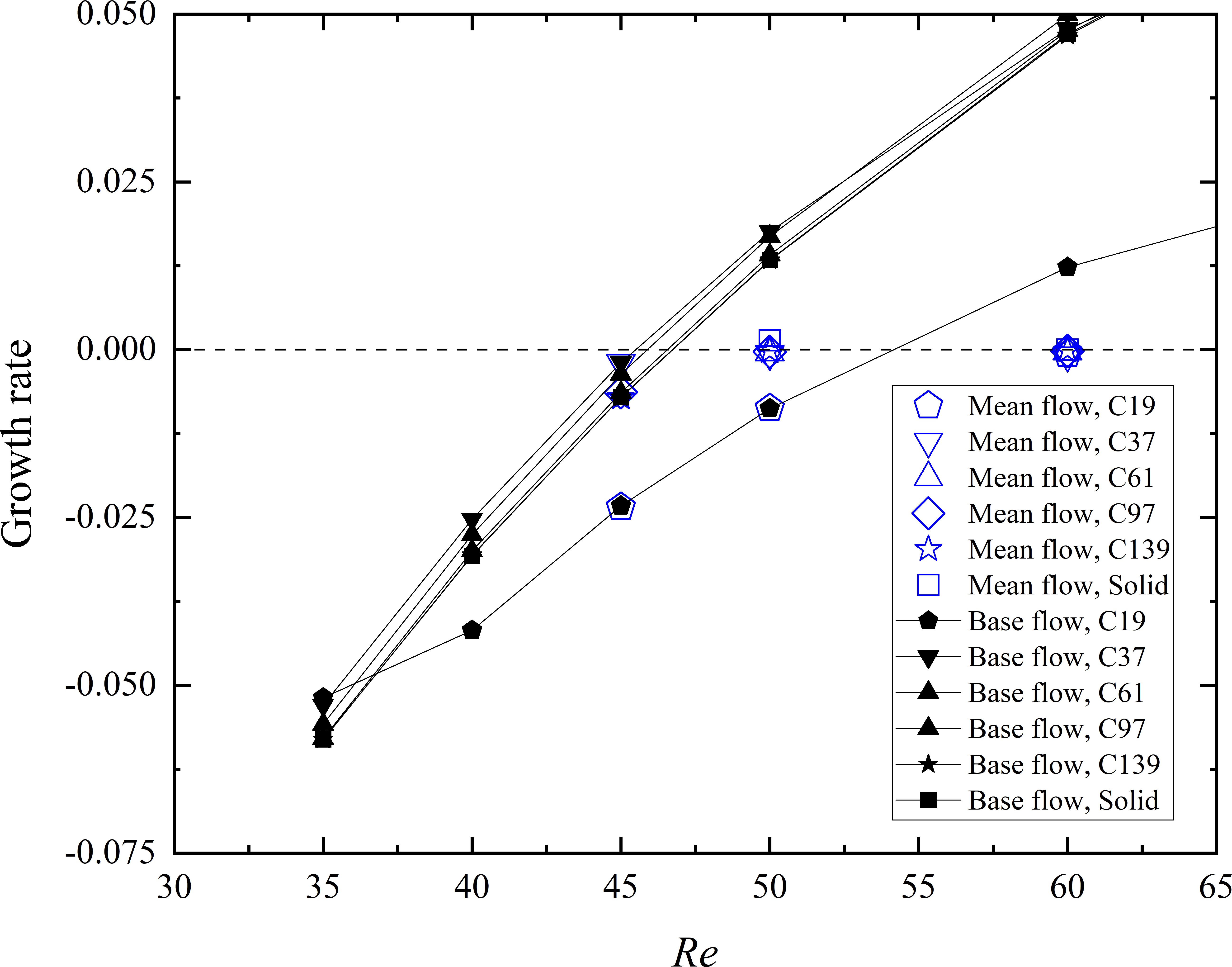}}}		
		\caption{\label{fig3:g1fig} Comparison of growth rate between present solutions and the results in the literature for the 2D flow through and around the cylinder array at Re = 100.  }
	\end{center}
\end{figure}

 Three distinct flow regimes were identified. For low void fractions ($\phi<$  0.084), $Re_c$ rapidly increases with decreased $\phi$.  

 For high void fractions ($\phi$ > 0.3), $Re_c = 47$ indicating that the array behaves as a solid body of the same scale.  

 
\begin{figure}
	\begin{center}
		\subfigure[\label{fig5:11}{}]{
			\resizebox*{6.5cm}{!}{\includegraphics{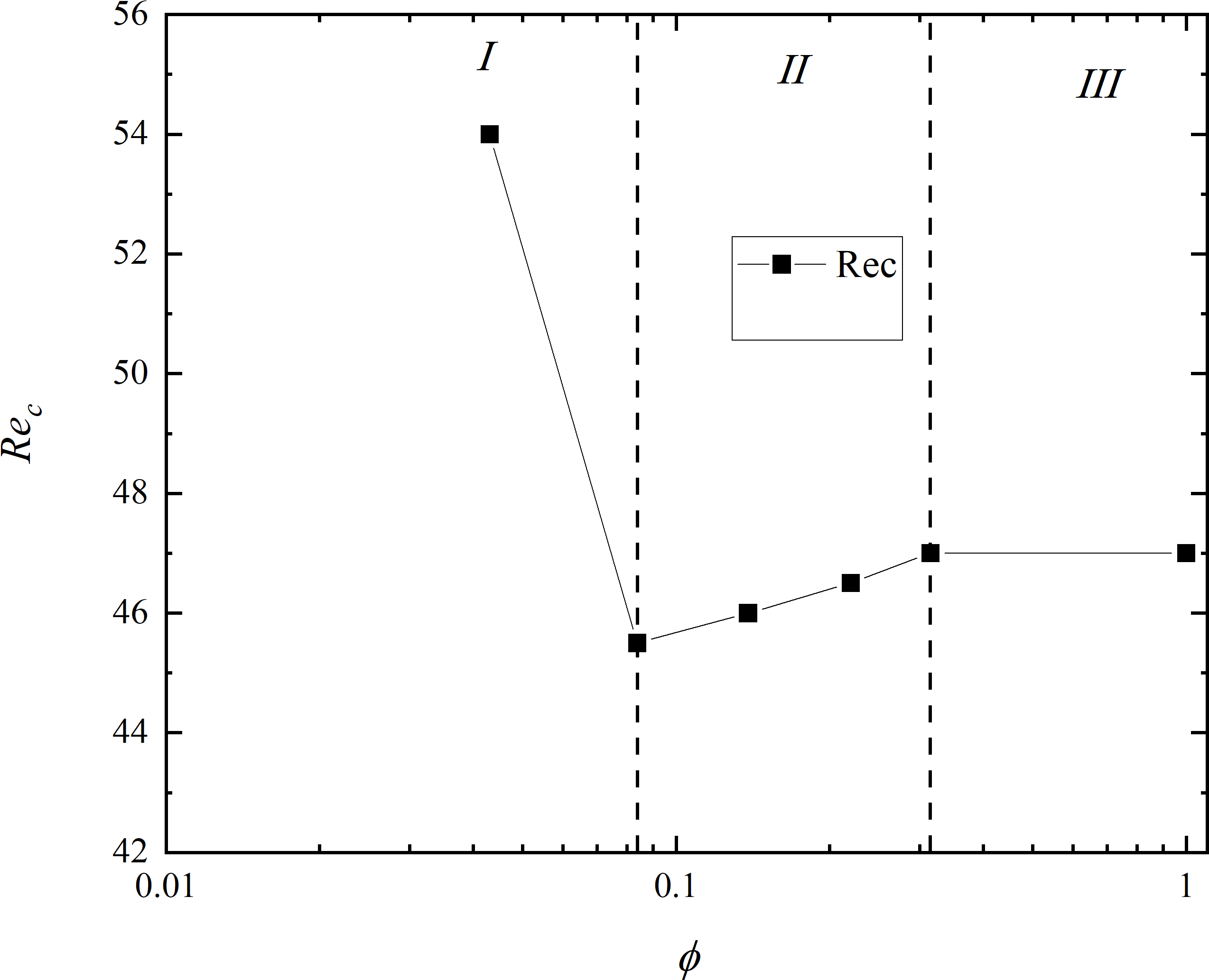}}}
		\subfigure[\label{fig5:12}{}]{
			\resizebox*{6.5cm}{!}{\includegraphics{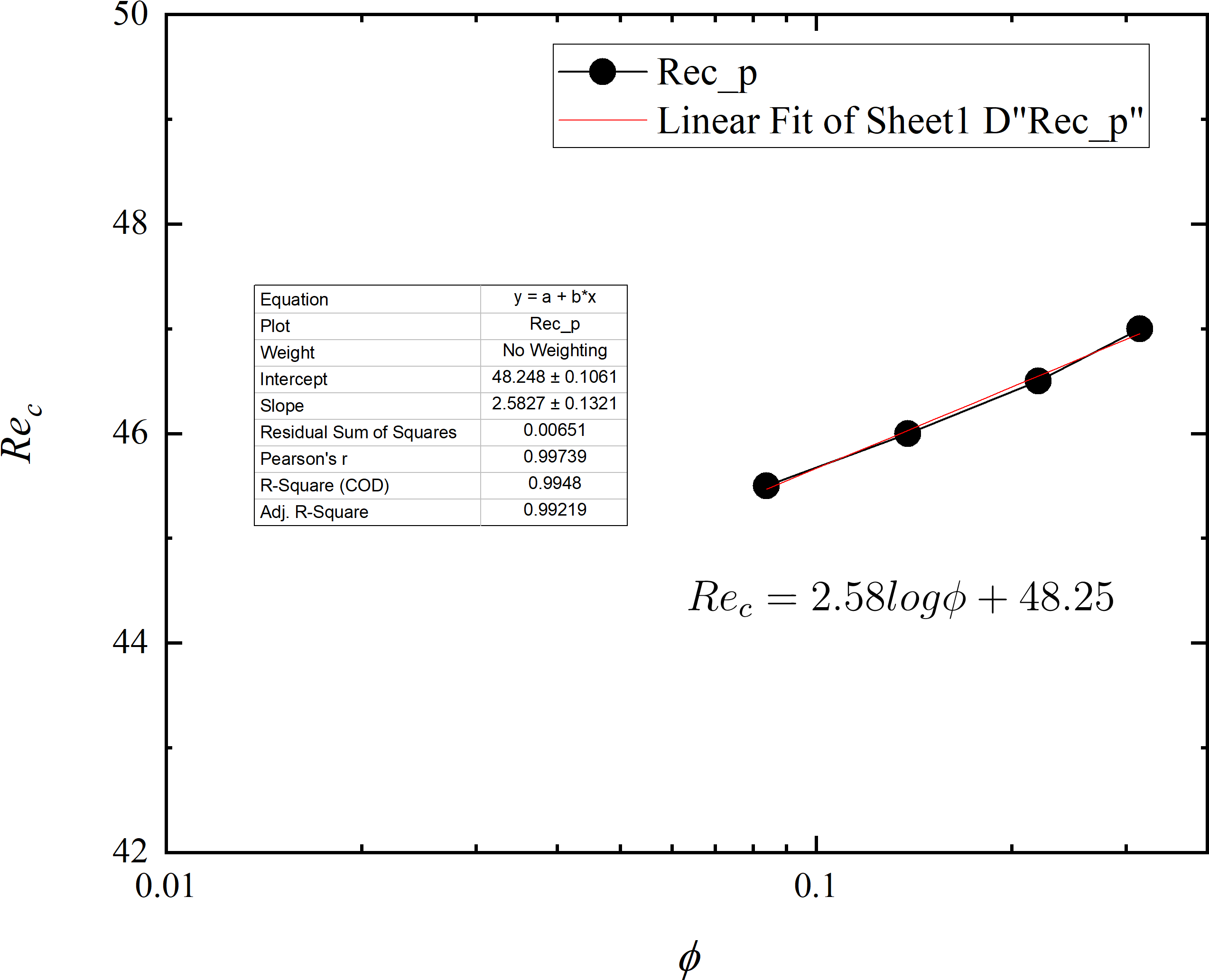}}}		
		\caption{\label{fig5:g1fig2} Critical Reynolds number ($Re_c$) at various solid faction ($\phi$). Three distinct flow regimes were identified. $I$ represents that of several isolated cylinders; $II$ represents the porous media; $III$ represents that of a single solid cylinder.   }
	\end{center}
\end{figure}
 At intermediate void fractions ($0.084<\phi<0.315$),  the group of cylinders resembles a porous medium. In this range, $Re_c$ is found to satisfy a linear relationship
\begin{equation}\label{eqn:fit}
	Re_c  = a \text{lg}\phi  + b 
\end{equation}
with the coefficients of $a$ and $b$ being 2.58 and 48.25, respectively. The goodness of fit $R2$ is 0.9948.

\subsection{WaveMaker Region}

The sensitivity of the flow to the perturbation can be studied via the structural sensitivity analysis. Wavemaker region delimits the wavemaker region, which is the superposition of the
leading global mode and its adjoint mode.

\textbf{Here I Quote } "In general, the wavemaker region is located in the near wake region
of the cylinder (in the top-right region of the xy plane). Its structure remains the same
symmetry as the base state, being symmetric with respect to the Oxy plane. The spatial
distributions of wakemakers based on mean flow and base flow are similar. Since the
wavemaker region indicates the most sensitive region in the flow, one can infer from these
observations that (i) the region responsible for the instability is located in the recirculation
region behind the cylinder, and (ii) the instability mainly amplifies the perturbations near
the cylinder surface (Citro et al. 2016)."

The product between the direct and adjoint fields gives the maximum possible coupling among the velocity components. The identification of the core region of the instability can help to understand the instability mechanism (~\cite{Giannetti2007, Giannetti2009, Luchini2014a, Gomez2014, Liu2016, Pini2019, Mitrut2021}). According to (~\cite{Giannetti2007, Liu2016}). The sensitivity wavemaker $\zeta$ can be identified by overlapping the direct eigenvector $\hat{\boldsymbol{u}}$ and adjoint eigenvector $\hat{\boldsymbol{v}}$,

\begin{equation}\label{eqn:wavenumber}
	\zeta=\frac{|\hat{\boldsymbol{u}}|\left|\hat{\boldsymbol{v}}\right|}{\left\langle\hat{\boldsymbol{u}}, \hat{\boldsymbol{v}}\right\rangle}  = \frac{|\hat{\boldsymbol{u}}|\left|\hat{\boldsymbol{v}}\right|}{ \int  \hat{\boldsymbol{u}} \cdot \hat{\boldsymbol{v}}  d\Omega }.
\end{equation}

Quote "Figure 17 shows that large values of $\zeta (x, y)$ are attained in two lobes located symmetrically across the separation bubble. Note that both close to the cylinder and far from it,
the product of the adjoint and direct modes is small, showing that these areas of the
flow are not really important for the instability dynamics."

Quote "If the Reynolds number is increased, the spatial separation between the maxima of
the direct and adjoint modes is reduced, but the main characteristics of $\zeta (x, y)$ remain
unaltered. As figure 18 shows, in fact, the maxima of $\zeta (x, y)$ are always located in
two symmetric lobes across the separation bubble and slowly move downstream when
the value of Re is increased. In the cases considered, however, the maxima lie at
a distance from the cylinder wall smaller than the recirculation length Lw (see also
figure 6)."

\begin{figure}
	\begin{center}

		\subfigure[$C_{19}$ \label{figw:C19-subfig1}]{
			\resizebox*{4.4cm}{!}{\includegraphics{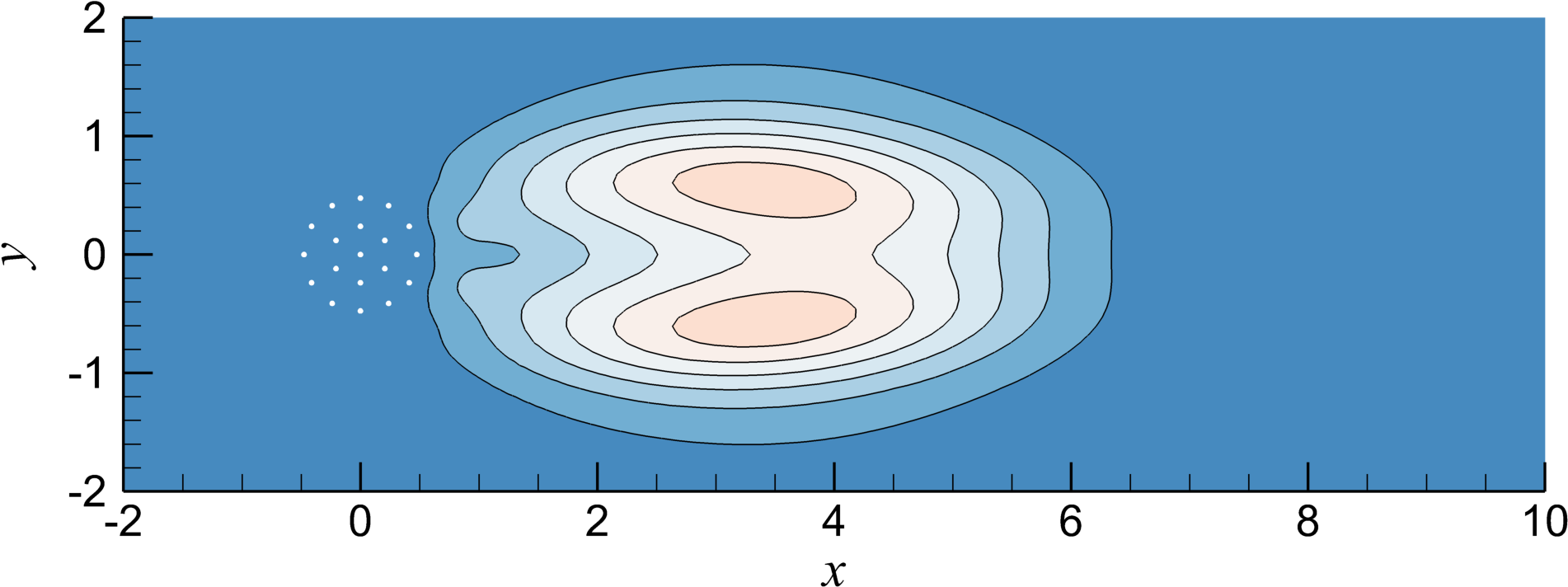}}}\hfill
		\subfigure[$C_{19}$ \label{figw:C19-subfig2}]{
			\resizebox*{4.4cm}{!}{\includegraphics{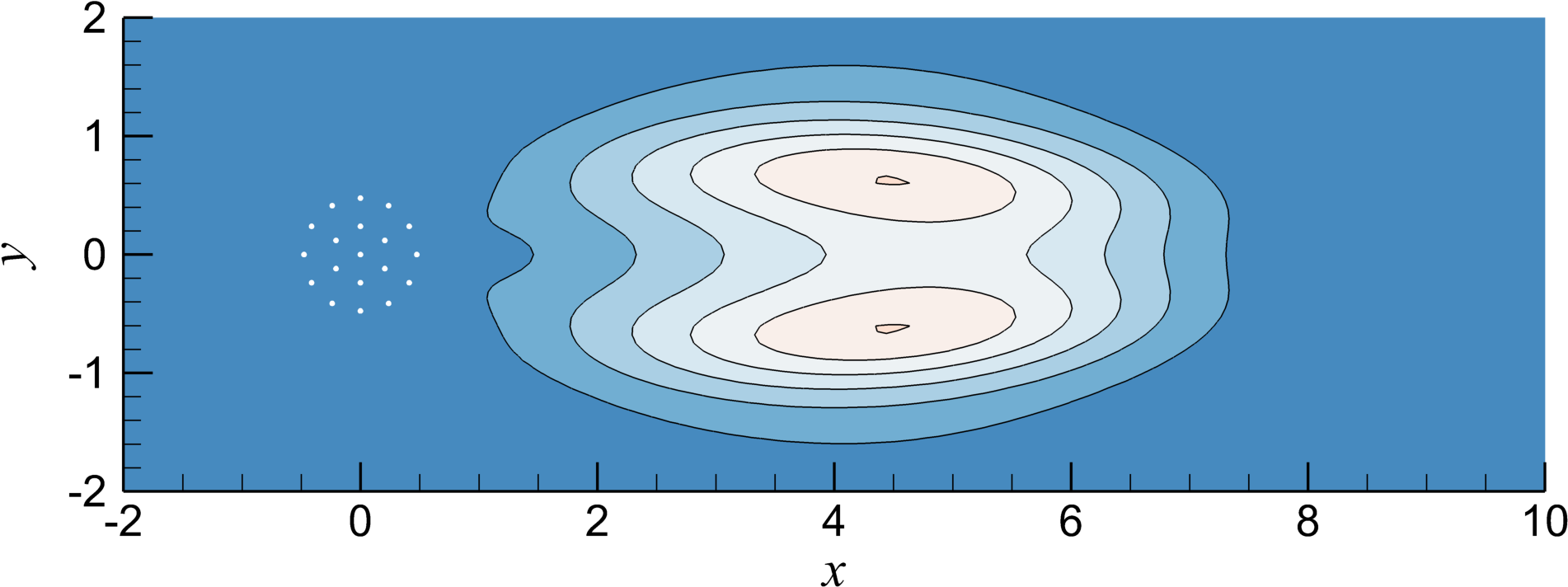}}}\hfill
		\subfigure[$C_{19}$ \label{figw:C19-subfig3}]{
			\resizebox*{4.4cm}{!}{\includegraphics{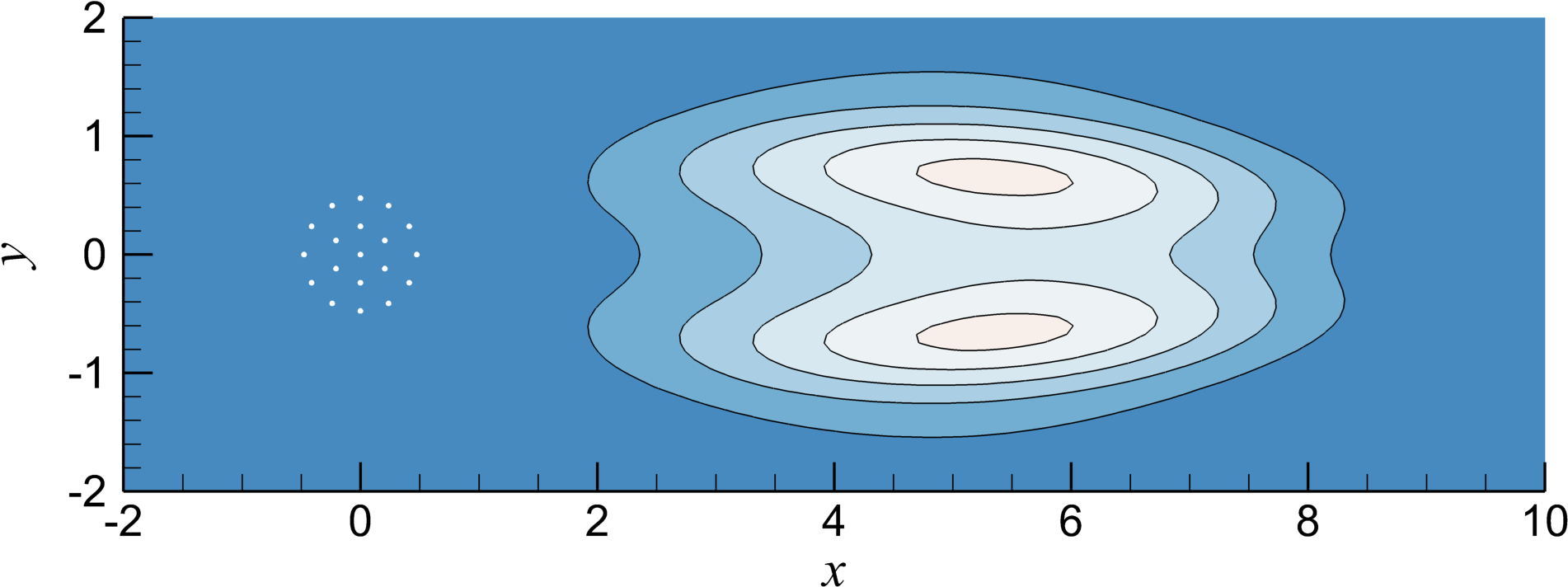}}}\\
			
		\subfigure[$C_{37}$ \label{figw:C37-subfig1}]{
			\resizebox*{4.4cm}{!}{\includegraphics{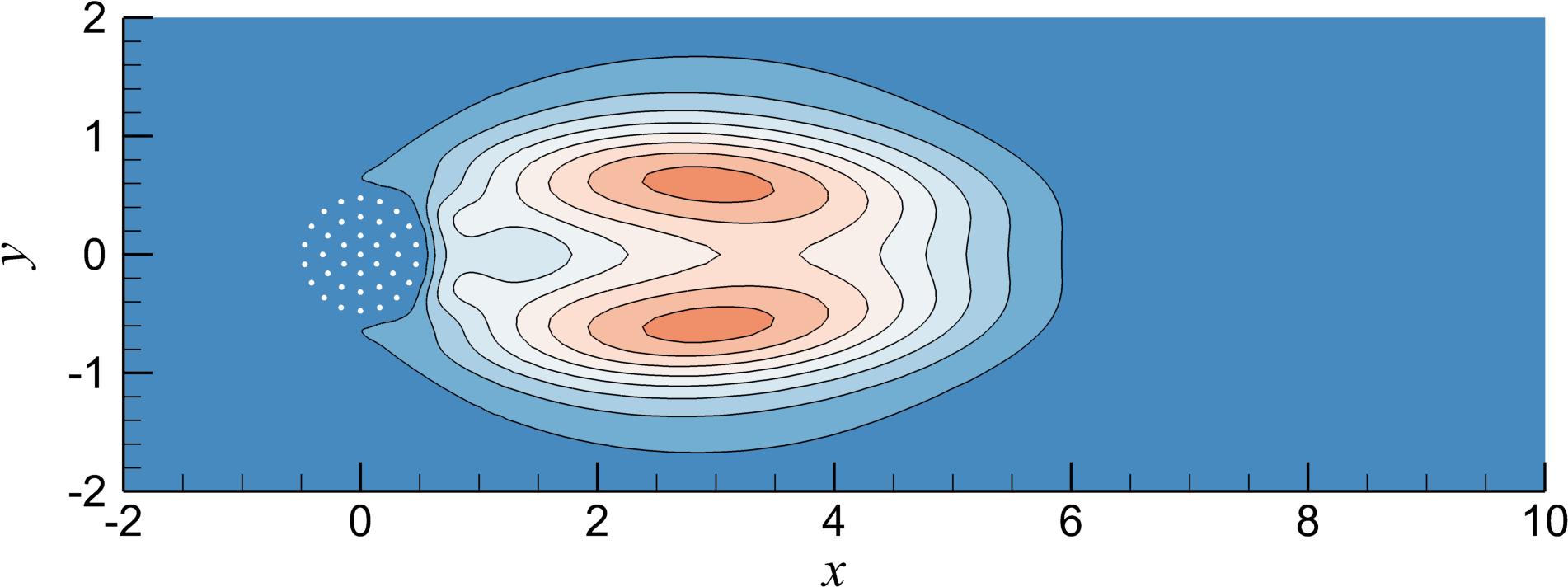}}}\hfill
		\subfigure[$C_{37}$ \label{figw:C37-subfig2}]{
			\resizebox*{4.4cm}{!}{\includegraphics{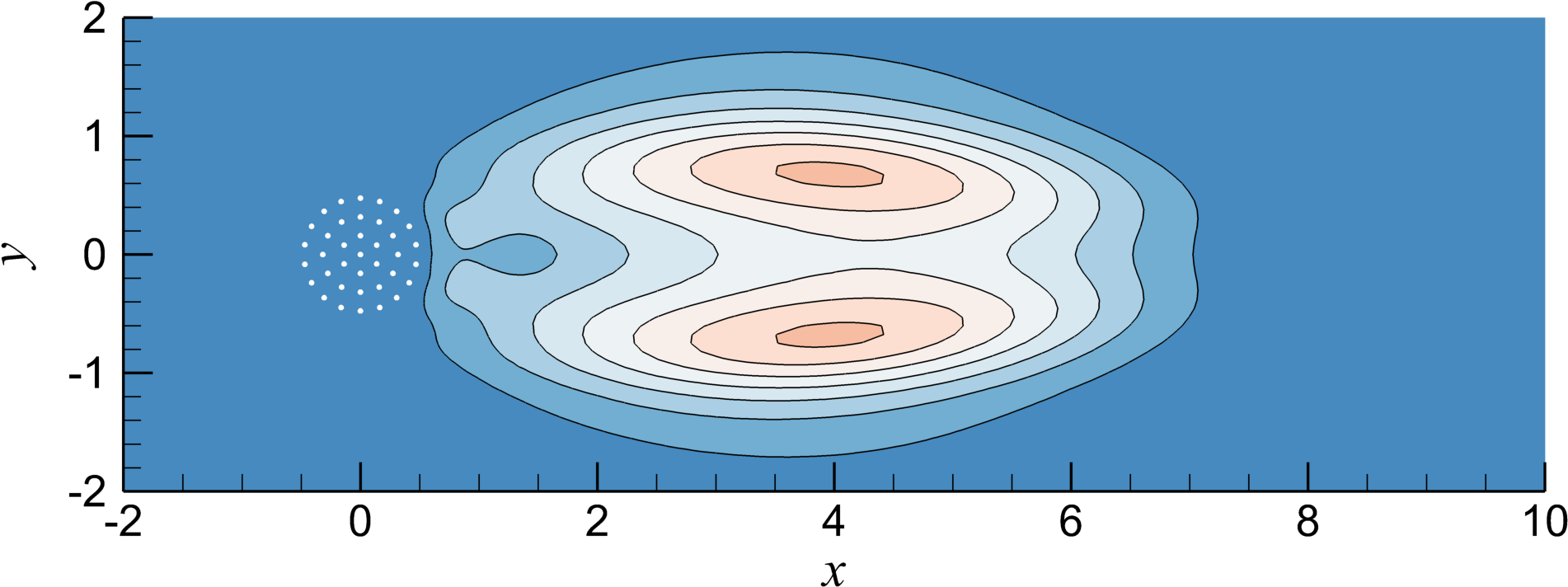}}}\hfill
		\subfigure[$C_{37}$ \label{figw:C37-subfig3}]{
			\resizebox*{4.4cm}{!}{\includegraphics{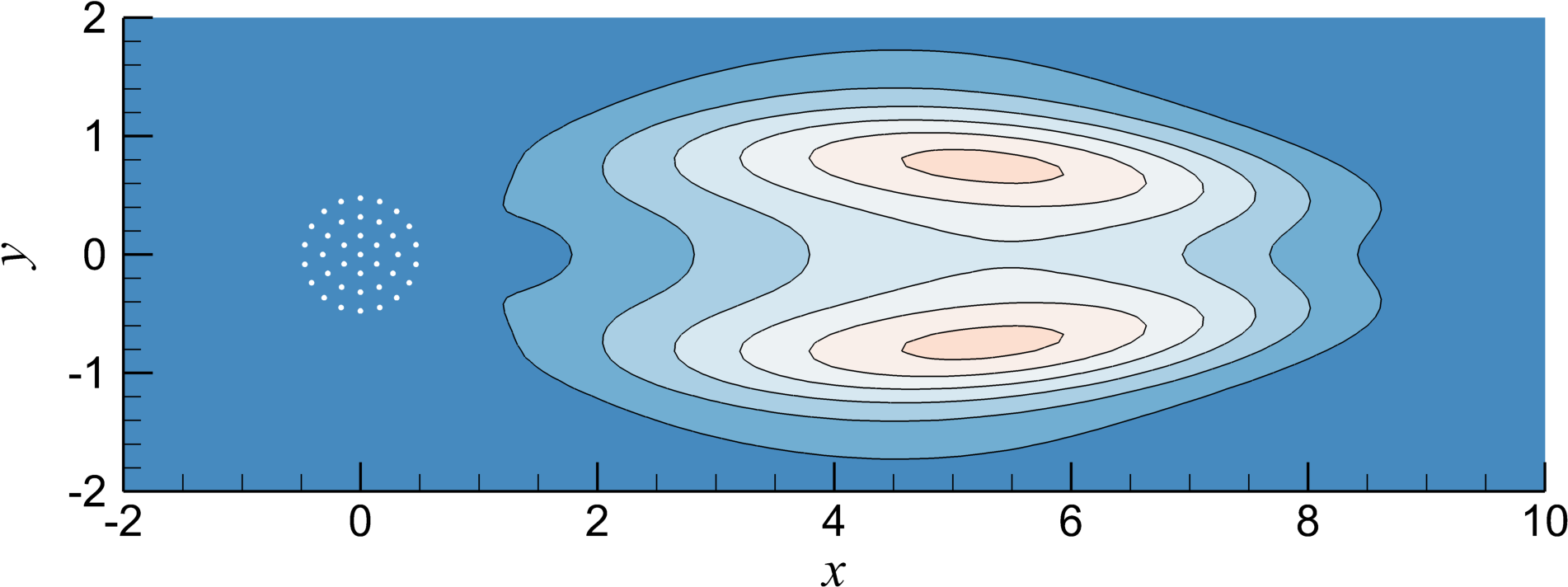}}}\\
			
		\subfigure[$C_{61}$ \label{figw:C61-subfig1}]{
			\resizebox*{4.4cm}{!}{\includegraphics{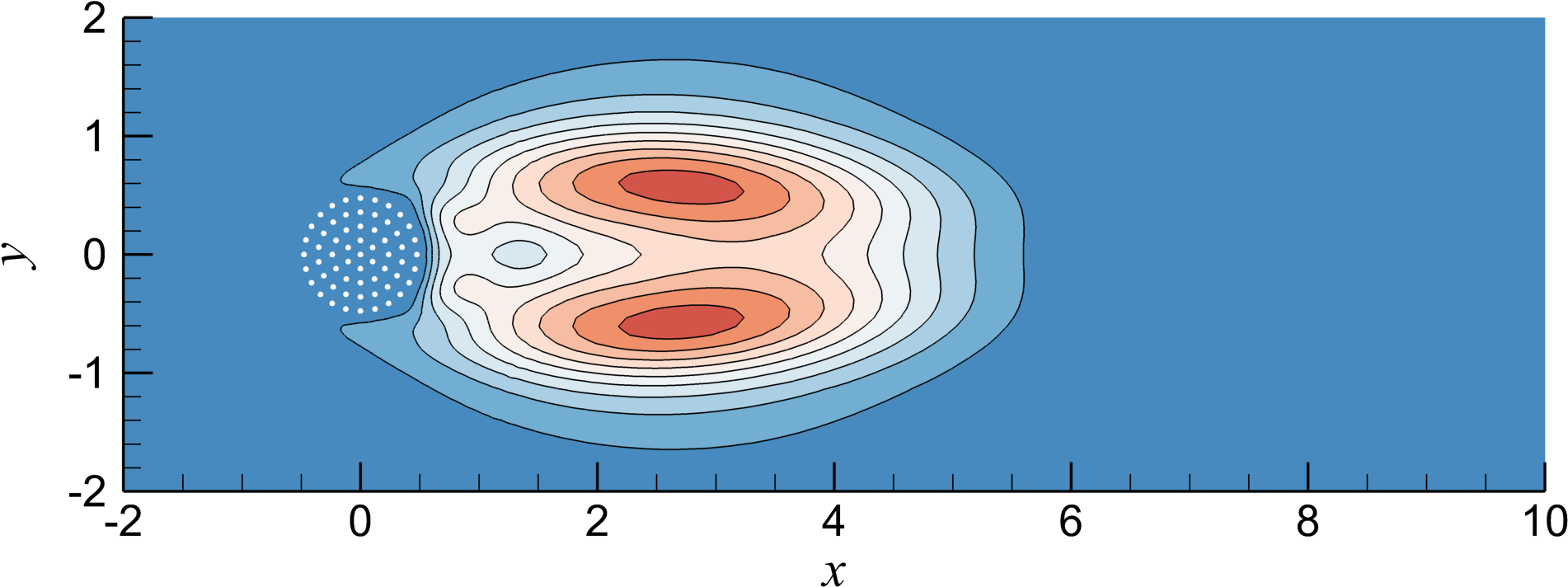}}}\hfill
		\subfigure[$C_{61}$ \label{figw:C61-subfig2}]{
			\resizebox*{4.4cm}{!}{\includegraphics{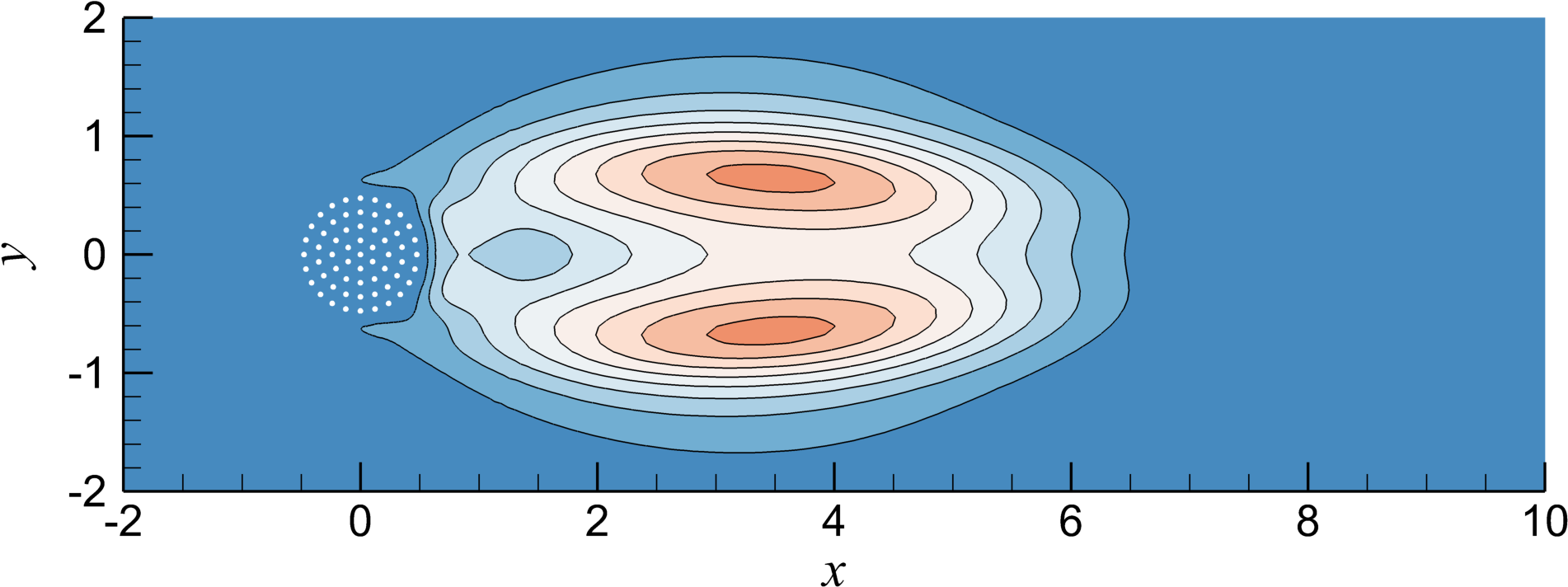}}}\hfill
		\subfigure[$C_{61}$ \label{figw:C61-subfig3}]{
			\resizebox*{4.4cm}{!}{\includegraphics{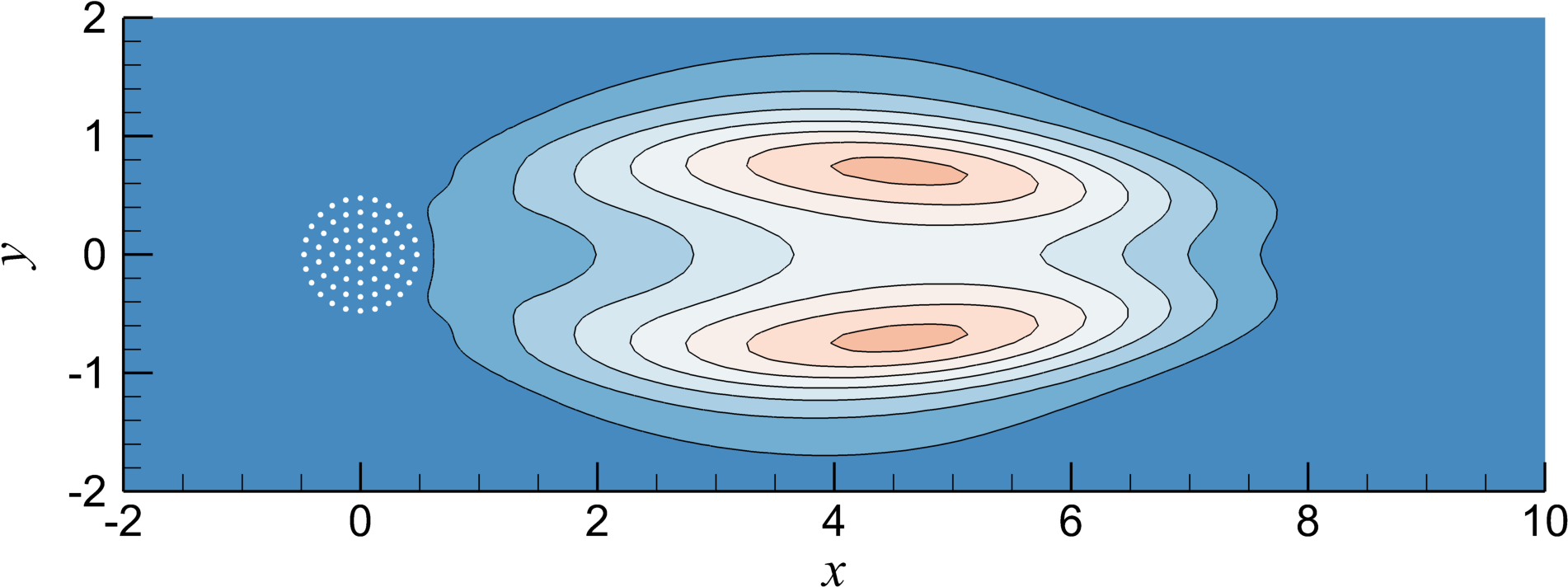}}}\\
			
		\subfigure[$C_{97}$ \label{figw:C97-subfig1}]{
			\resizebox*{4.4cm}{!}{\includegraphics{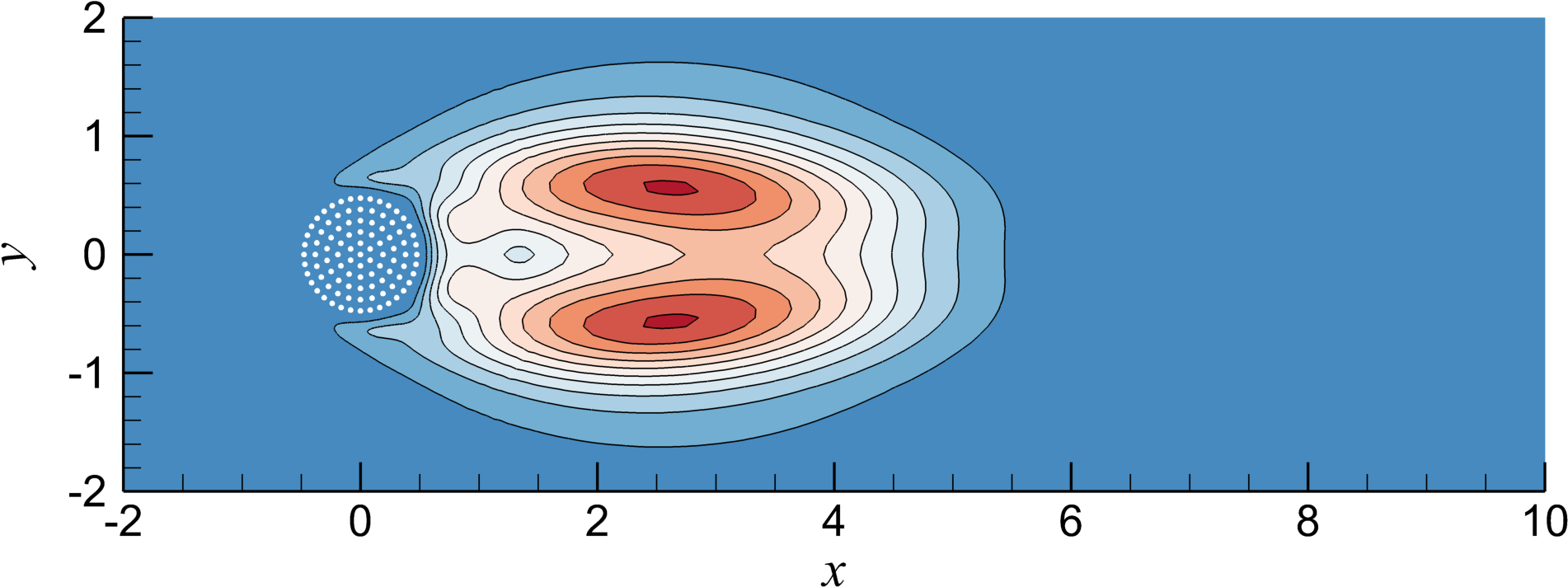}}}\hfill
		\subfigure[$C_{97}$ \label{figw:C97-subfig2}]{
			\resizebox*{4.4cm}{!}{\includegraphics{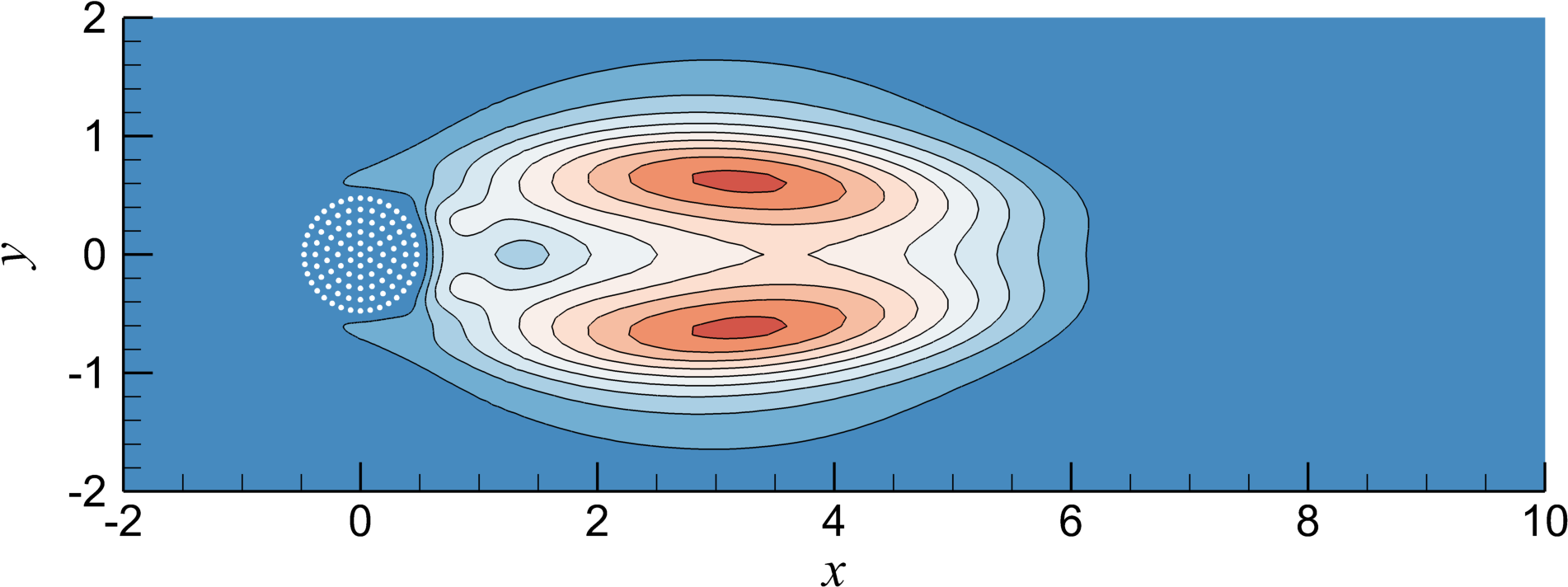}}}\hfill
		\subfigure[$C_{97}$ \label{figw:C97-subfig3}]{
			\resizebox*{4.4cm}{!}{\includegraphics{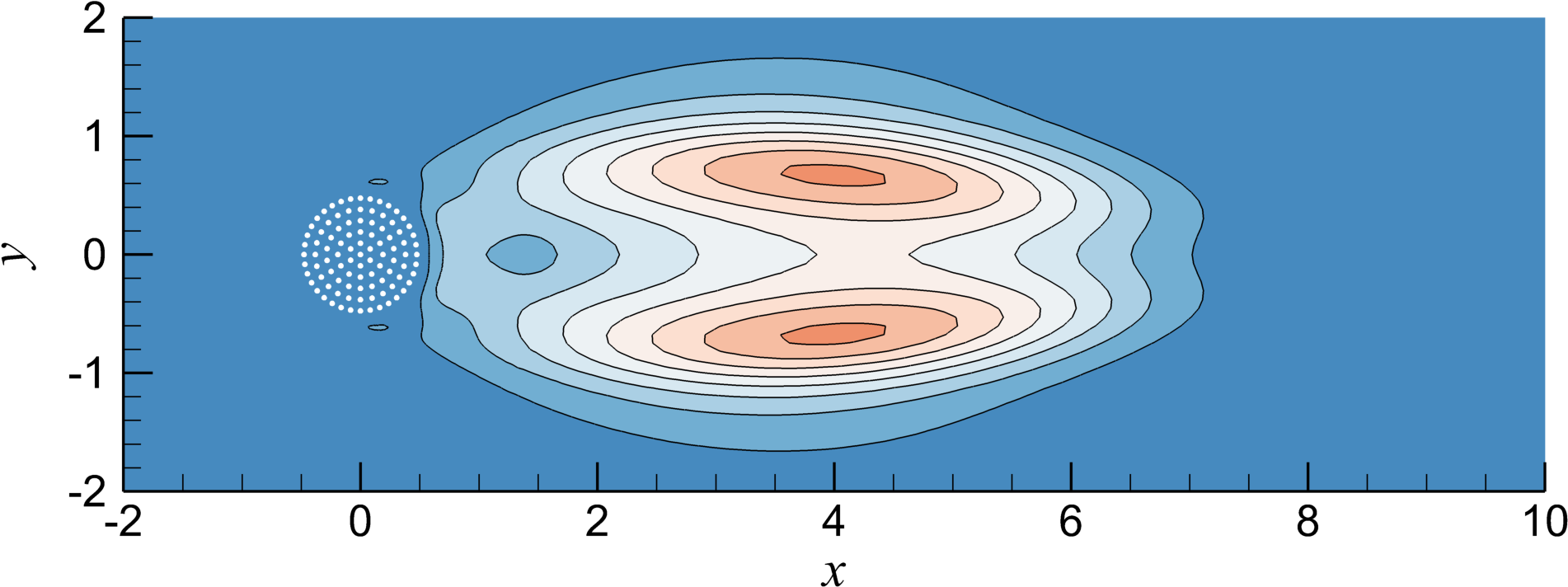}}}\\
			
		\subfigure[$C_{139}$ \label{figw:C139-subfig1}]{
			\resizebox*{4.4cm}{!}{\includegraphics{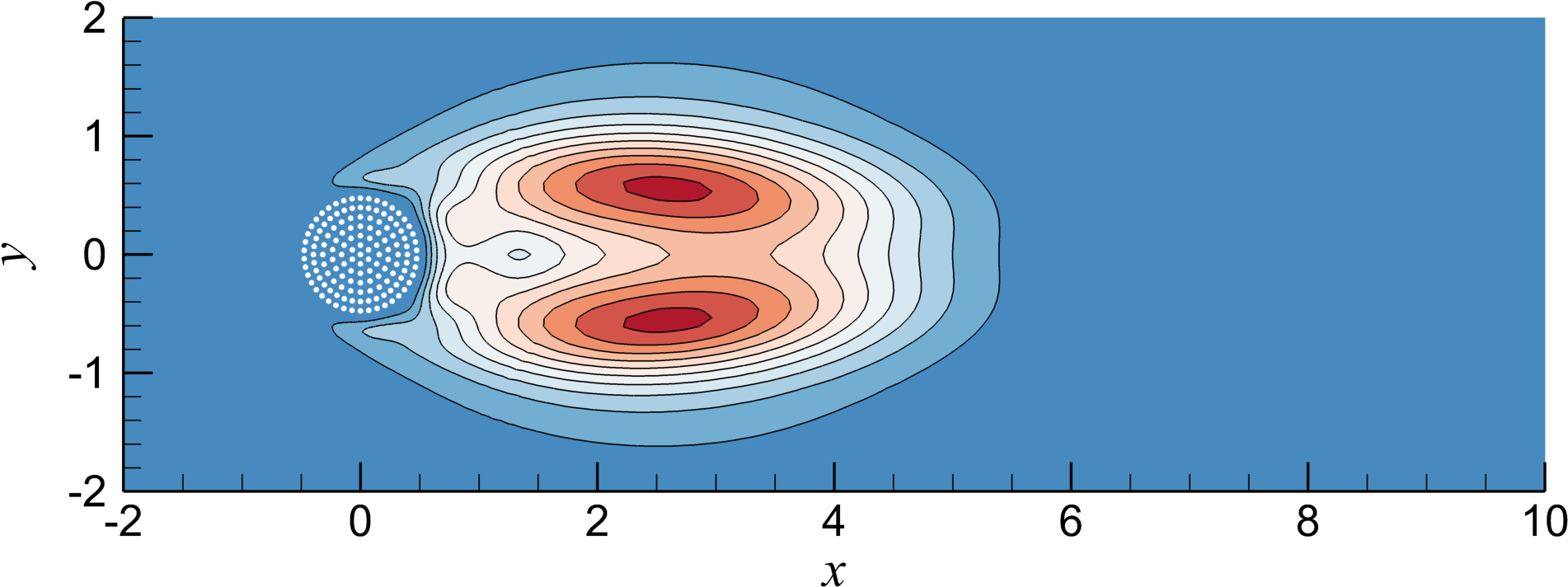}}}\hfill
		\subfigure[$C_{139}$ \label{figw:C139-subfig2}]{
			\resizebox*{4.4cm}{!}{\includegraphics{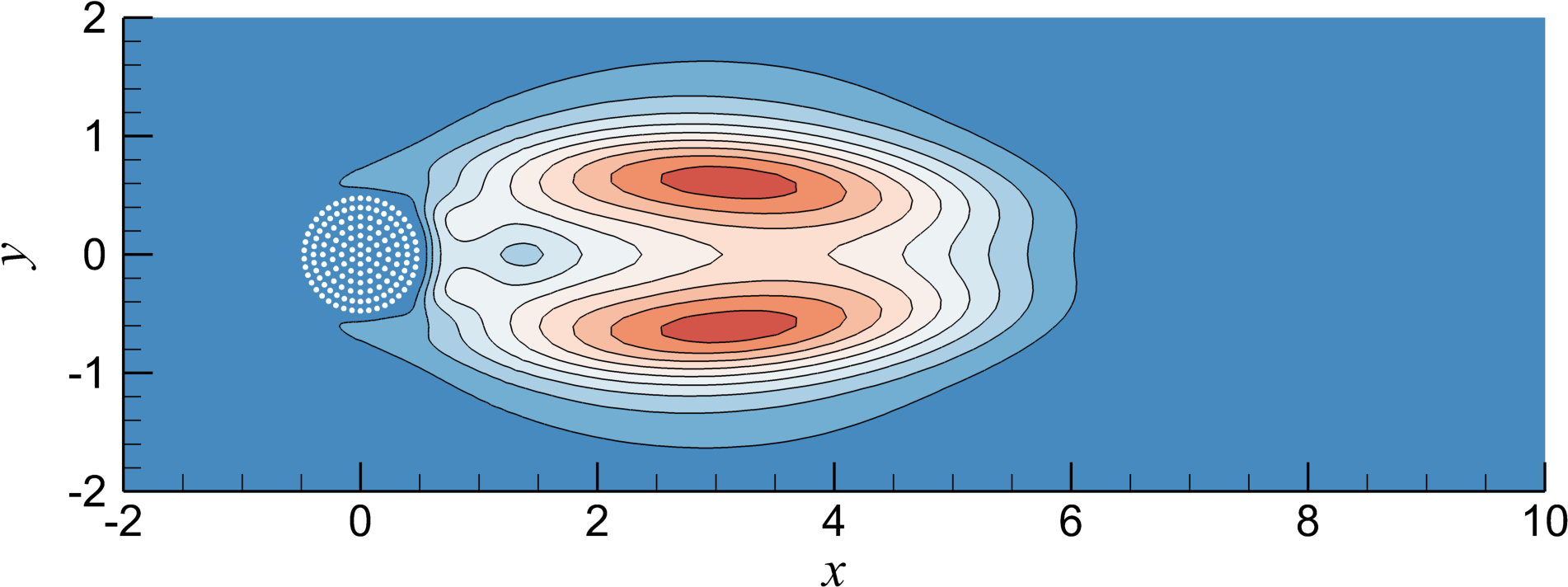}}}\hfill
		\subfigure[$C_{139}$ \label{figw:C139-subfig3}]{
			\resizebox*{4.4cm}{!}{\includegraphics{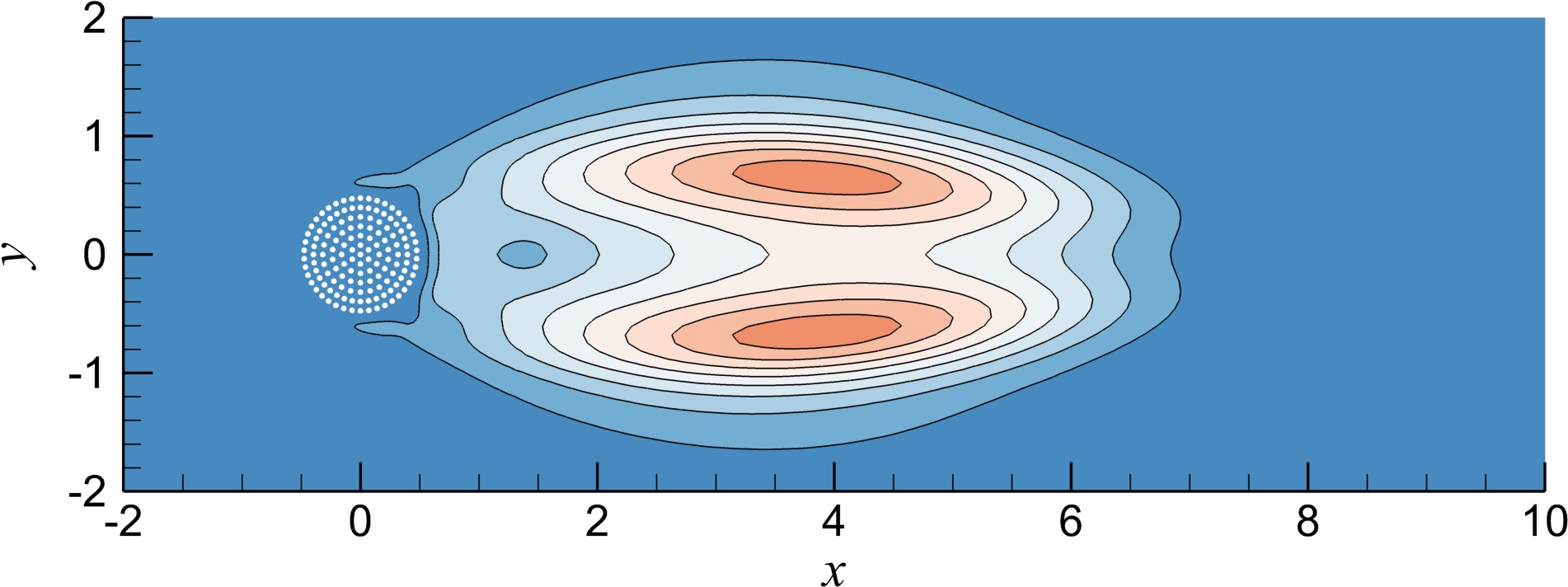}}}\\
			
		\subfigure[$C_{\text{solid}}$ \label{figw:solid-subfig1}]{
			\resizebox*{4.4cm}{!}{\includegraphics{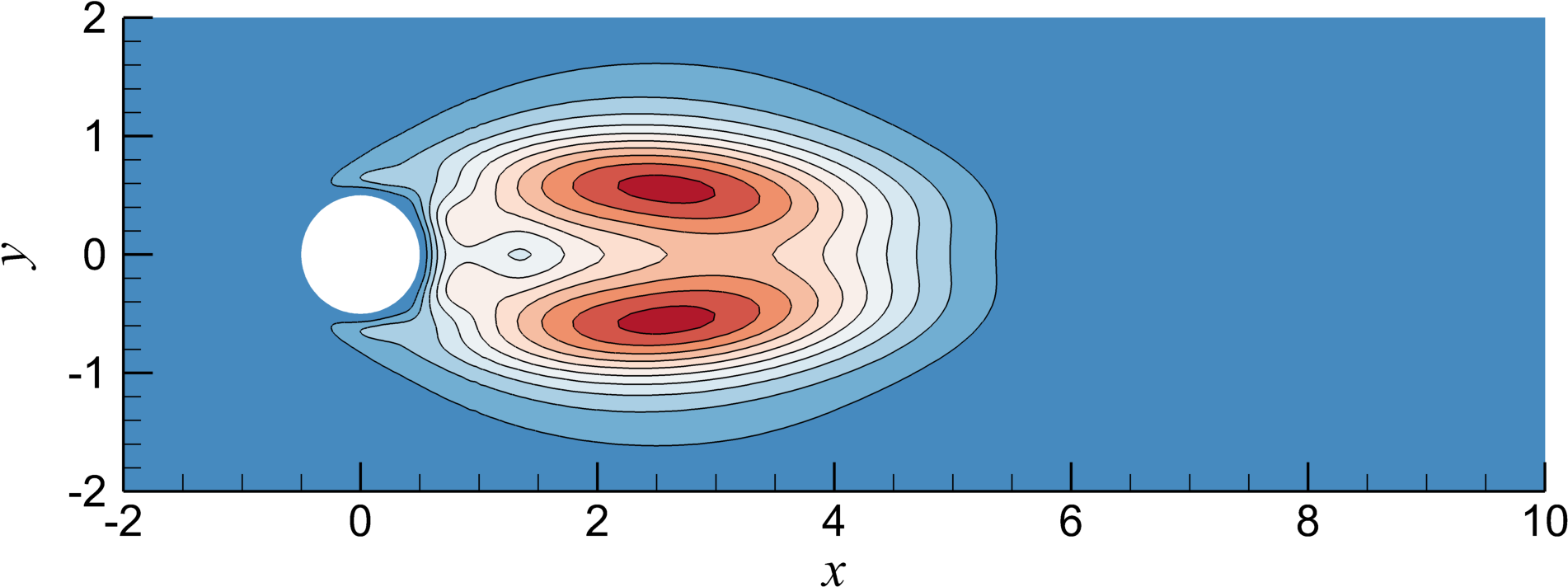}}}\hfill
		\subfigure[$C_{\text{solid}}$ \label{figw:solid-subfig2}]{
			\resizebox*{4.4cm}{!}{\includegraphics{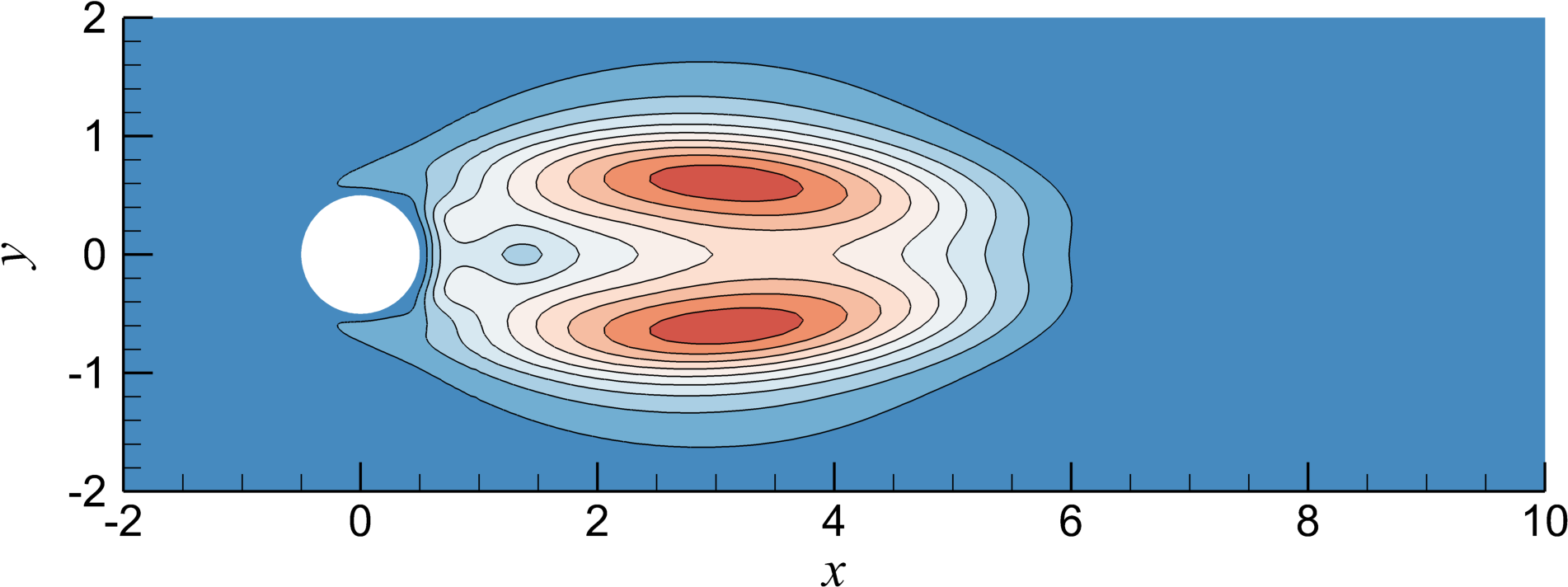}}}\hfill
		\subfigure[$C_{\text{solid}}$ \label{figw:solid-subfig3}]{
			\resizebox*{4.4cm}{!}{\includegraphics{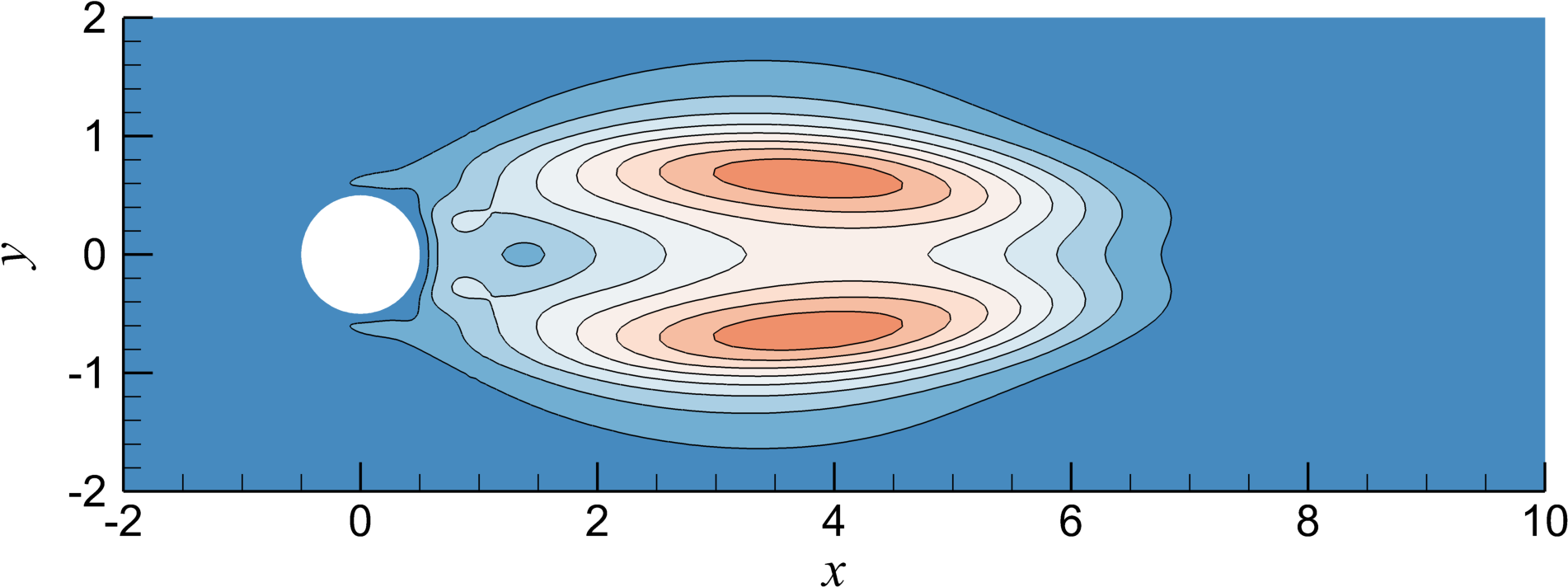}}}
		
		\caption{\label{figw:wavemaker}  Wavemaker region $\zeta$ for 2-D cylinder flow at $Re = 60, 80$, and 100. Eleven contour levels are uniformly spaced between $0$ and $0.2$ for $\zeta$, indicating values from low (blue) to high (red).  }
	\end{center}
\end{figure}

\begin{itemize}
\item (1) Global instability \cite{Theofilis2011}
\item (2) Effect of porosity
\item (3) Linear and nonlinear instability mechanisms
\item  (4) Impact on mixing?
\end{itemize}

\section{Base-Flow Sensitivity Analysis}

In this section we derive and present the framework for analyzing the 
sensitivity of global stability eigenvalues to modifications of the 
base flow. This type of analysis reveals the spatial regions where 
changes in the base flow exert the strongest influence on the stability 
properties of the system. The result is often referred to as 
\emph{structural sensitivity} or \emph{base-flow sensitivity}. 

\subsection{Recap of the Global Eigenvalue Problem}

Consider the incompressible Navier--Stokes equations, linearized about a 
steady base flow $U_b(x,y)$. The evolution of infinitesimal perturbations 
$q(x,y,t)$ (with velocity and pressure components) can be written in 
abstract form as
\begin{equation}
\frac{\partial q}{\partial t} = \mathcal{L}(U_b)\,q,
\end{equation}
where $\mathcal{L}(U_b)$ is the linearized operator, which depends 
parametrically on the base flow $U_b$. Seeking normal modes of the form
\begin{equation}
q(x,y,t) = \phi(x,y)\,e^{\lambda t},
\end{equation}
leads to the generalized eigenvalue problem
\begin{equation}
\mathcal{L}(U_b)\,\phi = \lambda\,\phi.
\label{eq:eigenproblem}
\end{equation}
Here, $\phi(x,y)$ is the \emph{direct eigenmode}, and $\lambda \in \mathbb{C}$ 
is the eigenvalue whose real part corresponds to the growth rate and 
whose imaginary part corresponds to the oscillation frequency.

The adjoint eigenproblem is defined with respect to the 
$L^2$ inner product
\begin{equation}
\langle p, q \rangle = \int_\Omega p^* \cdot q \, d\Omega,
\end{equation}
and reads
\begin{equation}
\mathcal{L}^\dagger(U_b)\,\phi^+ = \lambda^*\,\phi^+,
\end{equation}
where $\mathcal{L}^\dagger$ is the adjoint operator and $\phi^+$ is the 
adjoint eigenmode. The bi-orthogonality condition
\begin{equation}
\langle \phi^+, \phi \rangle \neq 0
\end{equation}
is used for normalization.

\subsection{Eigenvalue Drift Formula}

Suppose the base flow is perturbed by a small modification 
$\delta U_b$. This induces a perturbation in the linear operator
\begin{equation}
\mathcal{L}(U_b + \delta U_b) = \mathcal{L}(U_b) + \delta \mathcal{L},
\end{equation}
and consequently a drift in the eigenvalue $\lambda$:
\begin{equation}
\lambda \to \lambda + \delta \lambda.
\end{equation}
Classical perturbation theory gives the first-order eigenvalue drift as
\begin{equation}
\delta \lambda = \frac{\langle \phi^+, \delta \mathcal{L} \,\phi \rangle}
                      {\langle \phi^+, \phi \rangle}.
\label{eq:eigen-drift}
\end{equation}
Thus, the key quantity to evaluate is the variation $\delta \mathcal{L}$ 
induced by $\delta U_b$.

\subsection{Derivative of the Operator with Respect to the Base Flow}

The linearized Navier--Stokes operator contains convective terms of the form
\[
(U_b \cdot \nabla)\,\phi + (\phi \cdot \nabla)\,U_b.
\]
When the base flow is perturbed, only the second term changes 
linearly with $\delta U_b$. Therefore, the operator derivative is
\begin{equation}
\frac{\partial \mathcal{L}}{\partial U_b} \,\delta U_b
= (\phi \cdot \nabla)\,(\delta U_b).
\end{equation}
Inserting into Eq.~\eqref{eq:eigen-drift} gives
\begin{equation}
\delta \lambda = \frac{1}{\langle \phi^+, \phi \rangle}
\int_\Omega \phi^{+*} \cdot \Big[ (\phi \cdot \nabla)\,\delta U_b \Big]\, d\Omega.
\end{equation}
By integration by parts, this expression can be recast in the form
\begin{equation}
\delta \lambda = \int_\Omega \nabla_U \lambda(x,y) \cdot \delta U_b(x,y)\, d\Omega,
\end{equation}
where $\nabla_U \lambda(x,y)$ is the local \emph{base-flow sensitivity field}.

\subsection{Definition of the Base-Flow Sensitivity Field}

After algebraic manipulations, the sensitivity field can be written as
\begin{equation}
\nabla_U \lambda(x,y) 
= - \frac{1}{\langle \phi^+, \phi \rangle}
\Big[ (\nabla \phi)\,\phi^{+*} + (\nabla \phi^{+*})\,\phi \Big].
\label{eq:sensitivity}
\end{equation}
Here
\[
\nabla \phi =
\begin{bmatrix}
\partial_x u & \partial_y u \\
\partial_x v & \partial_y v
\end{bmatrix}, \qquad
\nabla \phi^{+*} =
\begin{bmatrix}
\partial_x u^{+*} & \partial_y u^{+*} \\
\partial_x v^{+*} & \partial_y v^{+*}
\end{bmatrix}.
\]
Equation \eqref{eq:sensitivity} yields a vector field at every grid point 
$(x,y)$ indicating the sensitivity of the eigenvalue to local base flow 
perturbations.

\subsection{Implementation Details}

In practice, the following steps are required:
\begin{enumerate}
  \item \textbf{Compute direct and adjoint modes:} Solve the eigenvalue 
        problems for $\phi = (u,v)$ and $\phi^+ = (u^+,v^+)$ associated 
        with the eigenvalue of interest.
  \item \textbf{Compute spatial derivatives:} Evaluate the derivatives 
        $\partial_x$ and $\partial_y$ of both $\phi$ and $\phi^+$, 
        using spectral differentiation matrices (e.g.\ Fourier/Chebyshev) 
        or finite differences, depending on the discretization.
  \item \textbf{Assemble contractions:} At each grid point compute
        \begin{align*}
        \big[ (\nabla \phi)\,\phi^{+*} \big]_x &= (\partial_x u)\,u^{+*} + (\partial_y u)\,v^{+*}, \\
        \big[ (\nabla \phi)\,\phi^{+*} \big]_y &= (\partial_x v)\,u^{+*} + (\partial_y v)\,v^{+*}, \\
        \big[ (\nabla \phi^{+*})\,\phi \big]_x &= (\partial_x u^{+*})\,u + (\partial_y u^{+*})\,v, \\
        \big[ (\nabla \phi^{+*})\,\phi \big]_y &= (\partial_x v^{+*})\,u + (\partial_y v^{+*})\,v.
        \end{align*}
  \item \textbf{Normalize:} Divide by the overlap integral 
        $\langle \phi^+, \phi \rangle$ to ensure invariance with respect 
        to scaling of the modes.
  \item \textbf{Post-process:} Visualize $\nabla_U \lambda(x,y)$ as a vector 
        field, or its magnitude $|\nabla_U \lambda|$. The real part corresponds 
        to sensitivity of the growth rate, while the imaginary part corresponds 
        to sensitivity of the oscillation frequency.
\end{enumerate}

\begin{figure}
	\centering
	\includegraphics[width=0.75\linewidth]{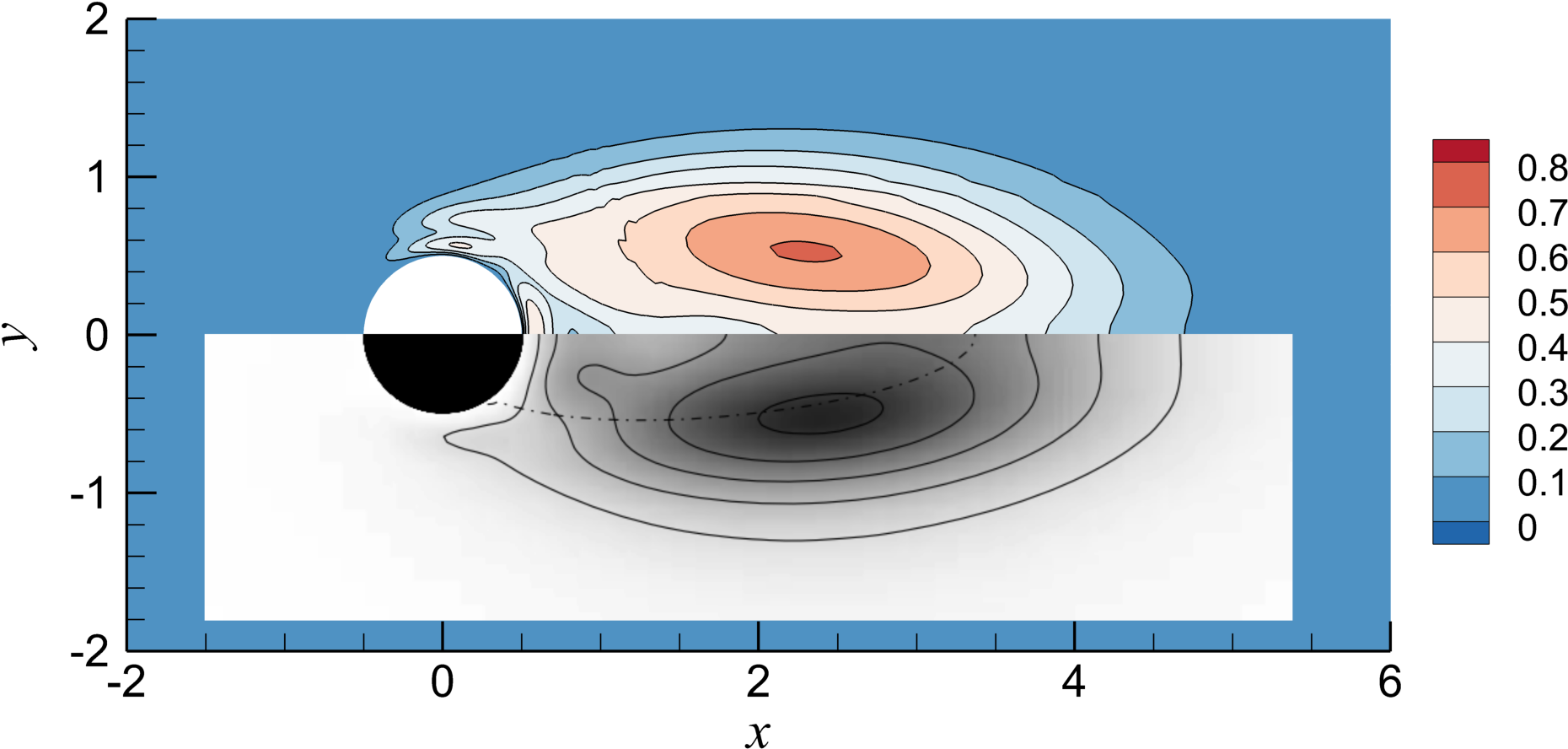}
	\caption{Comparison of the structural sensitivity field $|\nabla_U \lambda|$  against the wavemaker region $\zeta $ for 2-D cylinder flow at $Re = 50$: computational domain schematic from the present study (upper) versus reference results from Giannetti \& Luchini (2007) (lower). Nine uniformly spaced contour levels are shown in the range $0 \leq  |\nabla_U \lambda|\leq 0.8$ while six uniformly spaced contour levels are shown in the range $0 \leq  \zeta \leq 0.2$, with low (blue) to high (red) values. }	\label{fig:cylinder-region22222}
\end{figure}

\begin{figure}
	\begin{center}

		\subfigure[$C_{19}$ \label{figws:C19-subfig1}]{
			\resizebox*{4.4cm}{!}{\includegraphics{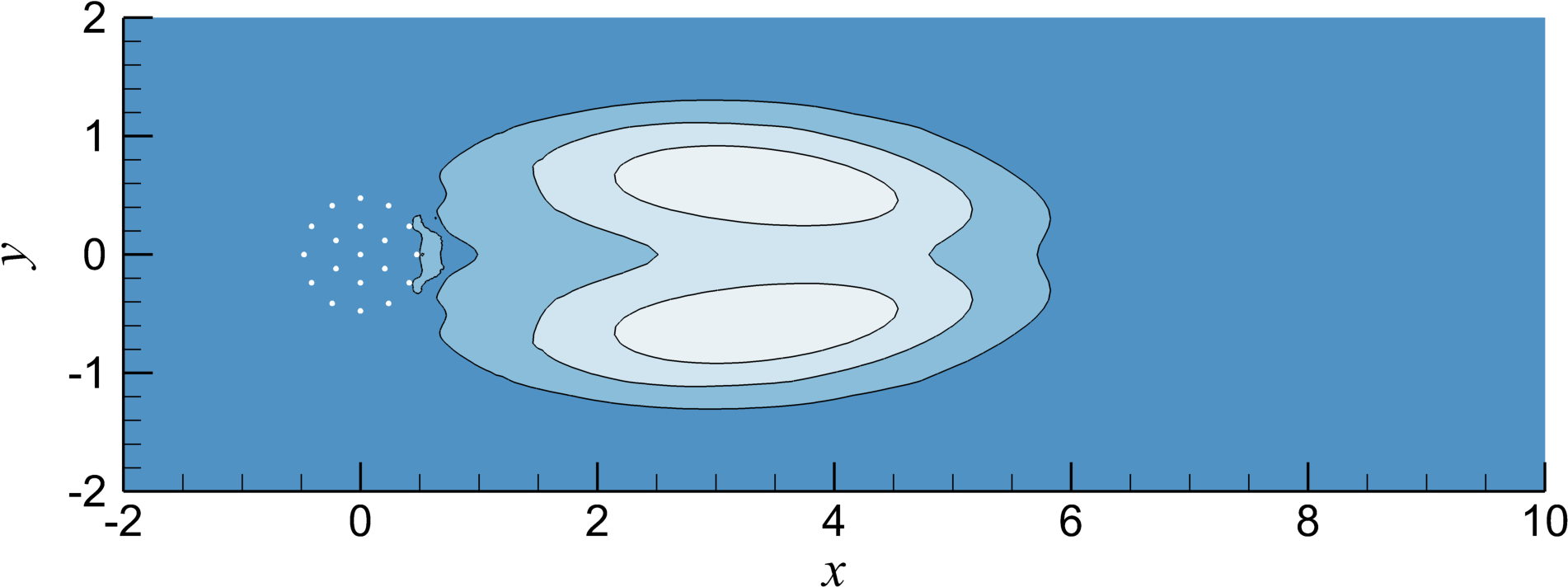}}}\hfill
		\subfigure[$C_{19}$ \label{figws:C19-subfig2}]{
			\resizebox*{4.4cm}{!}{\includegraphics{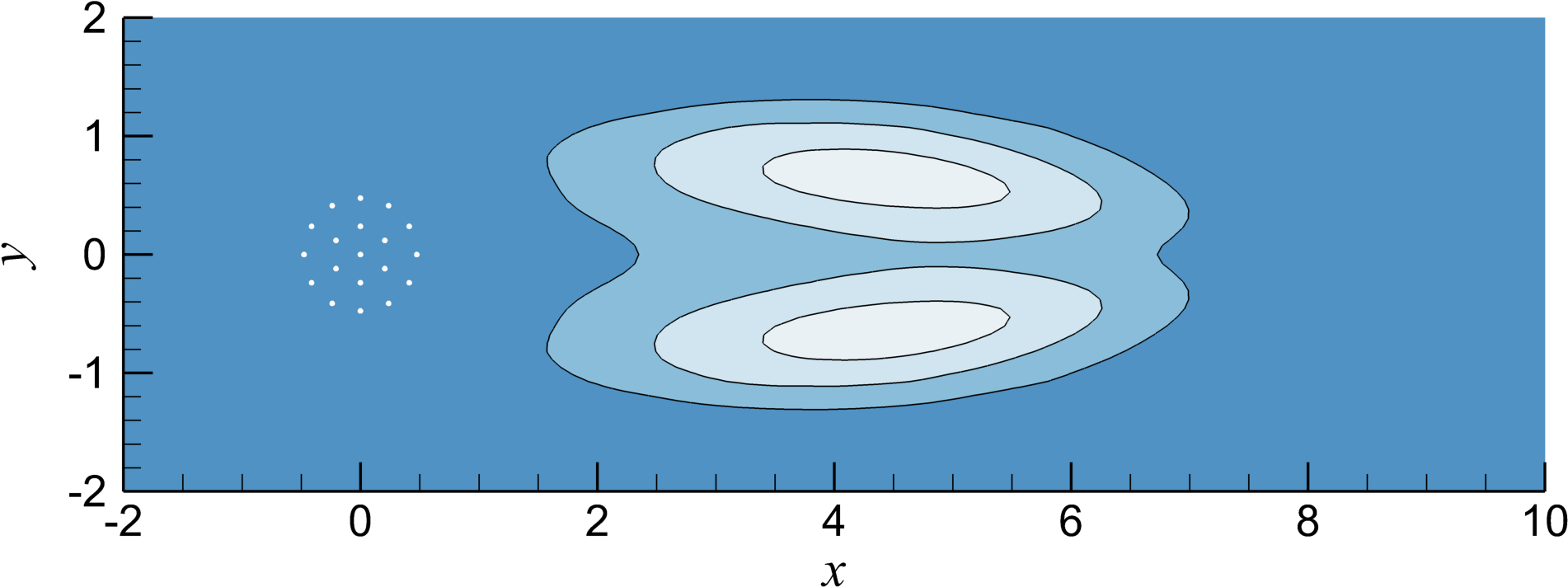}}}\hfill
		\subfigure[$C_{19}$ \label{figws:C19-subfig3}]{
			\resizebox*{4.4cm}{!}{\includegraphics{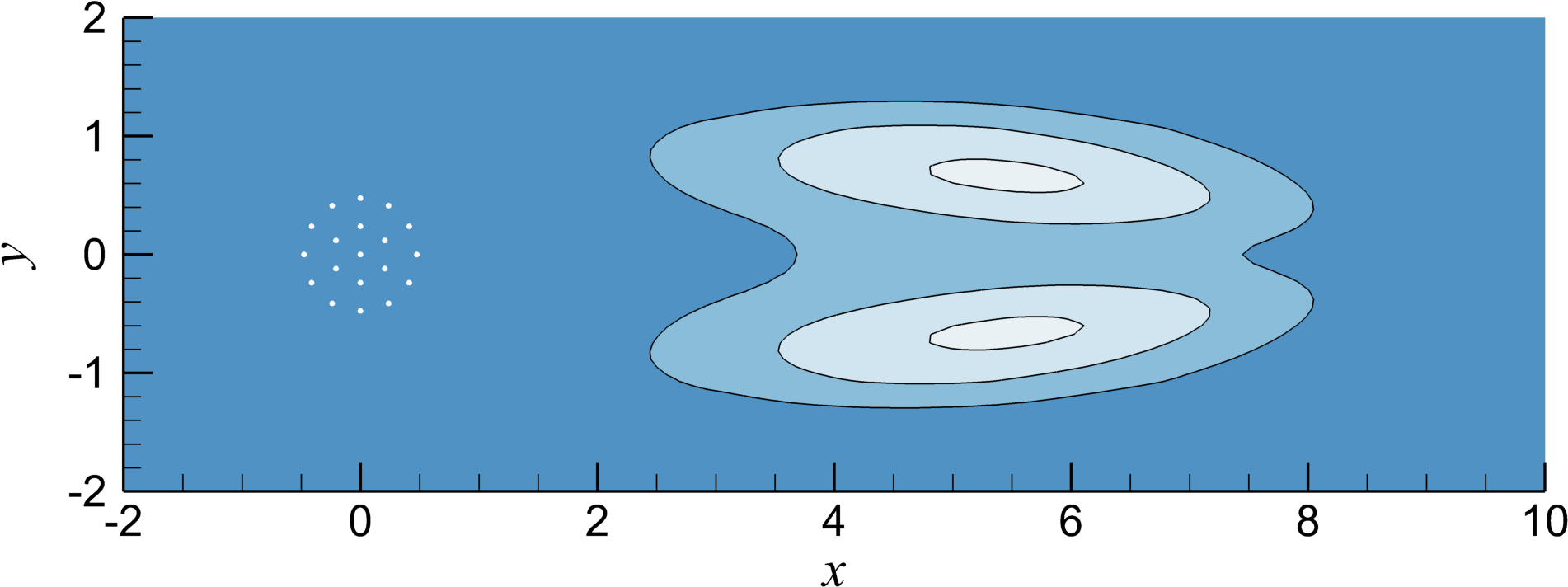}}}\\
			
		\subfigure[$C_{37}$ \label{figws:C37-subfig1}]{
			\resizebox*{4.4cm}{!}{\includegraphics{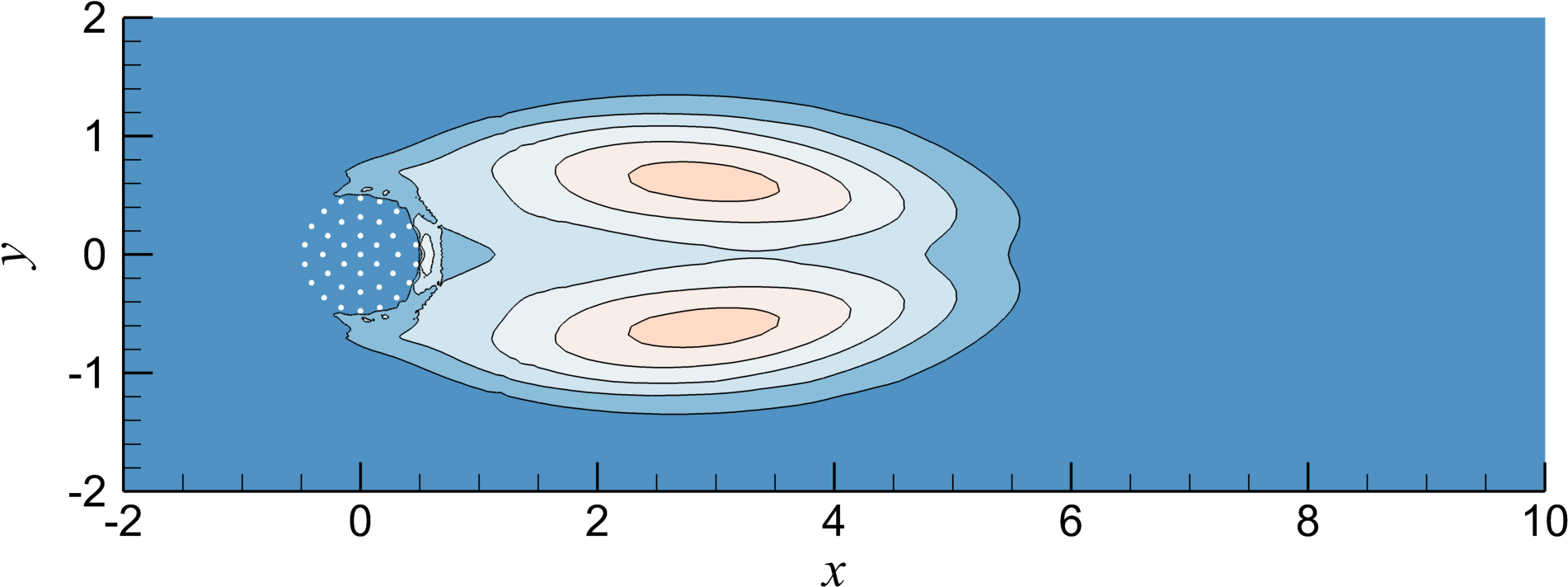}}}\hfill
		\subfigure[$C_{37}$ \label{figws:C37-subfig2}]{
			\resizebox*{4.4cm}{!}{\includegraphics{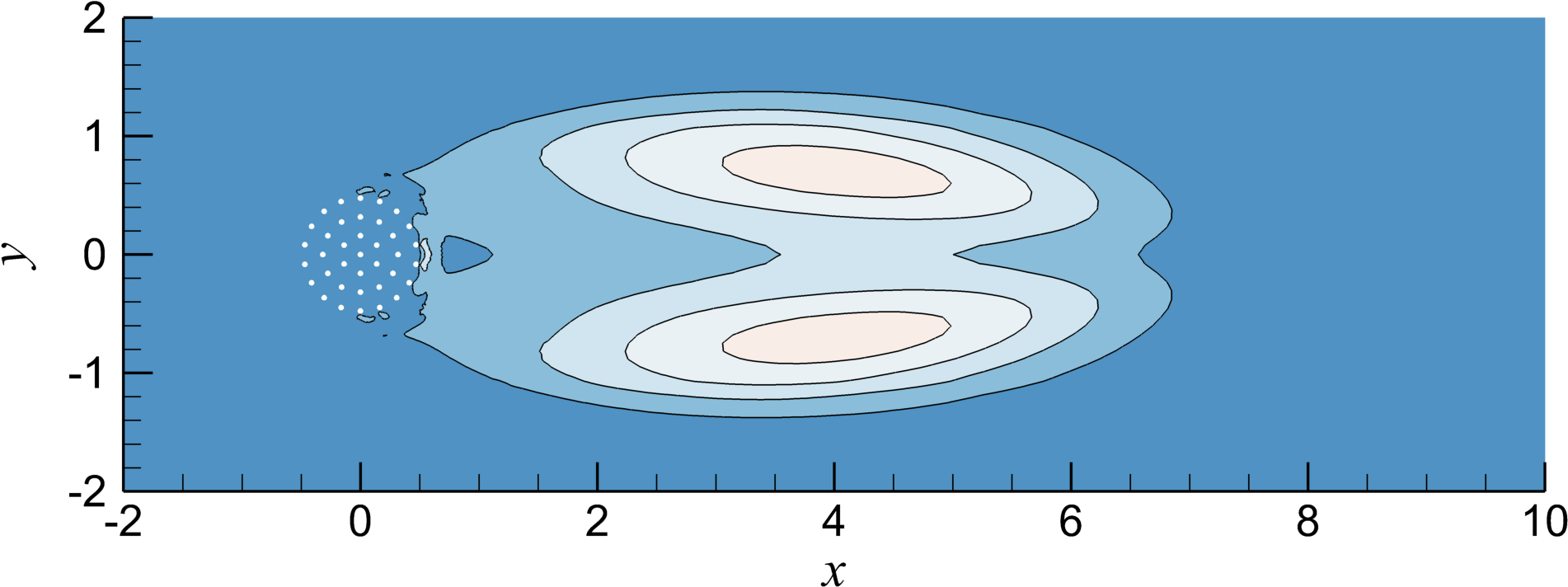}}}\hfill
		\subfigure[$C_{37}$ \label{figws:C37-subfig3}]{
			\resizebox*{4.4cm}{!}{\includegraphics{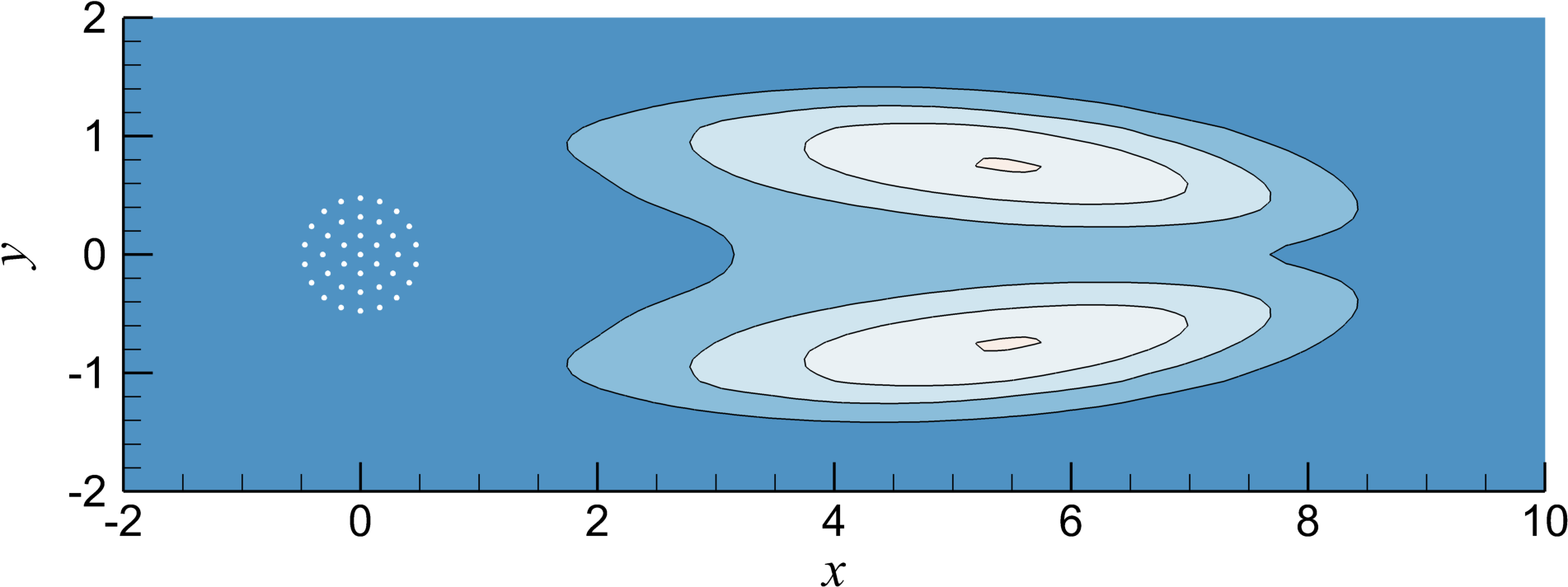}}}\\
			
		\subfigure[$C_{61}$ \label{figws:C61-subfig1}]{
			\resizebox*{4.4cm}{!}{\includegraphics{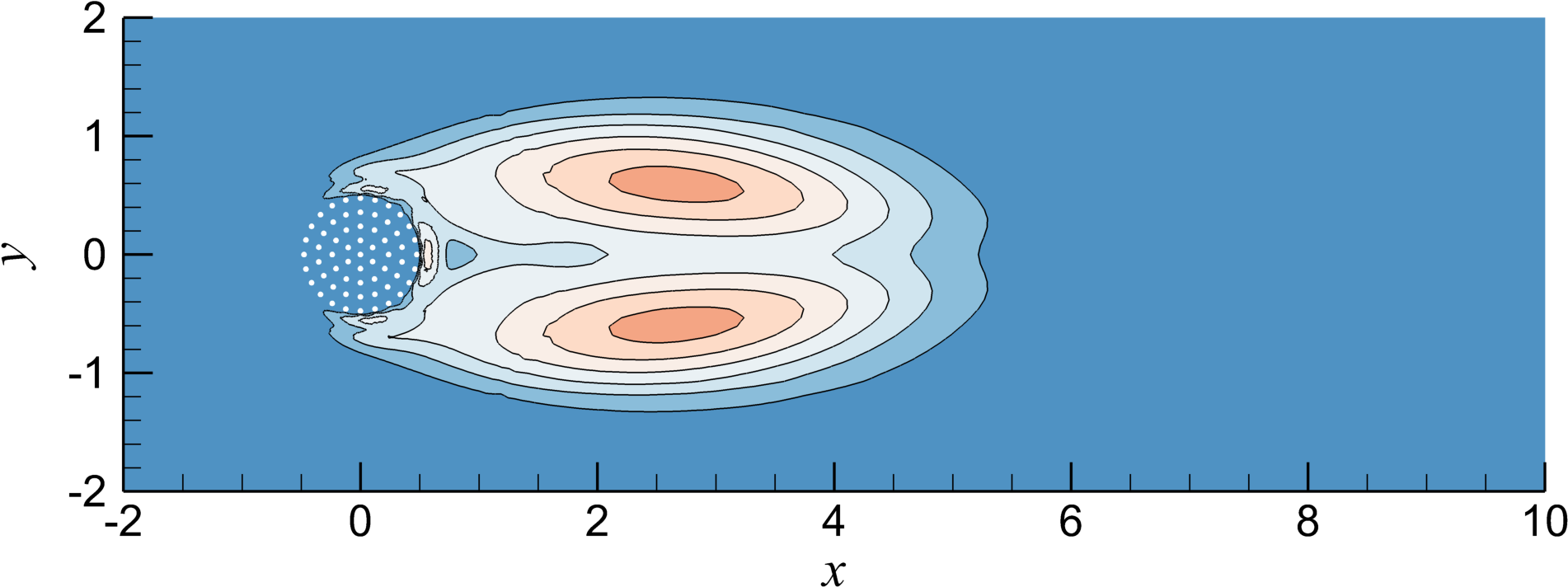}}}\hfill
		\subfigure[$C_{61}$ \label{figws:C61-subfig2}]{
			\resizebox*{4.4cm}{!}{\includegraphics{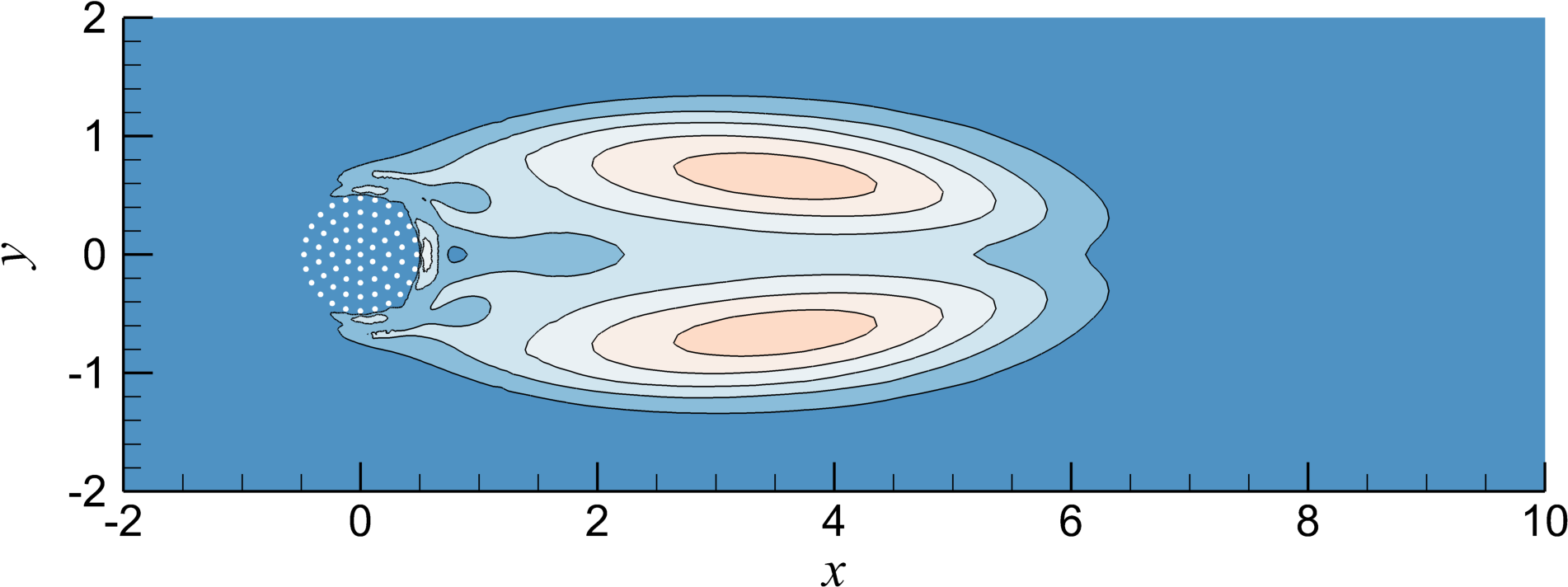}}}\hfill
		\subfigure[$C_{61}$ \label{figws:C61-subfig3}]{
			\resizebox*{4.4cm}{!}{\includegraphics{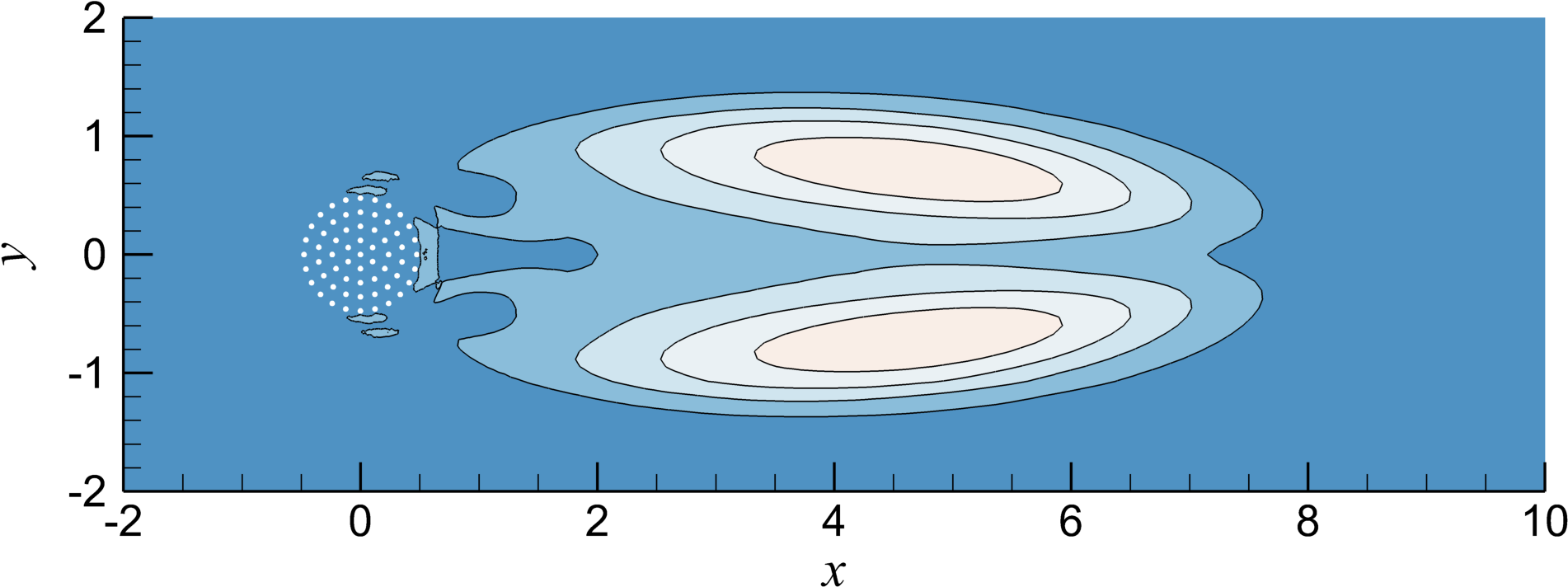}}}\\
			
		\subfigure[$C_{97}$ \label{figws:C97-subfig1}]{
			\resizebox*{4.4cm}{!}{\includegraphics{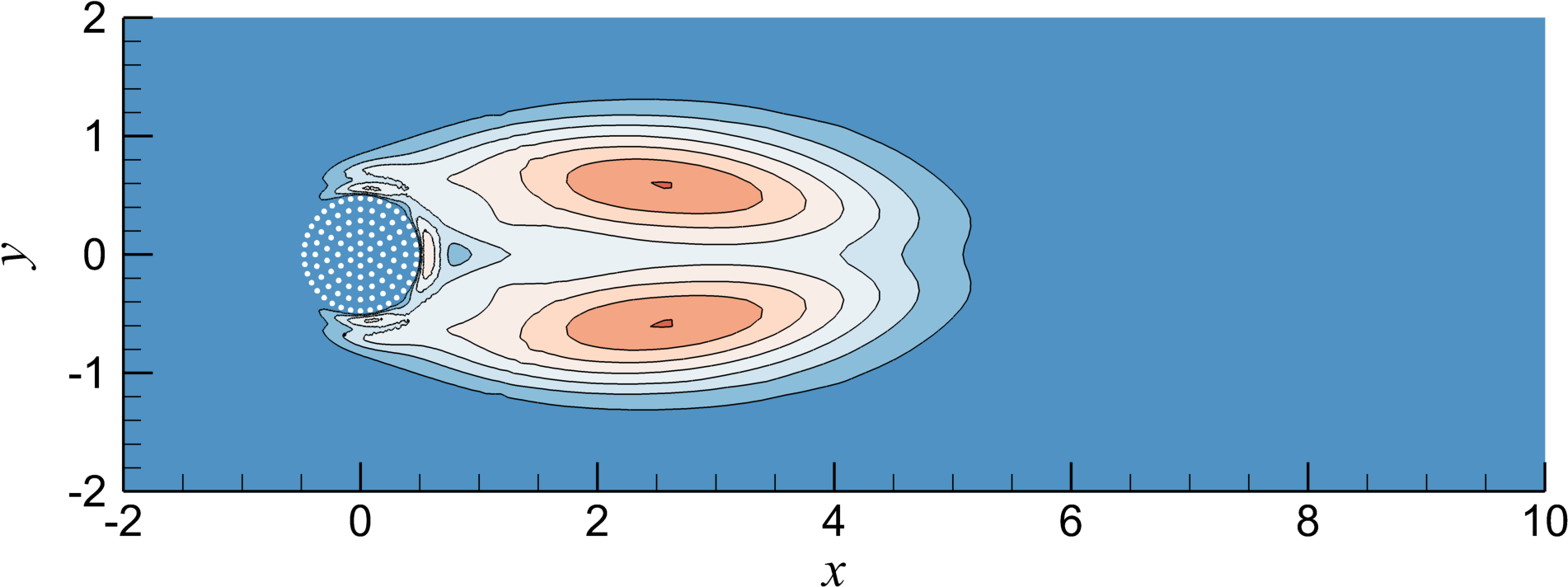}}}\hfill
		\subfigure[$C_{97}$ \label{figws:C97-subfig2}]{
			\resizebox*{4.4cm}{!}{\includegraphics{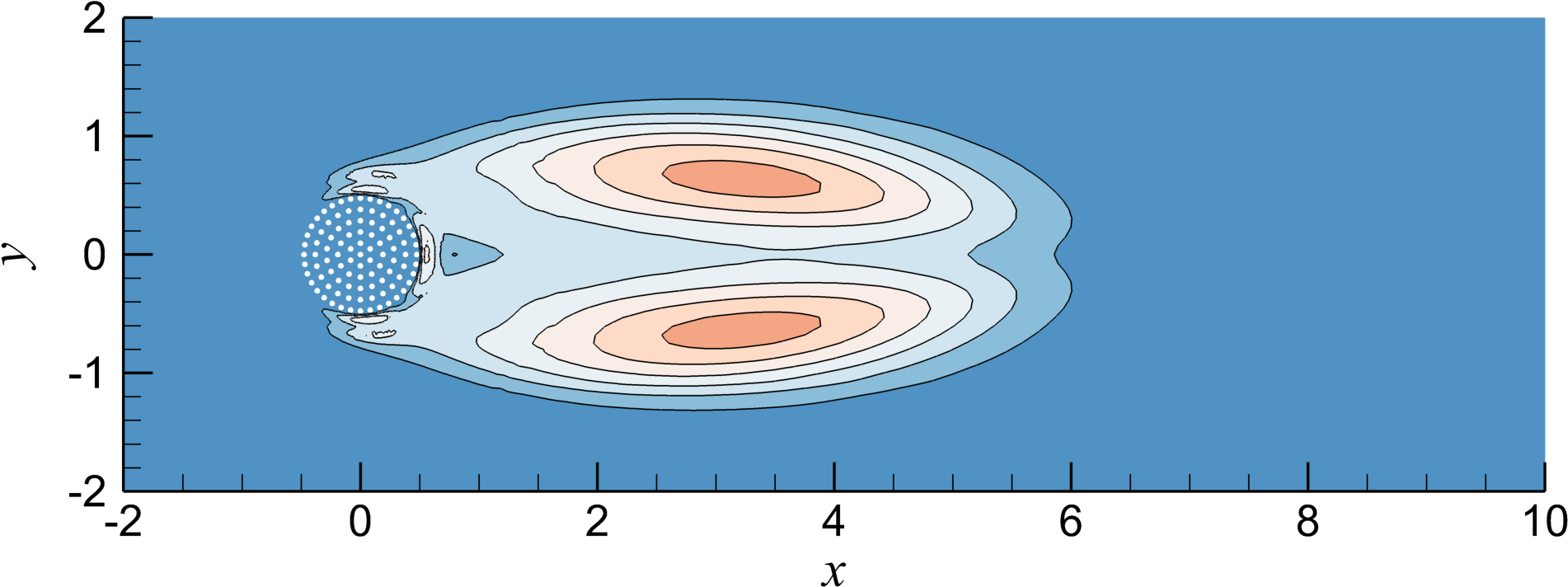}}}\hfill
		\subfigure[$C_{97}$ \label{figws:C97-subfig3}]{
			\resizebox*{4.4cm}{!}{\includegraphics{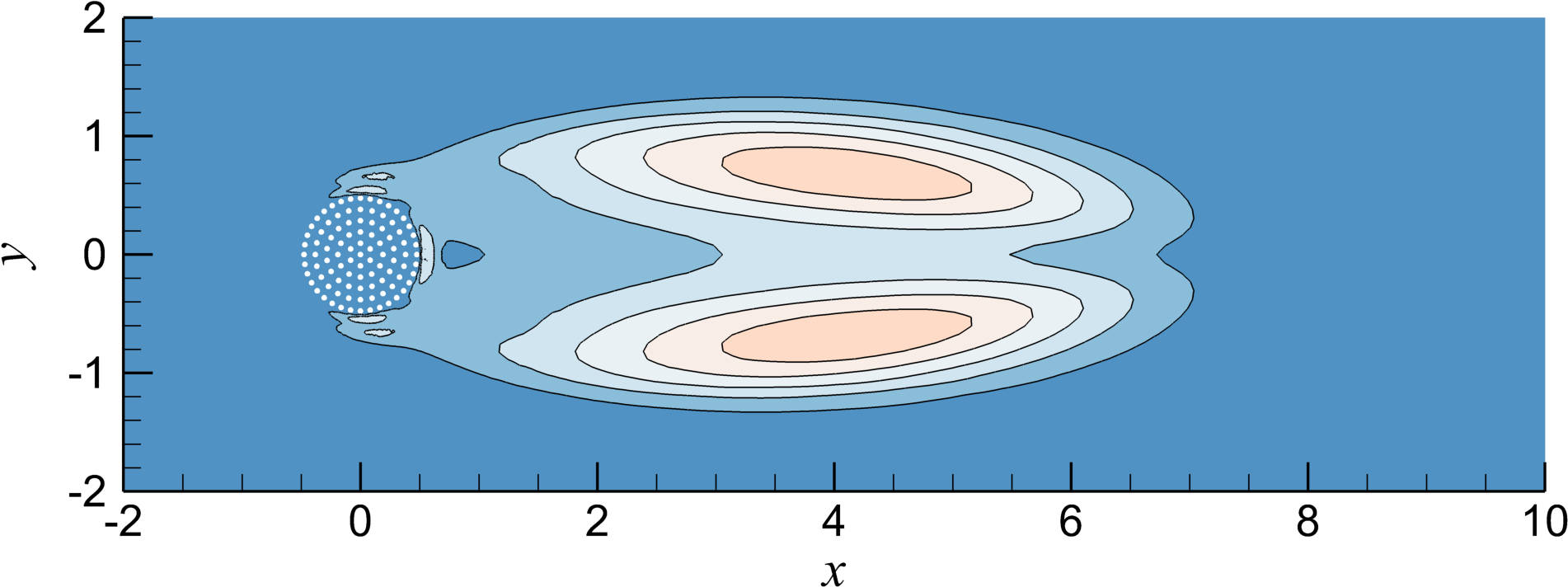}}}\\
			
		\subfigure[$C_{139}$ \label{figws:C139-subfig1}]{
			\resizebox*{4.4cm}{!}{\includegraphics{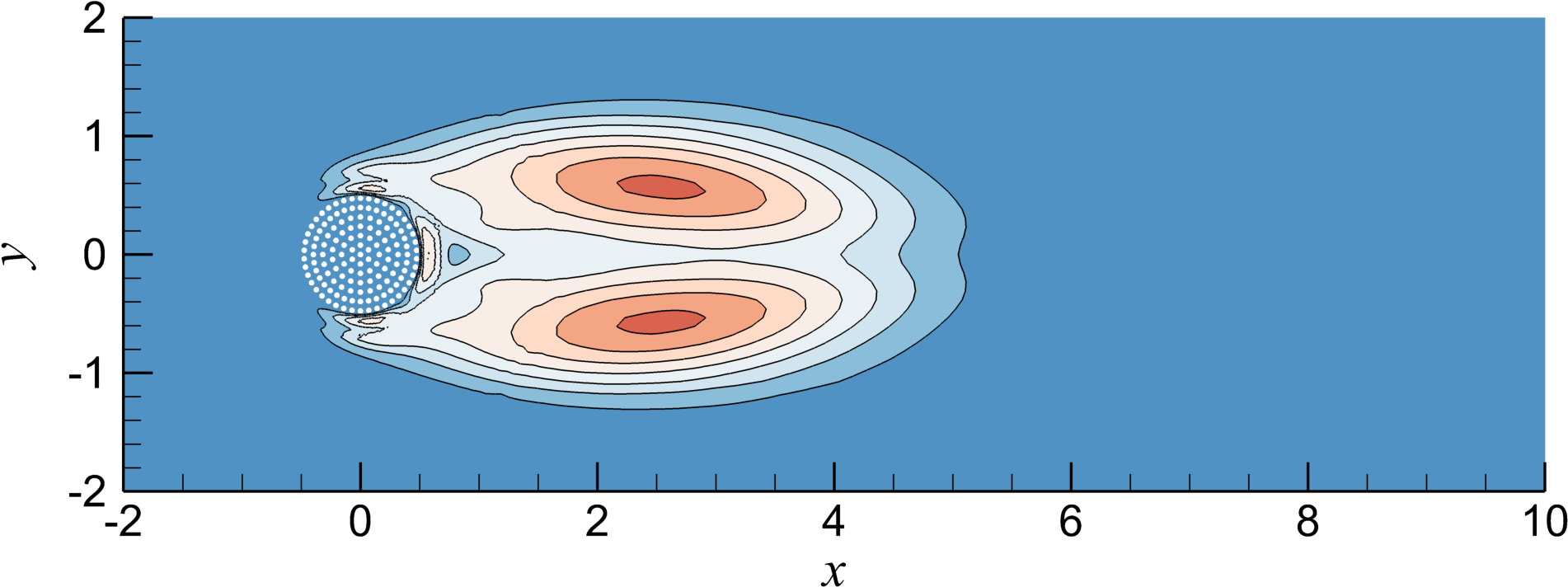}}}\hfill
		\subfigure[$C_{139}$ \label{figws:C139-subfig2}]{
			\resizebox*{4.4cm}{!}{\includegraphics{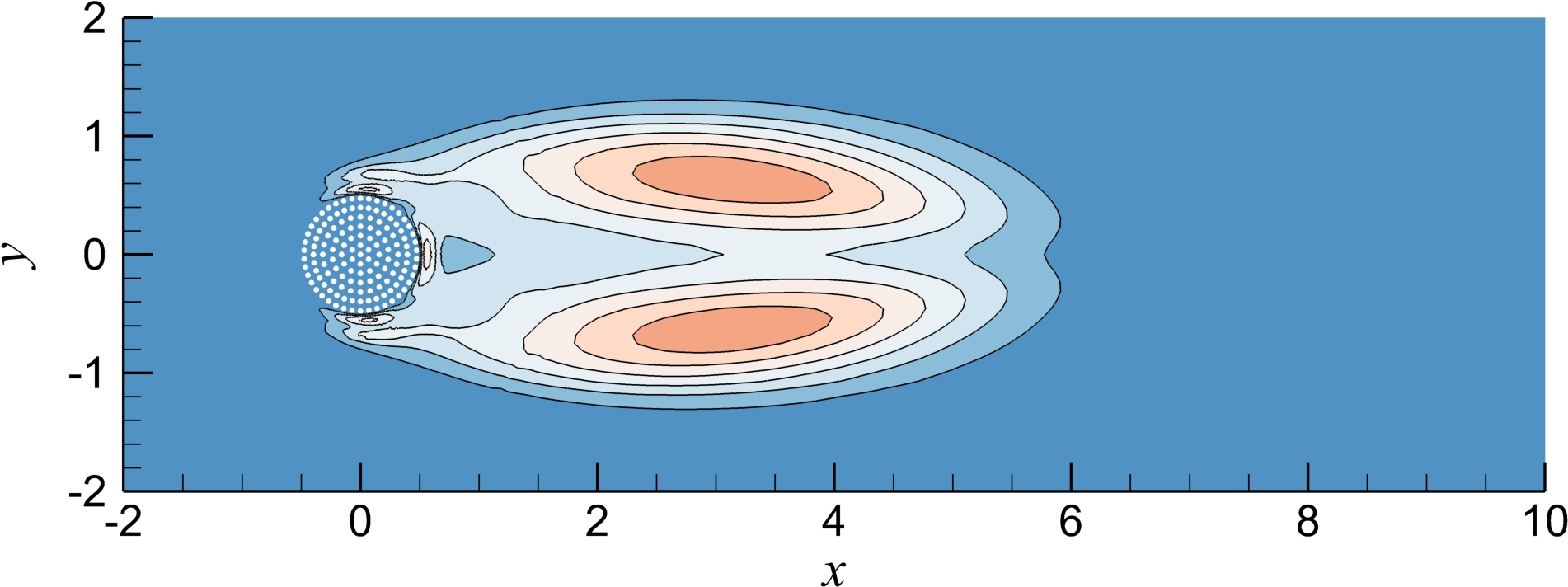}}}\hfill
		\subfigure[$C_{139}$ \label{figws:C139-subfig3}]{
			\resizebox*{4.4cm}{!}{\includegraphics{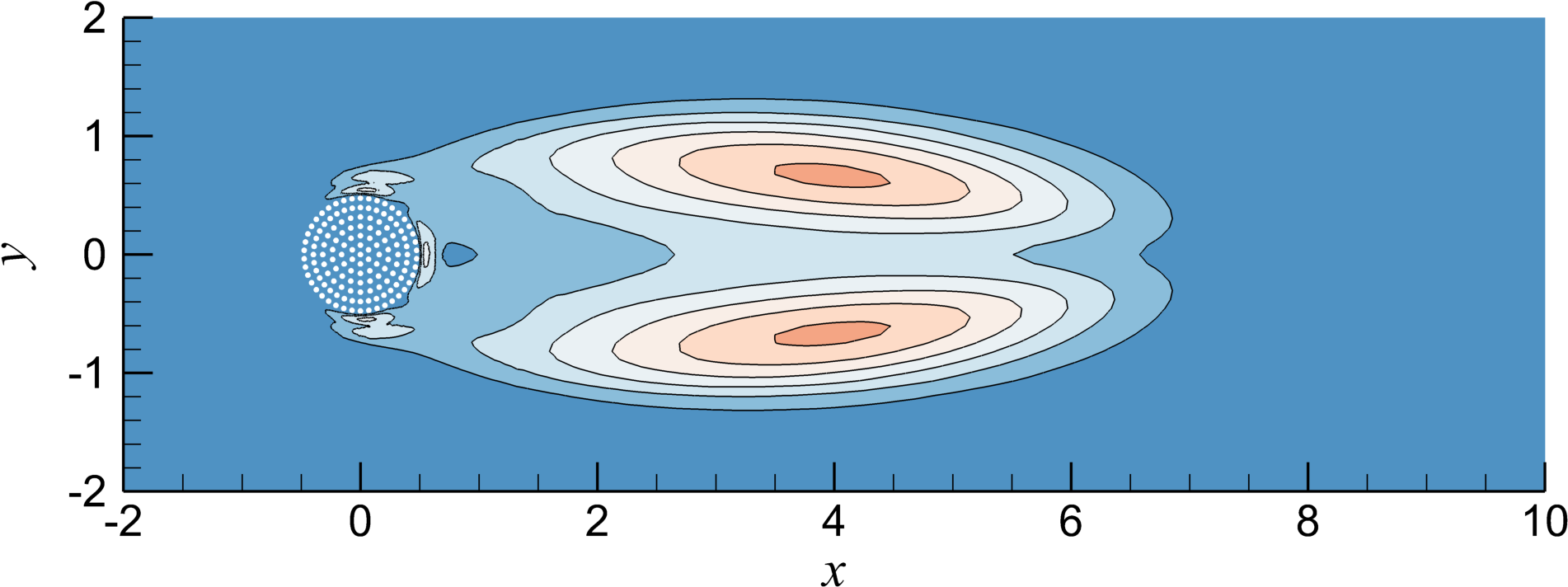}}}\\
			
		\subfigure[$C_{\text{solid}}$ \label{figws:solid-subfig1}]{
			\resizebox*{4.4cm}{!}{\includegraphics{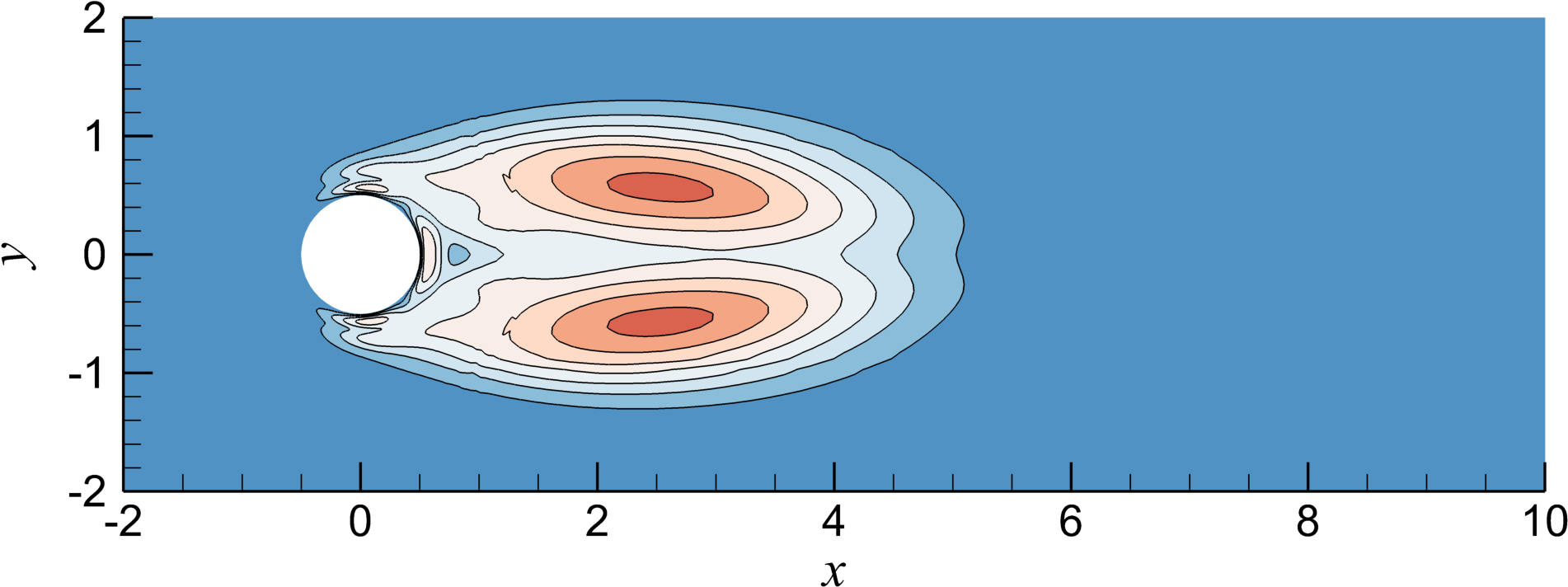}}}\hfill
		\subfigure[$C_{\text{solid}}$ \label{figws:solid-subfig2}]{
			\resizebox*{4.4cm}{!}{\includegraphics{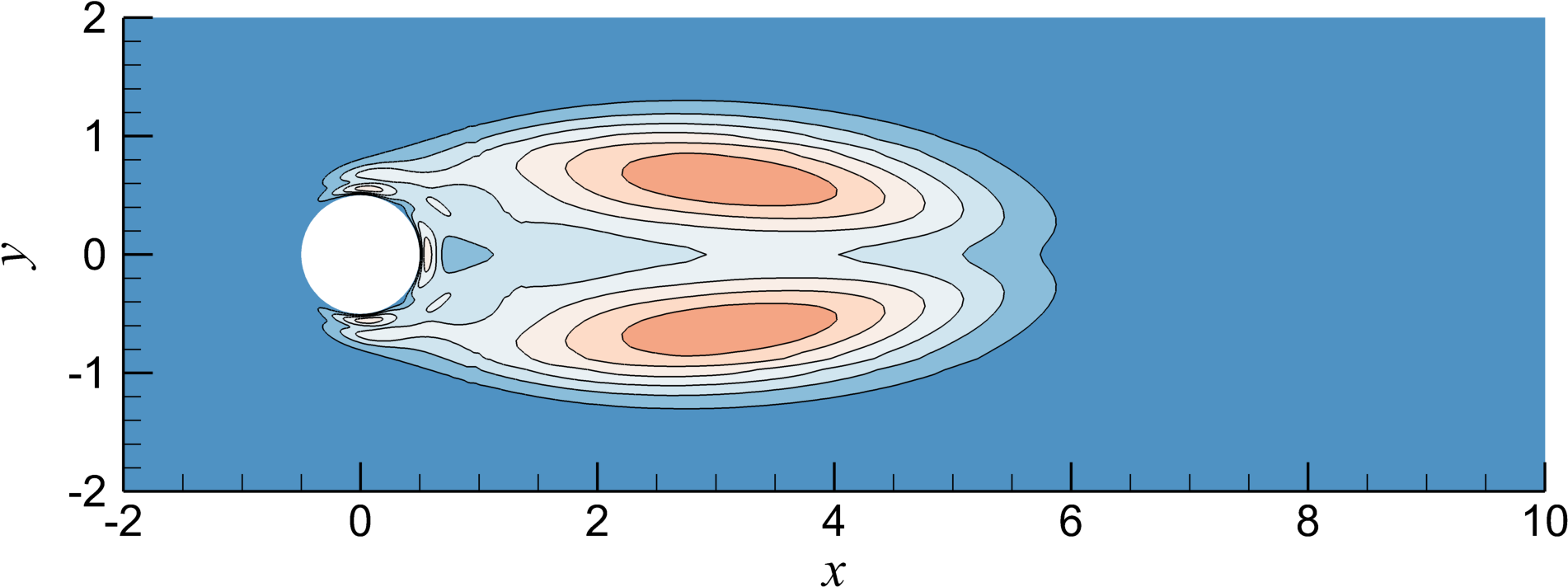}}}\hfill
		\subfigure[$C_{\text{solid}}$ \label{figws:solid-subfig3}]{
			\resizebox*{4.4cm}{!}{\includegraphics{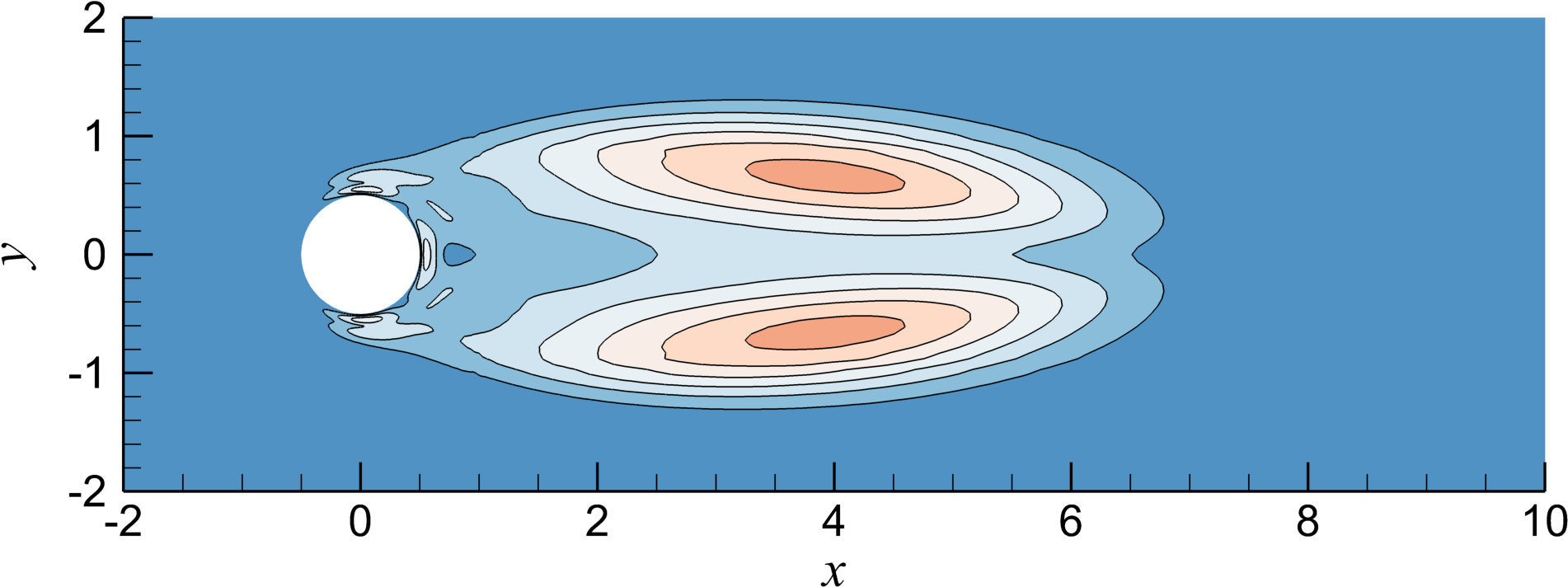}}}
		
\caption{\label{figws:Structural} Structural sensitivity field  $|\nabla_U \lambda|$  for two-dimensional flow past a porous cylinder array ($C_{19}$ to $C_{139}$) and a solid cylinder ($C_{\text{solid}}$) at Reynolds numbers $Re = 60, 80$, and $100$. Nine uniformly spaced contour levels are shown in the range $0 \leq \zeta \leq 0.8$, with low (blue) to high (red) values.}

	\end{center}
\end{figure}

\subsection{Physical Interpretation}

Regions of large $|\nabla_U \lambda|$ identify flow structures that are 
most influential in determining the stability characteristics.

\section{Conclusions} \label{Conclusions}
We conducted DNS and global linear stability analysis (LSA) of the flow around the circular cylinder array, based on mean flow and base flow, to analyse the instability and bifurcation in the flow past several 2-D cylinder arrays. 

We have determined the instability threshold at which the unsteadiness occurs using the global stability approach which has not been reported in the literature.

To identify the instability region responsible for the unsteadiness, we also probe the structural sensitivity of the flow. Our work will contribute to a understanding of the wake flow around cylinder array and provide insights into the dynamics of this flow configuration, which has practical implications for various engineering applications.

The results elucidate the fundamental link between discrete microstructure and the emergent stability properties of the array, providing a predictive framework for transition in multi-body systems relevant to offshore structures, wind farms, and vegetated flows.

\vspace{3 pt}

\section{Appendix A.}
In the present study, DNS are carried out for $Re$ up to 400.
The mesh, time step and domain independence tests are performed under $Re=300$. 

\subsection{The 2D mesh dependence study }
Several key parameters for the standard and refined meshes are listed in table 1.
\begin{table}
	\centering	
	\begin{tabular}{ccccc}
		Parameter & Value   \\ 
		Distance from the cylinder centre to the inlet   & $20D$ \\
		Distance from the cylinder centre to the cross flow boundaries   & $30D$ \\		
		Distance from the cylinder centre to the outlet   & $30D$ \\
		Number of nodes around the cylinder perimeter    & $132$ \\		
		Height of the first layer of mesh next to the cylinder    & $1\times 10^{-3}D$ \\				
		Cell expansion ration in the domain    & $\le 1.1$ \\						
	\end{tabular}
	\caption{Computational parameters for the reference mesh B1, following the configuration of \cite{Jiang2016} for flow past a single circular cylinder at $Re=300$.}	  
	\label{table1:G1pars}
\end{table}

\begin{table}
	\centering	
	\begin{tabular}{cclll}
		Case & $\delta t^*$ & $\overline{C_D}$~(Err$\%$) & $C_{L}'$~(Err$\%$) & $St$~(Err$\%$)   	\\ 
		Base mesh B1    & 0.001        & 1.371~(-0.41)                       & 0.6391~(-0.25)                       & 0.2102~(-0.22) \\
		\cite{Jiang2016}    &  0.00642   & 1.3768             & 0.6407      & 0.21068  \\ 				
	\end{tabular}
\caption{\label{tab:validation}
		Validation of the present numerical method for flow past a single circular cylinder at $Re=300$. Results obtained with the present solver on mesh B1 are compared against the benchmark data of \cite{Jiang2016}. Relative errors of the present results, calculated against \cite{Jiang2016}, are shown in parentheses.}
	\label{table2:G1}
\end{table}

\begin{table}
	\centering
	\small 
	\begin{tabular}{lllclllllll}
Case & Grid &  Blockage Ratio (\%) & $L_y$ & $L_{x1}$ & $L_{x2}$  &$\overline{C_D}$ (Err$\%$) & $C_{L}'$ (Err$\%$) & $St$ (Err$\%$)  & $G_R$\\ 		
1 &B1 & 1.67 & 30  & 20 & 30  & 1.371~(1.82)~ & 0.6391~(2.03)~   & 0.2102~(1.23)~  & -7.78E-04 \\
2 &S1 & 0.83 & 60  & 30 & 60  & 1.356~(0.66)~ & 0.6323~(0.95)~   & 0.2084~(0.35)~   & -1.06E-03 \\
3 &S2 & 0.42 & 120 & 40 & 120 & 1.350~(0.24)~ & 0.6284~(0.32)~   & 0.2079~(0.13)~   & -9.05E-04 \\
4$^*$ &S3 & 0.21 & 240 & 50 & 240 & 1.347     & 0.6264       & 0.2076       & -1.25E-03        
	\end{tabular}
\caption{\label{tab:domain_size_study}
		Domain and blockage ratio independence study for the solid cylinder at $Re=300$. The computational domain size is defined by the upstream length $L_{x1}$, downstream length $L_{x2}$, and spanwise width $L_y$, all normalized by the cylinder diameter. Case 4 (Grid S3) with the largest domain and lowest blockage (0.21\%) is used as the reference for error calculation.}
\end{table}

\subsubsection{Time step independence tests } 

\begin{table}

	\centering	
	\begin{tabular}{llllll}
		Case & $\delta t^*$ & $\overline{C_D}$ (Err$\%$) & $C_{L}'$ (Err$\%$) & $St$ (Err$\%$)   &$G_R$ (Err$\%$)	\\ 
1   &0.002     &1.345 (-0.11 ) 	&0.6231 (-0.52) &0.2074 (-0.12) &-1.60E-04 \\
2$^*$ &0.001     &1.347          	&0.6264 	    &0.2076 	    &-1.25E-03 \\
3   &0.0005   &1.347 (0.05 ) 	    &0.6276 (0.20) 	&0.2079 (0.11) 	&-1.45E-03 \\		
	\end{tabular}
	\caption{\label{tab:time_step_study}
		Time-step independence study for the solid cylinder array at $Re=300$ on mesh $S_3$. Case 2 ($\delta t^* = 0.001$) is the chosen baseline, with relative errors computed against it.}
\end{table}



\begin{figure}
	\centering
	\includegraphics[width=10 cm]{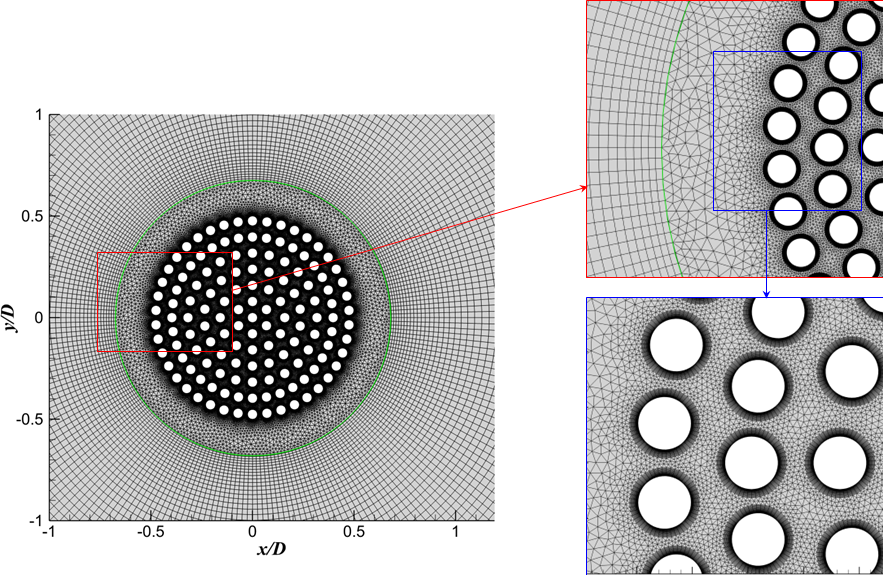}
	\caption{ \label{fig:g2fig}
Hybrid mesh for the $C_{139}$ cylinder array. The complex internal region is discretized using a hybrid mesh with quadrilateral elements resolving the boundary layers around the cylinders and triangular elements filling the interstitial regions.}
\end{figure}

\begin{table}
	\centering
	\small 
	\begin{tabular}{lllllll}
Case & Grid  & Max Edge Length &$\overline{C_D}$ (Err$\%$) & $C_{L}'$ (Err$\%$) & $St$ (Err$\%$)  & $G_R$\\ 		
1 & A1 & 1.5  & 0.034 (-0.12)  & 0.0149 (-0.26) & 0.2082 (-0.02) & -1.75E-03 \\
2$^*$ & A2 & 1.0    & 0.034   & 0.0149   & 0.2082   & -8.84E-04 \\
3 & A3 & 0.6  & 0.034 (0.22)   & 0.0150 (0.45)  & 0.2082 (-0.02) & -4.92E-04 \\       
	\end{tabular}
\caption{\label{tab:mesh_study}	
Grid independence study for the $C_{139}$ cylinder array at $Re=300$. Case 2 (A2) is the chosen baseline, with relative errors computed against it. The non-dimensional time step is $\delta t^* = 0.001$.}	 
\end{table}

\begin{figure}
	\centering
	\includegraphics[width=0.75\linewidth]{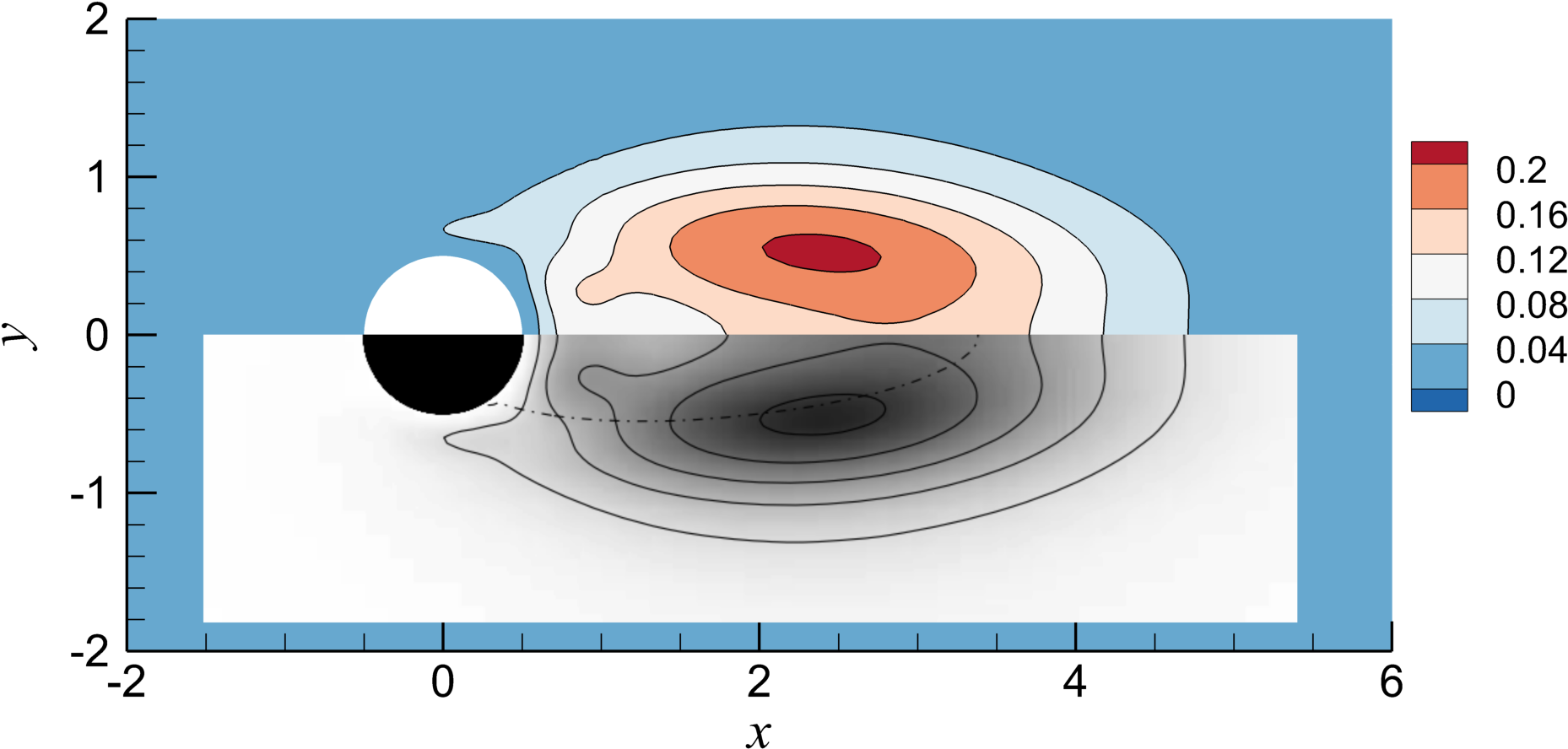}
	\caption{Comparison of the wavemaker region $\zeta$ for 2-D cylinder flow at $Re = 50$: computational domain schematic from the present study (upper) versus reference results from Giannetti \& Luchini (2007) (lower).}	\label{fig:cylinder-region22}
\end{figure}

Among all cases, the relative errors of the quantities for each case are well within $0.6\%$, which demonstrates that further increase of the domain size and wake resolution has negligible effect on the results. The value of the growth rate is close to zero at $Re = 300$. The error in $G_R$ appears more sensitive to the domain size than the other variables. The above results suggest that the mesh $S3$ is adequate for the present study with $Re\le 300$.

%

%

\begin{acknowledgments}
\end{acknowledgments}

\bibliographystyle{jfm}
\bibliography{BibRef}

\end{document}